\documentclass[11pt]{cernyrep}
\usepackage{graphicx,epsfig}
\graphicspath{{./Pics/}}
\usepackage{amsmath}
\usepackage{amssymb}
\usepackage{wasysym}

\usepackage{cite}

\def\GeV{\textnormal{GeV}}
\def\bma#1{\mbox{\boldmath{$#1$}}}
\def\Refs#1{Refs.~\cite{#1}}
\def\Ref#1{Ref.~\cite{#1}}
\def\Eq#1{Eq.~(\ref{#1})}
\def\Fig#1{Fig.~\ref{#1}}
\def\ttjet{\ensuremath{t\bar t +\rm 1-Jet+X}}
\def\sigmattjet{\ensuremath{\sigma_{\ttjet}}}
\def\mt{\ensuremath{m_t}}
\def\rhos{\ensuremath{\rho_s}}

\def\as{\ensuremath{\alpha_s}}
\def\GeV{\mbox{GeV}}

\newcommand{\QTY}[2]{\mbox{\(#1\rm\,#2\)}}

\providecommand{\alps}{\ensuremath{\alpha_s}\xspace}
\providecommand{\alpsmz}{\ensuremath{\alpha_s(M_Z)}\xspace}
\providecommand{\alpsq}{\ensuremath{\alpha_s(Q)}\xspace}
\providecommand{\chisq}{\ensuremath{\chi^2}\xspace}
\providecommand{\chisqndof}{\ensuremath{\chi^2/n_\mathrm{dof}}\xspace}

\providecommand{\rbthm}{\rule[-2ex]{0ex}{5ex}}
\providecommand{\rbtrr}{\rule[-0.8ex]{0ex}{3.2ex}}

\begin{document}
\noindent
MITP/14-036
\hfill
May 2014 \\
\vspace{0.5cm}

\title{High precision fundamental constants at the TeV scale\thanks{
Proceedings of the 
Mainz Institute for Theoretical Physics (MITP) scientific program on 
{\it High precision fundamental constants at the TeV scale}, March 10-21, 2014. 
}}

\author{
  S.~Moch$^{1,2}$, 
  S.~Weinzierl$^3$, 
  S.~Alekhin$^{2,4}$,
  J.~Bl\"umlein$^2$,
  L.~de la Cruz$^5$,
  S.~Dittmaier$^6$,
  M.~Dowling$^{2}$, 
  J.~Erler$^{3,5}$,
  J.R.~Espinosa$^{7,8}$,
  J.~Fuster$^9$,
  X.~Garcia~i~Tormo$^{10}$, 
  A.H.~Hoang$^{11}$,
  A.~Huss$^6$,
  S.~Kluth$^{12}$,
  M.~Mulders$^{13}$,
  A.S.~Papanastasiou$^{14}$,
  J.~Piclum$^{15,16}$,
  K.~Rabbertz$^{17}$,
  C.~Schwinn$^6$,
  M.~Schulze$^{13}$,
  E.~Shintani$^3$,
  P.~Uwer$^{18}$,
  N.~Zerf$^{19}$
}

\institute{
{\footnotesize
  $^1$II. Institut f\"ur Theoretische Physik, Universit\"at Hamburg, Luruper Chaussee 149, D--22761 Hamburg, Germany \\
  $^2$Deutsches Elektronensynchrotron DESY, Platanenallee 6, D--15738 Zeuthen, Germany \\
  $^3$PRISMA Cluster of Excellence, Mainz Institute for Theoretical Physics, Johannes Gutenberg-Universit\"at, D--55099 Mainz, Germany \\
  $^4$Institute for High Energy Physics, 142281 Protvino, Moscow region, Russia \\
  $^5$Departamento de F\'isica Te\'orica, Instituto de F\'isica, Universidad Nacional Aut\'onoma de M\'exico, M\'exico D.F. 04510, M\'exico \\
  $^6$Albert-Ludwigs Universit\"at Freiburg, Physikalisches Institut, D--79104 Freiburg, Germany \\
  $^7$ICREA, Instituci\'{o} Catalana de Recerca i Estudis Avan\c{c}ats, Barcelona, Spain \\
  $^8$IFAE, Universitat Aut\`{o}noma de Barcelona, Bellaterra E--08193, Barcelona, Spain \\
  $^9$IFIC, Universitat de Val\`encia and CSIC, Catedr\'atico Jose Beltr\'an 2, E--46980 Paterna, Spain \\
  $^{10}$Albert Einstein Center for Fundamental Physics, Institut f\"ur Theoretische Physik, Universit\"at Bern, Sidlerstra{\ss}e 5, CH--3012 Bern, Switzerland \\
  $^{11}$Fakult\"at f\"ur Physik, Universit\"at Wien, Boltzmanngasse 5, A--1090 Vienna, Austria \\
  $^{12}$Max-Planck-Institut f\"ur Physik, F\"ohringer Ring 6, D--80805 M\"unchen, Germany \\
  $^{13}$CERN, PH-TH, CH--1211 Geneva 23, Switzerland \\
  $^{14}$Deutsches Elektronensynchrotron DESY, Notkestra{\ss}e 85, D--22607 Hamburg, Germany \\
  $^{15}$Institut f\"ur Theoretische Teilchenphysik und Kosmologie, RWTH Aachen, D--52056 Aachen, Germany \\
  $^{16}$Physik Department T31, James-Franck-Stra\ss{}e~1, Technische Universit\"at M\"unchen, D--85748 Garching, Germany \\
  $^{17}$Karlsruher Institut f\"ur Technologie, Institut f\"ur Experimentelle Kernphysik, Campus S\"ud, Postfach 6980, D-76128 Karlsruhe, Germany \\
  $^{18}$Humboldt-Universit\"at zu Berlin, Institut f\"ur Physik, Newtonstra{\ss}e 15, D--12489 Berlin, Germany \\
  $^{19}$Department of Physics, University of Alberta, Edmonton AB T6G 2J1, Canada
}
}

\maketitle

\begin{abstract}

This report summarizes the proceedings of the 2014 Mainz Institute for Theoretical Physics (MITP) scientific program on 
{\it High precision fundamental constants at the TeV scale}.
The two outstanding parameters in the Standard Model dealt with during the MITP scientific program 
are the strong coupling constant $\alpha_s$ and the top-quark mass $m_t$.
Lacking knowledge on the value of those fundamental constants 
is often the limiting factor in the accuracy of theoretical predictions.
The current status on $\alpha_s$ and $m_t$ has been reviewed and 
directions for future research have been identified.

\end{abstract}

\newpage
\tableofcontents

\newpage
\chapter{The strong coupling constant}

\section{Summary on $\alpha_s$}

The strong coupling $\alpha_s$ is a fundamental parameter of quantum chromodynamics.
Its numerical value affects many cross sections at the Large Hadron Collider
and in consequence our current uncertainty on the numerical value
has a direct impact on precision tests of the Standard Model and on the interpretation of potential 
deviations in experimental measurements in terms of new physics phenomena.

\noindent The strong coupling can be extracted from various measurements.
Examples are
\begin{itemize}
\item Deep inelastic scattering
\item Hadronic $Z$-decays
\item Event shapes and jets in electron-positron annihilation
\item Jets in hadron-hadron collisions
\item Hadronic $\tau$-decays
\item Heavy quarkonia decay
\item Lattice QCD
\end{itemize}
Most of these methods have been discussed at the workshop.
The numerical value of $\alpha_s$ is determined by comparing experimental data from measurements like in the list above 
with the corresponding theoretical predictions including perturbative and non-perturbative corrections.
In this way the error on the theory predictions translates into an error on $\alpha_s$.
In the past years, tremendous progress in higher-order perturbative corrections has been made, resulting in
very small theoretical uncertainties.
We are now facing a situation, where individual determinations of $\alpha_s$ report 
a numerical value of the strong coupling with a rather small error,
however different determinations of $\alpha_s$ are, at best, marginally compatible within their errors.
This phenomena is seen inside a specific method for the extraction of $\alpha_s$ from the list above
as well as by comparing the extracted numerical value from different methods.
The potential sources of these discrepancies have been discussed intensively at the workshop.
In some cases the origins of the discrepancies can be identified, in particular if one compares the extraction of $\alpha_s$ from the same measurements.
It is then possible to decide, if one analysis is correct, while the other one is not, or if both analyses are valid and the numerical difference reflects
our ignorance about unknown contributions.
In the first case one would take the correct analysis to contribute to the world average of $\alpha_s$ (and exclude the incorrect one), 
while in the second case one concludes that the uncertainties in both analyses have been underestimated and both analyses should contribute
with a larger error to the world average of $\alpha_s$.
In the latter case this corresponds to the range averaging procedure used by the PDG group.
In the former case the current practice is to use range averaging as well.
However this procedure will punish a correct analysis with a larger error if a competing group does something wrong.
This situation can be improved by scientific discussions among the involved groups, identifying the causes of discrepancies 
and correcting -- if need be -- erroneous results.
In the pursuit of getting to a more precise value of $\alpha_s$ this is a worthwhile effort.
In the discussions of the workshop it emerged that the appropriate format for these discussions, comparisons and reviews would be a working group on $\alpha_s$.
The participants therefore recommend after the successful Munich workshop~\cite{Bethke:2011tr} on ``Precision Measurements of $\alpha_s$'' in 2011, the current workshop
on ``High Precision Fundamental Constants at the TeV Scale'' at Mainz in 2014 to continue the effort within a working group on $\alpha_s$.

\section*{2~~~~\boldmath $\alpha_s(M_Z)$ at NNLO and N$^3$LO~\footnote{J.~Bl{\"u}mlein}}
\addcontentsline{toc}{section}{\protect\numberline{2}{\boldmath $\alpha_s(M_Z)$ at NNLO and N$^3$LO}}
\vspace*{1mm}
\noindent
Precision determinations of parton distribution functions  (PDFs) and $\alpha_s(M_Z^2)$ 
are currently being performed at NNLO ($\alpha_s^3$). NLO fits suffer from scale uncertainties 
being of $O(\pm 5\%)$  \cite{Blumlein:1996gv} and are therefore too large compared to the precision of deep-inelastic data. The 
heavy flavor corrections are available at NLO and threshold corrections and the calculation of the NNLO corrections is making 
progress, cf.~\cite{Behring:2013dga}. 
\begin{center}
\begin{tabular}{|l|l|l|}
\hline
\multicolumn{1}{|c|}{ } &
\multicolumn{1}{c|}{$\alpha_s({M_Z^2})$} &
\multicolumn{1}{c|}{  } \\
\hline
Alekhin [2001]  & $0.1143 \pm 0.013$ & DIS  \cite{Alekhin:2001ih}\\
BBG [2004]      & $0.1134 {\tiny{\begin{array}{c} +0.0019 \\
           -0.0021 \end{array}}}$
         & {\rm valence~analysis, NNLO}  \cite{Blumlein:2004ip,Blumlein:2006be}
\\
GRS      & $0.112 $ & {\rm valence~analysis, NNLO}  \cite{Gluck:2006yz}
\\
ABKM           & $0.1135 \pm 0.0014$ & {\rm HQ:~FFNS~$N_f=3$} \cite{Alekhin:2009ni}
\\
JR14       & $0.1136 \pm 0.0004$ & {\rm dynamical~approach} \cite{Jimenez-Delgado:2014twa}
\\
JR14       & $0.1162 \pm 0.0006$ & {\rm including NLO-jets}  \cite{Jimenez-Delgado:2014twa}
\\
MSTW & $0.1171\pm 0.0014$ &  (2009) \cite{Martin:2009bu}
\\
Thorne & $0.1136$ &  [DIS+DY, HT$^*$] (2014) \cite{Thorne:2014toa}
\\
ABM11$_J$         & $0.1134-0.1149 \pm 0.0012$ &  Tevatron jets (NLO) incl.
\cite{Alekhin:2010iu}
\\
ABM13 & $0.1133\pm 0.0011$ &  \cite{Alekhin:2013nda}\\
ABM13 & $0.1132\pm 0.0011$ & (without jets) \cite{Alekhin:2013nda}\\
CTEQ           & $0.1159...0.1162$ & \cite{Gao:2013xoa} \\
CTEQ           & $0.1140$ & (without jets)  \cite{Gao:2013xoa} \\
\hline
NN21          & $0.1174 \pm 0.0006 \pm 0.0001$ &  \cite{Ball:2011us}\\
\hline
Gehrmann et al.& {{$0.1131~^{+~0.0028}_{-~0.0022}$}} & {\rm
$e^+e^-$~thrust}~ \cite{Gehrmann:2012sc}
\\
Abbate et al.& {{$0.1140 \pm 0.0015$}} & {\rm
$e^+e^-$~thrust}~\cite{Abbate:2012jh}
\\
CMS & {{$0.1151 \pm 0.0033$}} & $t\bar{t}$ \cite{Chatrchyan:2013haa} \\
NLO Jets ATLAS &  $0.111 {\tiny{\begin{array}{c} +0.0017 \\
           -0.0007 \end{array}}}$ &
                       \cite{Rabbertz:2013vxa} \\
NLO Jets CMS   &  $0.1148 \pm 0.0055$  &                    \cite{Rabbertz:2013vxa} \\
\hline
BBG [2004] & {{$
0.1141 {\tiny{\begin{array}{c} +0.0020 \\
-0.0022 \end{array}}}$}}
& {\rm valence~analysis, N$^3$LO}  \cite{Blumlein:2004ip,Blumlein:2006be}
\\
\hline
3-jet rate      & $0.1175 \pm 0.0025$
         & Dissertori et al. 2009 \cite{Dissertori:2009qa}
\\
Z-decay rate & $0.1189 \pm 0.0026$ & BCK 2008/12 (N$^3$LO) \cite{Baikov:2008jh,Baikov:2012er}
\\
\hline
$\tau$-decay rate & $0.1212 \pm 0.0019$ & BCK 2008 (N$^3$LO) \cite{Baikov:2008jh,Baikov:2012er}
\\
$\tau$-decay rate & $0.1204 \pm 0.0016$ & Pich 2011 \cite{Bethke:2011tr}
\\
$\tau$-decay rate & $0.325 \pm 0.018 $~~~(at $m_\tau$) & FOTP: \cite{JAMIN:2013gpa}
\\
$\tau$-decay rate & $0.374 \pm 0.025 $~~~(at $m_\tau$) & CIPT: \cite{JAMIN:2013gpa}
\\
\hline
Lattice & $0.1205 \pm 0.0010$ & PACS-CS 2009 (2+1 fl.) \cite{Aoki:2009tf}
\\
Lattice & $0.1184 \pm 0.0006$ & HPQCD 2010 \cite{McNeile:2010ji}
\\
Lattice & $0.1200 \pm 0.0014$ & ETMC 2012 (2+1+1 fl.) \cite{Blossier:2012ef}
\\
Lattice & $0.1156 \pm 0.0022$ & Bazavov et al. (2+1 fl.) \cite{Bazavov:2012ka}
\\
Lattice & $0.1130 \pm 0.0010 (stat)$ & RBC-UKQCD (preliminary, 2014) \cite{UKQCD}
\\
\hline
world average & $0.1184 \pm 0.0007$ & (2012) \cite{Bethke:2012jm}
\\
\hline
\end{tabular}
\end{center}

\noindent
They are needed ultimately to perform a fully consistent NNLO analysis. Sensitive data, capable to constrain 
the known PDFs better, have to be selected for analysis, rather than performing global fits using data with problematic 
systematics. At lower scales it is necessary to fit $\alpha_s$ together with the higher twist terms to obtain correct results.
Current precision data include the DIS World data, including the H1 and ZEUS combined  data sets, 
the di-muon data and Drell-Yan data to constrain the different sea-quark densities, as well as the $pp$-jet data from Tevatron 
and  LHC data on $W^\pm, Z$-production and off-resonance Drell-Yan data. Very soon also the  LHC jet data can be analyzed at 
NNLO. 

In this note we give a brief summary, mostly on NNLO and N$^3$LO analyses, on the status of $\alpha_s(M_Z^2)$, including also
results from lattice simulations. At present, no unique picture on $\alpha_s(M_Z^2)$ from precision determinations has been 
obtained yet. Over the years the errors have significantly diminished. Yet there seems to be still some systematics 
between different
analyses and/or methods of measurement which has to be understood in the future.

Deep-inelastic analyses 
\cite{Alekhin:2001ih,Blumlein:2004ip,Blumlein:2006be,Gluck:2006yz,Alekhin:2009ni,Jimenez-Delgado:2014twa,Thorne:2014toa,Alekhin:2010iu,Alekhin:2013nda,Gao:2013xoa} 
mostly yield low NNLO values of $\alpha_s(M_Z^2) \sim 0.114$, even including jet data, with the exception of NNPDF
\cite{Ball:2011us}. Very recently MSTW improved its formerly high value \cite{Martin:2009bu} to $\alpha_s(M_Z^2) = 0.1136$ in 
\cite{Thorne:2014toa}. The value $\alpha_s(M_Z^2) = 0.1141 \pm 0.0022$ obtained in the valence N$^3$LO analysis 
\cite{Blumlein:2006be} compared to the NNLO value $0.1134$ may serve as an error estimate of the remainder theory error.
Heavy flavor uncertainties are of the same oder (0.0007) \cite{Alekhin:2009ni}.
Also the values from precision measurements on thrust in $e^+e^-$ annihilation are of this size  
\cite{Gehrmann:2012sc,Abbate:2012jh}. Recent NLO analyses of LHC jet data \cite{Rabbertz:2013vxa} also report low values of 
$\alpha_s(M_Z^2)$.

Larger values around $\alpha_s(M_Z^2) = 0.1184$ and higher are obtained from the inclusive measurement on the $Z$-peak 
\cite{Baikov:2008jh,Baikov:2012er} and the 3-jet rate \cite{Dissertori:2009qa} in $e^+e^-$ annihilation.
Also the $\alpha_s$-measurements from $\tau$-decay yield high values, with some theoretical systematics depending on the 
method used, despite of the N$^3$LO QCD corrections used. A wider spread of values is currently reported by lattice 
measurements. (After this workshop a detailed account on $\alpha_s$-determinations using lattice methods was given 
in \cite{Aoki:2013ldr}v2.) 
Here the central values differ more than the errors quoted, needing further clarification.
Over the years the systematics has steadily improved. The pion mass $m_\pi$ used is  still lowering towards the physical 
value and 
recently 2+1+1-flavor simulations have been performed. It also seems to be necessary to perform  the renormalization 
non-perturbatively. 

Among the hard processes, the next important analyses will be those of the LHC jet data on single jets.
Later NNLO calculations will allow to analyze also two-jet data and 3-jet/2-jet ratios at NNLO. 
An important inclusive measurement to determine $\alpha_s(M_Z^2)$ in $e^+e^-$ annihilation would consist
of the measurement of $R(s)$ using the Giga-Z option at an ILC in the more distant future.
Given the different theory errors of O(0.5 \% - 1\%) the accuracy on $\alpha_s(M_Z)$ cannot be quoted to be 
better than 1\% in individual measurements. The world-average on $\alpha_s(M_Z)$ has to account for the uncertainty
of theoretical and systematic errors in different determinations, which have not yet been understood completely.

\section*{3~~~~Precision $\alpha_S$: experimental aspects~\footnote{S.~Kluth}}
\addcontentsline{toc}{section}{\protect\numberline{3}{Precision $\alpha_S$: experimental aspects}}
In this contribution we concentrate on exclusive observables,
e.g.\ jet production rates or event shape distributions since here the
potential for improvements from more complete calculations is larger
compared with inclusive observables.  Also, theory improvements for
exclusive observables can have a close relation with experimental
procedures.

\subsection*{Particle- and parton-level}

In the analysis of jet or event shape observables the typical
procedure is to correct the data calculated from selected event
samples for so-called experimental effects.  With these the combined
effect of acceptance of selection cuts, detector inefficiencies and
resolution is summarised.  These issues have been discussed
e.g.\ in~\cite{Buttar:2008jx} and a concrete set of recommendations
for the definition of {\em particle- or hadron- level} as reference
for the experimental corrections was given.  Essentially, for jets,
the full hadronic final state including possible underlying event
activity and with particle lifetimes above 10~ps but without prompt
leptons or photons is recommended. However, no recommendation for the
identification of heavy bosons or heavy flavour jets is given.

The {\em parton-level} is discussed as well in~\cite{Buttar:2008jx},
defined as the result of a theory calculation in fixed order possibly
augmented with resummed leading logarithms.  No recommendation is made
for defining the corresponding parton-level in Monte-Carlo generators
in order to extract the hadronisation corrections needed to compare
theory calculations with data corrected to the hadron-level.

A brief and rather incomplete search in recent publications shows that
the recommendations from~\cite{Buttar:2008jx} are not followed
universally. For example, in~\cite{Aad:2012hg} for the measurement of
top quark pair differential cross sections ``The kinematic properties
of the generated t and $\bar{\mathrm{t}}$ partons in simulated
t$\bar{\mathrm{t}}$ events define the ???true??? properties of the tt
events'', which seems in contradiction with the recommendations to
use only hadronic final state particles.  

The Rivet manual~\cite{Buckley:2010ar} adds to the recommendations to
identify on-shell heavy bosons via their stable (or long-lived) decay
leptons and imposing a mass window.
In~\cite{Hoeche:2014qda,Aad:2013fba} the particle-level is defined in
accordance with the recommendations and in addition new methods are
introduced for heavy flavour jets such as searching for jet
constituent partons or B-Hadrons near to the jet axis. Top quarks are
identified using jets, heavy flavour jets and leptons at
particle-level using the same kinematic selection as for the data.

A synthesis of the recommendations of~\cite{Buttar:2008jx} and the
additional methods discussed above to define the particle-level as
continuation of the discussion would be:
\begin{itemize}
\item Use a common standard for the Monte Carlo generator event record
  such as HepMC~\cite{Dobbs:2001ck}
\item Define a cut on lifetime, e.g.\ 300~ps, for stable particles in
  generator
\item Apply the same jet algorithms and kinematic cuts as for data
\item Identify b-jets by searching for B-hadrons near the jets
\item Use Monte Carlo particle identification for leptons and photons
\item Apply mass window on lepton, photon or jet combinations to
  identify on-shell Z, W or H bosons
\item Apply same kinematic cuts on lepton b-jet or b-jet light jets
  combinations as for data to identify top quark decays
\end{itemize}
For off-shell Z, W or H bosons the mass window method will fail and in
this case one may have to resort to information on the decay parents
of final state leptons or jets.

In addition one could try to define a parton-level suitable for
extracting hadronisation corrections from the Monte Carlo generators
inspired by the above discussion.
\begin{itemize}
\item Define parton-level final state by all partons entering the
  hadronisation model and all prompt leptons and photons from the hard
  scattering or Z, W or H boson decays
\item Apply the same jet algorithms and kinematic cuts as for data
\item Identify b-jets by searching for b-partons near the jets
\item Apply mass window on lepton, photon or jet combinations to
  identify on-shell Z, W or H bosons
\item Apply same kinematic cuts on lepton b-jet or b-jet light jets
  combinations as for data to identify top quark decays
\end{itemize}
For off-shell Z, W or H bosons one will have to use Monte Carlo
particle ID as above for the particle-level.

\subsection*{Using the predictions}

In order to use theory prediction for precision measurements of the
strong coupling constant with essentially all exclusive observables
one has to consider hadronisation corrections which relate the
calculation involving partons to the measurements corrected to the
particle level.  There are currently three different procedures
available.

The first and most frequently used method consists of folding the
perturbative QCD calculations for an observable with the hadronisation
corrections extracted from a Monte Carlo generator, see
e.g.~\cite{Kluth:2006bw}.  In the generator the observable is calculated at
parton-level and at particle-level and the correction is obtained from
a ratio of distributions or from a migration matrix derived from
parton- and particle-level values for the same events.  This method is
universal, but is limited in precision, because the parton-level in
the generator is not identical to the perturbative QCD calculation.

The second method combines the perturbative QCD calculation with
analytic models for the hadronisation effects available for a selected
set of observables, see
e.g.~\cite{Davison:2008vx,Pahl:2009aa,Abbate:2010xh} for recent
examples.  The extraction of the strong coupling proceeds via a
simultaneous fit together with the free parameters of the analytic
hadronisation model.  This method is not universal since a dedicated
analytic model for each observable is needed. The advantage lies in
potentially better accuracy since the simultaneous fit can exploit
correlations between perturbative and non-perturbative parameters.

The third method is more recent and is based on combining consistently
the perturbative QCD calculations for the hard scattering with the
parton shower algorithms in the Monte Carlo generators, see
e.g.~\cite{Gehrmann:2012yg}.  This improves the accuracy and
reliability of the generator predictions.  The extraction of
parameters such as the strong coupling could be done by a global
adjustment of the free parameters of the generator program to the data
at particle-level.  In this way the parameters of the hadronisation
model of the generator and of the perturbative calculation including
the parton shower are fitted simultaneously similar to the second
method with potentially similar advantages.

\subsection*{Averaging several measurements}

Once measurements of the strong coupling have been performed using
various observables, in different processes, using different
procedures and with different levels of theoretical accuracy one
is faced with the task of combining the results to a common best value.

The best known procedure to achieve a world average of the strong
coupling is used by the PDG group~\cite{Beringer:1900zz}.  Closely
related measurements from the same or similar datasets whose
uncertainties are inconsistent with the range covered by the results
are combined using so-called {\em range averaging} where the overall
error is given by the range.  For largely independent results from the
same class of observables a $\chi^2$-based method is used, which is
based on calculating an error weighted average of the selected
individual measurements.  If the resulting $\chi^2$ based on the
average and the individual total errors is smaller than the degrees of
freedom $N_{dof}$ of the average a global overestimation of the errors
is assumed.  The calculation of the error of the average is repeated
with a covariance matrix $C$ constructed from the individual errors
$\sigma_i$ and $\sigma_j$ with $C_{ij}=f\sigma_i\sigma_j$.  The common
factor $f$ is adjusted such that $\chi^2=N_{dof}$.  In the opposite
case of $\chi^2>N_{dof}$ underestimated errors are assumed.  A common
factor $g$ is applied to all individual errors to get $\chi^2=N_{dof}$
and the error of the average is calculated again.  For $f>0$ and $g>1$
the $\chi^2$-based method yields conservative uncertainties. It is a
pragmatic and stable procedure.  All pre-averages are combined using
the $\chi^2$-based method.

Other fundamental parameters of the Standard Model such as the mass of
the W boson~\cite{Aaltonen:2013iut} or the top
quark~\cite{ATLAS:2014wva} are averaged using different procedures.
These procedures calculate an error weighted average with a complete
covariance matrix for all individual errors.  The covariance matrix is
constructed from a detailed decomposition of the individual errors
into related categories and a model for the correlations between the
error components.

In the case of the average of strong coupling measurements the
application of the procedure used for other Standard Model
measurements should in principle be possible.  In order to put this
into practice one would have find an appropriate decomposition of the
errors as well as a model for the correlations of the individial error
components.  For the average of measurements of the strong coupling
from event shape distributions by the LEP experiments and also a world
average value such a procedure has been used~\cite{Kluth:2006bw,Kluth:2006vf}.

\section*{4~~~~$\alpha_s$ determination in the PDF fits~\footnote{S.~Alekhin}}
\begin{figure}
    \begin{center}
      \includegraphics[width=0.5\textwidth]{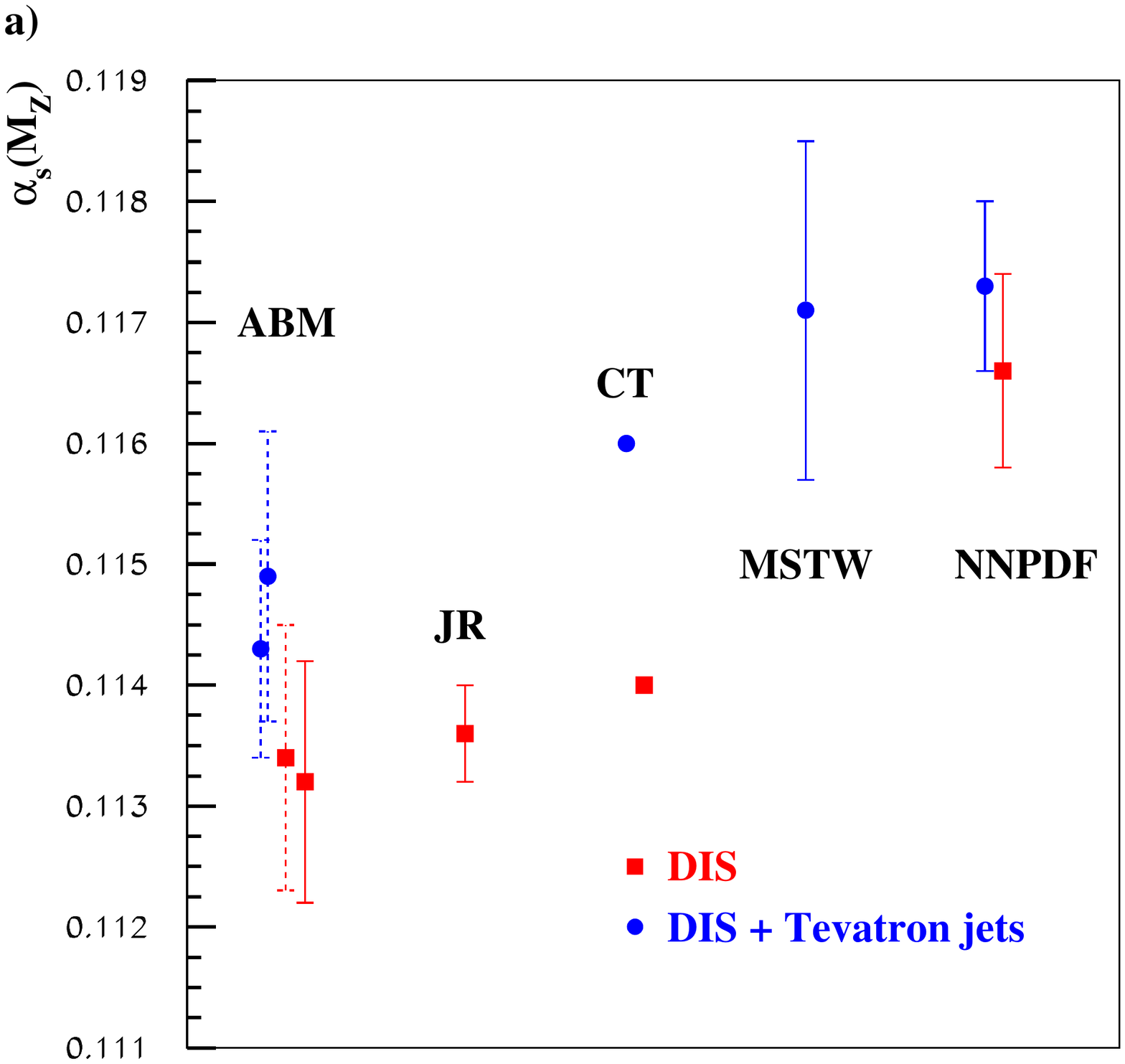}~
      \includegraphics[width=0.5\textwidth]{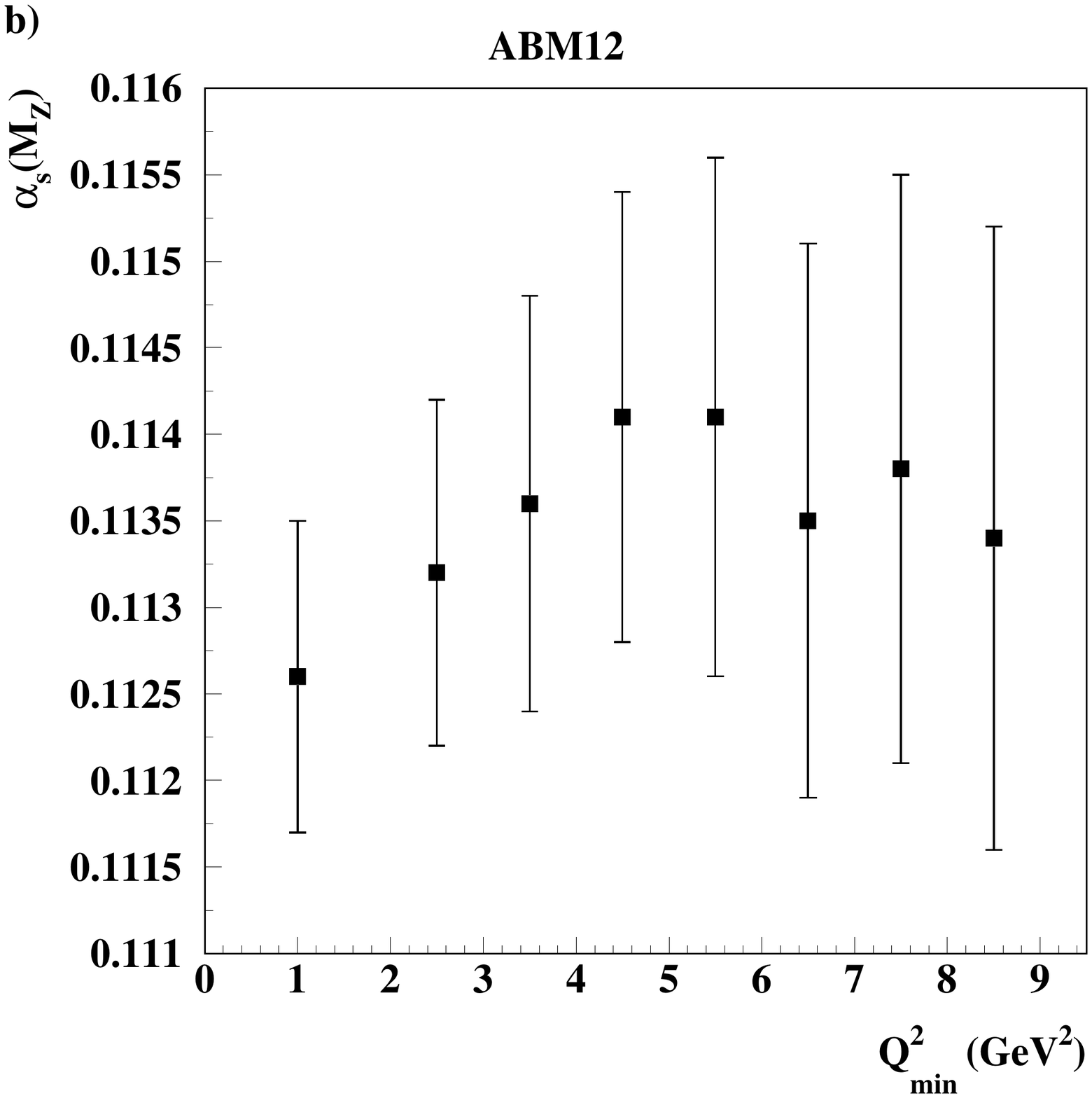}
      \caption{a): The values of $\alpha_s(M_Z)$ at NNLO obtained in the PDF fits of ABM
    (solid bars: ABM12~\cite{Alekhin:2013nda}, 
dashed bars: ABM11~\cite{Alekhin:2012ig}) in 
    comparison with the  
    CT~\cite{Gao:2013xoa}, 
    JR~\cite{Jimenez-Delgado:2014twa}, 
    MSTW~\cite{Martin:2009iq} and 
    NNPDF~\cite{Ball:2011us} results. Only experimental uncertainties are
    shown. 
b): The same as a) for the versions of ABM12 
fit~\cite{Alekhin:2013nda} with different cuts of $Q^2>Q^2_{\rm min}$ imposed 
on the DIS data.
}
      \label{alekhin:fig1}
    \end{center}
  \end{figure}

\addcontentsline{toc}{section}{\protect\numberline{4}{$\alpha_s$ determination in the PDF fits}}
The value of $\alpha_s$ is commonly considered as a free parameter in the PDF
global fits~\cite{Alekhin:2013nda,Gao:2013xoa,Jimenez-Delgado:2014twa,Martin:2009iq,Ball:2011us} 
that provides its independent estimate with the accuracy comparable
to the world average. Herewith the constraint on $\alpha_s$ comes essentially
from two processes involved: the deep-inelastic scattering (DIS) and the hadron
jet production. These processes have different theoretical footing. While
the QCD corrections to the DIS Wilson coefficients are basically known up 
to NNLO, the data on nucleon-nucleon jet production employed in the PDF fits
are commonly described to NLO and only the threshold soft-gluon resummation
corrections calculated to NNLO~\cite{Kidonakis:2000gi} 
are added on the top. Meanwhile, the NNLO terms
are evidently non-negligible and in particular they amount to 
$20\div 25$\% in the gluonic channel~\cite{Ridder:2013mf}. Furthermore the 
threshold resummation approach of Ref.~\cite{Kidonakis:2000gi} 
was shown to be applicable for a limited kinematics only~\cite{Kumar:2013hia}
also lacking the jet cone size dependence~\cite{deFlorian:2013qia}.
These shortcomings do not allow for sufficient theory control in the 
NNLO PDF analyses based on the jet data therefore an additional theoretical 
uncertainty should be assigned to the value of $\alpha_s$ obtained. 
This uncertainty can be estimated as $\sim 0.0015$ at the scale of 
$M_Z$~\cite{Watt:2013oha} 
that roughly corresponds to the experimental uncertainty
in $\alpha_s(M_Z)$, cf. Fig.~\ref{alekhin:fig1}a. 
More accurate determination of $\alpha_s$ is obtained in the variants of PDF
fits without use of the jet data. In these cases the value of $\alpha_s$ is
determined by the DIS data and typically they go lower by
$1\sigma$ than ones based on the jet data. 
At that the results obtained by different groups spread out by more than 
$1\sigma$ despite strong overlap in the data sets most sensitive to
$\alpha_s$.
Thereby this spread appears evidently due to details of the theoretical 
framework and the data treatment. 

One of the most important factor affecting determination of $\alpha_s$
from the DIS data are power corrections to the leading-twist terms appearing
in the operator-product-expansion (OPE) formalism. 
Qualitative estimates based on
the dimension counting suggest that the power corrections including the
dynamical high-twist (HT) terms are enhanced at small
momentum transfer $Q$ and/or invariant hadronic mass $W$. However their shape 
and magnitude cannot be obtained in this way and commonly they are determined
phenomenologically. In particular, the ABM analysis~\cite{Alekhin:2013nda}
is based on the simultaneous fit of the 
leading-twist PDFs, twist-4 terms, and $\alpha_s$ to the existing DIS data by
the SLAC, NMC, BCDMS, and HERA experiments. The SLAC and NMC data are
particularly sensitive to the HT terms since they spread down to 
$Q^2=O(1)~{\rm GeV}^2$ and quite stringent cut on $Q^2$ is necessary to reduce
this sensitivity. Indeed, if the HT terms are disregarded, 
the fitted value of $\alpha_s(M_Z)$ goes down by
0.0025, when the low-$Q^2$ margin of the DIS data used in the fit 
is moved up from $2.5~{\rm GeV}^2$ to $10~{\rm GeV}^2$~\cite{Alekhin:2012ig}.
At the same time in the standard ABM approach, with the HT terms fitted, the 
value of $\alpha_s$ obtained is stable w.r.t. the low-$Q^2$ margin, 
cf. Fig.~\ref{alekhin:fig1}b. For the cut of $Q^2>2.5~{\rm GeV}^2$ employed in
the recent ABM fit~\cite{Alekhin:2013nda} this gives 
  \begin{equation}
    \label{alekhin:eq1}
    \alpha_s(M_Z) = 0.1132 \pm 0.0011~({\rm exp.}) \, .
  \end{equation}
In the PDF analyses of Refs.~\cite{Gao:2013xoa,Martin:2009iq,Ball:2011us} the
HT terms are not considered and a cut of $W^2>W^2_{\rm min}$ is
imposed instead, where $W^2_{\rm min}\sim 12\div 15~{\rm GeV}^2$. However this
cut does not remove the low-$Q$ SLAC data at Bjorken 
$x\sim 0.1$, which still are sensitive to the HT 
contribution~\cite{Alekhin:2012ig}. It is worth noting in this context that 
the SLAC data are not used in the CT10 analysis~\cite{Gao:2013xoa} and the 
DIS value of $\alpha_s(M_Z)$ obtained in this case is quite comparable
with Eq.~(\ref{alekhin:eq1}), cf. Fig.~\ref{alekhin:fig1}a.

Another issue discriminating the PDF fits is treatment of the heavy-quark
contribution to the DIS. In the ABM analysis it is based on the
fixed-flavor-number (FFN) scheme with 3 light quarks in the initial state,
while in other cases different variants of the general-mass 
variable-flavor-number (GMVFN) scheme are 
employed~\cite{Gao:2013xoa,Martin:2009iq,Ball:2011us}. Note, the value of 
$\alpha_s$ found in the FFN version of the MSTW fit~\cite{Thorne:2014toa} is 
in a good agreement with Eq.~(\ref{alekhin:eq1}).
Meanwhile, the GMVFN scheme suffers from
the uncertainties due to modeling of the Wilson coefficients at low-$Q$, which
cannot be derived from the field-theory basis in the GMVFN
approach. Additional GMVFN-scheme uncertainties stemming from the choice of the
matching scale for 4(5)-flavor PDFs and missing NNLO corrections to the
massive operator matrix elements are estimated in a sum 
as 0.001~\cite{Alekhin:2013fua}. These uncertainties are comparable to the
experimental ones and thereby put a limit on the accuracy of $\alpha_s$ 
determination within the GMVFN scheme. 

\section*{5~~~~Determination of the strong coupling constant in a global fit~\footnote{K.~Rabbertz, P.~Uwer}}
\addcontentsline{toc}{section}{\protect\numberline{5}{Determination of the strong coupling constant in a global fit}}

Quantum chromodynamics (QCD) is a mature theory providing a
solid understanding of the strong interaction. While research on different
aspects of QCD is still ongoing, QCD has developed over the past into an
indispensable tool for collider phenomenology. This is particularly true
for hadron collisions where the colliding particles interact strongly
already in the first place. As SU(3) non-abelian gauge theory QCD has only one
free parameter if the quark masses are ignored: the gauge coupling
$g_s(\mu) = \sqrt{4\pi\as(\mu)}$, where $\mu$ denotes the
renormalization scale. Evidently, a precise knowledge of the gauge
coupling is mandatory for all theoretical predictions in QCD\@.
The importance of a precise determination of \alps is reflected in the
large numbers of different measurements. \alps is currently extracted
in a variety of different environments:
\begin{enumerate}
\item $\tau$ decays,
\item deep inelastic scattering,
\item inclusive hadron production in $e^+e^-$ annihilation,
\item inclusive jet production in hadron collisions,
\item three-jet rates in $e^+e^-$ annihilation and hadron collisions,
\item four- and five-jet production in $e^+e^-$ annihilation, and
\item lattice gauge theory / hadron spectroscopy
\end{enumerate}
to name only a few. A common feature of all extractions is, that the coupling
constant \alps appears as a free parameter in the theoretical
predictions, which then is fitted to data.
Obviously, reliable values can only be expected,
if theoretical predictions give a precise description of the
observable under consideration. In many cases this requires the
inclusion of next-to-leading order (NLO) or even higher
corrections. Schematically we may write
\begin{equation}
  \label{eq:ObsFixedOrder}
  {\cal O} = \as^k(\mu) ( c_0 + \as(\mu) c_1(\mu) + \as^2(\mu)
  c_2(\mu) +\ldots).
\end{equation}
Note that from $c_1$ onwards the expansion coefficients depend in general
on the renormalization scale $\mu$. The logarithmic $\mu$-dependence can be
reconstructed using
\begin{equation}
  {d\over d\mu} {\cal O} = 0
\end{equation}
and
\begin{equation}
  \label{eq:QCDbetaFunction}
  {d\as(\mu)\over d\mu} = -\beta(\as(\mu)) = - \as^2(\mu)(b_0 +
  \as(\mu)b_1
  +\ldots)\,.
\end{equation}
The QCD beta-function $\beta(\as(\mu))$ describes how the coupling
constant changes as a function of the unphysical renormalization
scale. When a fixed order prediction of the form as shown in
\Eq{eq:ObsFixedOrder} is used to extract \alps, the renormalization
scale is frozen to some reasonable value $\mu_0$, which should avoid
the appearance of large logarithmic corrections in the coefficients
$c_i$. The fit then leads to the determination $\as(\mu_0)$, i.e.\
\alps at a fixed scale $\mu_0$ (which is usually close to the typical
momentum transfer of the considered process to avoid the
aforementioned large logarithmic corrections). It is thus possible to
attribute the \alps measurement to a unique energy scale. In particular since
the running of $\as$ is not used in the theoretical predictions it is
possible to check \Eq{eq:QCDbetaFunction} experimentally by comparing
measurements at different scales.  In some cases fixed order
corrections are not sufficient and logarithmically enhanced corrections
have to be resummed. Resummation appears in a variety of
different flavors. For the following the details are not important
and the simplest case is sufficient to see the difference to fixed
order calculations. A phase space dependent \alps for example may
be used to resum large logarithms related to particular phase space
regions. The main difference compared to the fixed order predictions
mentioned above is, that the running of \alps is used in the
theoretical predictions and the determined \alps value is no longer
related to physics at one particular energy scale. Although an \alps
value at a fixed scale is quoted in such measurements as final result,
the extraction may be affected by physics at rather different energy
scales. Furthermore the running of \alps is used in the extraction
already. Using these results to test the running of \alps by comparing
with measurements at different scales is thus questionable. After
these preliminaries let us now address the combination of \alps
extracted from different measurements. In most cases the authors not only
quote $\as(\mu_0)$ but also the value \alpsmz evolved to
the Z-resonance where historically the most precise measurements were
performed. Thus it is possible to calculate the combined \alps value
using for example a weighted average. In principle also correlations
between different measurements could be taken into account. However,
this approach is not fully satisfying since in
general the evolution of the individual \alps measurements has been 
performed using different values for \alps. The uncertainty of \alps at low
scales leads to an additional uncertainty through the
evolution. (Since the evolution is non-linear it makes in general also
a difference whether uncertainties are investigated at
low scale or high scale). In principle one may also ask
what the right order of the evolution equation to be used is.
These effects can be taken into account by a proper
error analysis.
Evidently we do not expect that these effects lead to significant
effects, however, minor corrections are possible and one may ask
whether these issues can be avoided.

In this note we advocate to perform a global fit of \alps to the
available collider data. Instead of combining \alps as extracted in
different experiments we suggest that the measured observables are
used directly and \alps is determined using the evolution equation
\Eq{eq:QCDbetaFunction} together with the theoretical predictions of
the form \Eq{eq:ObsFixedOrder}. This approach guarantees that a
consistent running is used for all measurements and that the extracted
value of \alps describes the available data best. Correlations due to
theoretical assumptions may also be easier to trace through the fit. 
The drawback of this approach is that we need to know for all
observables the corresponding expressions \Eq{eq:ObsFixedOrder}. 
For measurements where resummation is used in the theoretical
predictions compact expressions may not be available in closed form. 

\begin{figure}[hbtp]
  \begin{center}
    \includegraphics[width=0.75\textwidth]{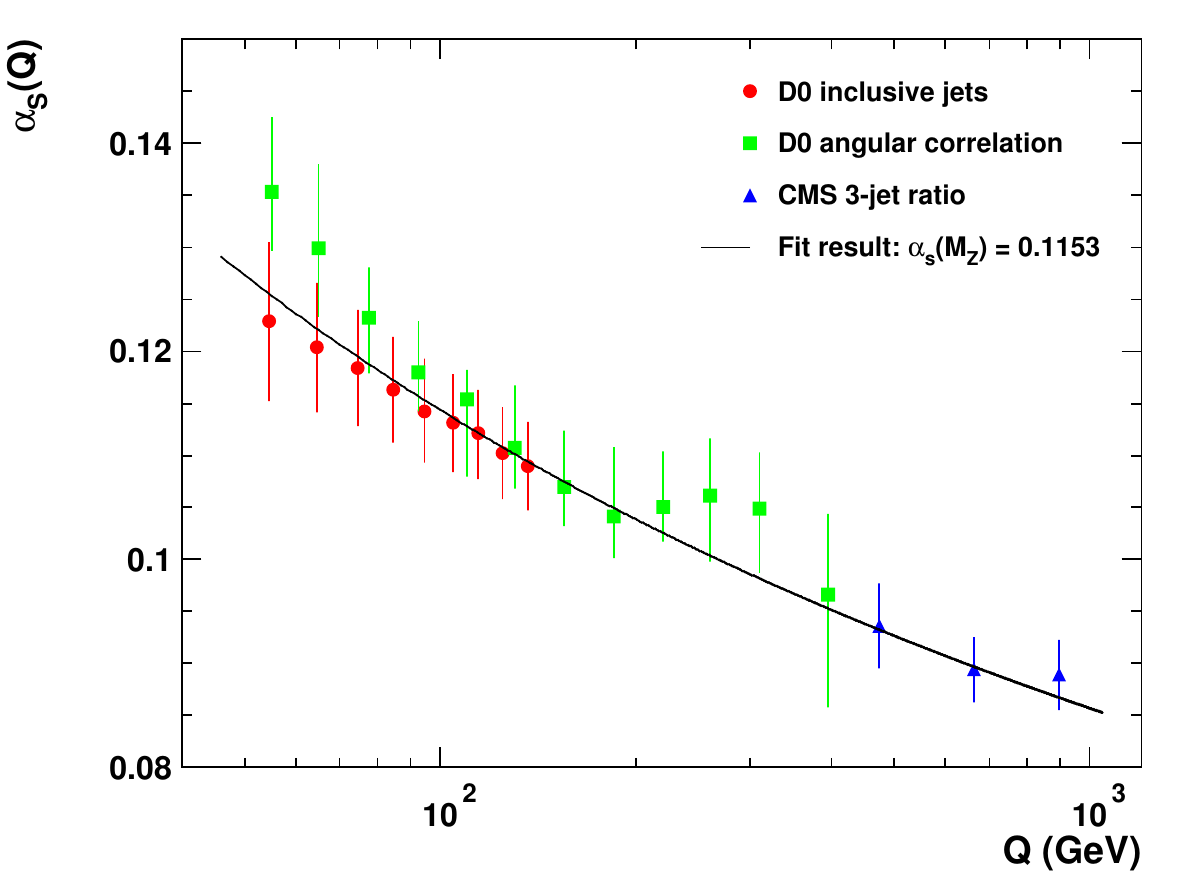}
    \caption{%
      \alpsq points with total uncertainties as published in
      Refs.~\cite{Abazov:2009nc,Abazov:2012lua,Chatrchyan:2013txa}
      together with the \alps evolution at 4-loop order and with $N_f
      = 5$ for the fitting test as described in the text.%
    }
    \label{fig:fitalphas}
  \end{center}
\end{figure}

As a first step in the aforementioned direction we present in this
note a \chisq fit, where we extract \alps from determinations of
\alpsq at different scales $Q$. The three investigated data sets, two
from the D0 collaboration~\cite{Abazov:2009nc,Abazov:2012lua} and one
from CMS~\cite{Chatrchyan:2013txa}, are listed in
Table~\ref{tab:asfits}.  For simplicity we use the published \alpsq
values and not the measured observables which is much preferable. The
fitted \alps value best describes the evolution
over the range in $Q$ of the measurements.  The \chisq fits are
performed considering all theory uncertainties, including the one from
scale variations, as correlated together with the correlated
experimental uncertainty. In the case of the CMS data, which do not
provide an uncorrelated experimental uncertainty, the increase in
experimental uncertainty from lowest to highest scale is considered as
uncorrelated.  Because it is observed that full correlation for the
given correlated uncertainties can lead to a significant increase in
the \chisq values of the fits even for individual data sets, the
correlation is reduced to a degree $\rho$ of 70--80\%.  The results of
these fits are presented in Table~\ref{tab:asfits} and
Fig.~\ref{fig:fitalphas}. Correlations between the different
measurements as required for full consistency are not accounted for at
this stage.  The final goal is to perform combined fits of multiple
measurements using directly the measured observables, where all
correlations are properly taken into account. This requires properly
documented systematic uncertainties of each measurement and
corresponding theory calculations in a format suitable for efficient
fitting procedures like for example in the \textsc{fastNLO}
framework~\cite{Britzger:2012bs} or
\textsc{APPLGRID}~\cite{Carli:2010rw}

\begin{table}[htbp]
  \caption{%
    \alpsmz fitted from the \alpsq values tabulated in the given references
    using a 4-loop evolution formula for $N_f = 5$ flavours as
    implemented in the CRunDec package~\cite{Schmidt:2012az}.
    Correlated uncertainties within one dataset are considered,
    but only to a degree $\rho$ of 70--80\%. Full correlation can lead
    to a significant increase in the \chisq values.%
  }
  \label{tab:asfits}
  \centering
  \begin{tabular}{lrlccc}
    \hline\hline
    Data set  & No.\ of data points & Refs. & \alpsmz & \chisq & \chisqndof \rbthm\\
    \hline
    D0 inclusive jets      &  9 & \cite{Abazov:2009nc}      & $0.1164 \pm 0.0042$ & 0.44 & 0.05 \rbtrr\\
    D0 angular correlation & 12 & \cite{Abazov:2012lua}     & $0.1173 \pm 0.0032$ & 14.3 & 1.3  \rbtrr\\
    CMS 3-jet ratio        &  3 & \cite{Chatrchyan:2013txa} & $0.1165 \pm 0.0051$ & 1.58 & 0.79 \rbtrr\\
    Combined               & 24 &           ---             & $0.1153 \pm 0.0029$ & 29.3 & 1.27 \rbtrr\\
    \hline\hline
  \end{tabular}
\end{table}

\paragraph*{Conclusion:} A precise knowledge of the QCD coupling
constant is of crucial importance for precision measurements at any
future collider. A sub percent accuracy on \alps can only be reached
through the combination of different measurements since at the moment
no individual measurement with that precision is available. In these
notes we have sketched how current \alps determinations are combined.
As an alternative approach we suggest to perform a global fit using
the respective theoretical predictions together with the running of
\as. In this way the same running is applied to all observables and
correlations of theoretical and experimental nature may be easier to
be taken into account. As a
first step in this direction we determined \alps from the evolution of
fixed-scale \alpsq determinations.

\section*{6~~~~$\alpha_s$ from the lattice~\footnote{E.~Shintani; 
thanks to Gregorio Herdoiza for checking OPE formulation.}}
\addcontentsline{toc}{section}{\protect\numberline{6}{$\alpha_s$ from the lattice}}

We report on a lattice calculation of the strong coupling constant from 
the vacuum polarization function (VPF) with $N_f=2+1$ domain-wall fermions. 
In this calculation, by using the all-mode-averaging (AMA) techniques, 
a precise lattice calculation of the VPF is obtained. 
By comparison with two cut-off scales, we estimate the lattice discretization effect. 
Fitting the VPF in the high momentum region, using the operator product expansion (OPE) and 
perturbative QCD, we obtain a more reliable estimate 
of the strong coupling constant. Furthermore, the
quark condensate is also evaluated. 

\subsection*{Introduction}

In these proceedings, 
we report on progress of the calculation of the strong coupling constant ($\alpha_s$) from 
the vacuum polarization function (VPF). 
The VPF as a function of the Euclidean momentum squared, $-q^2=Q^2$, 
can be extracted from the two-point correlation function of the vector current.
The VPF provides rich information for hadronic contributions 
to low-energy physics and perturbative QCD. 

The non-perturbative determination of $\alpha_s$ from first principles of QCD 
is important for a precise test of the Standard model (SM), since 
this is one of the large uncertainties of, for instance Higgs production \cite{Dittmaier:2011ti}. 
The operator product expansion (OPE) represents the VPF as a perturbative diagram of 
quark-gluon and the multi-dimensional operator condensate with inverse powers of $Q^2$. 
A non-perturbative comparison with lattice calculations plays an essential role for 
the check of other estimates using experimental values and estimates of the QCD scale $\Lambda_{\rm QCD}$. 

\begin{figure}
\begin{center}
\includegraphics[width=100mm]{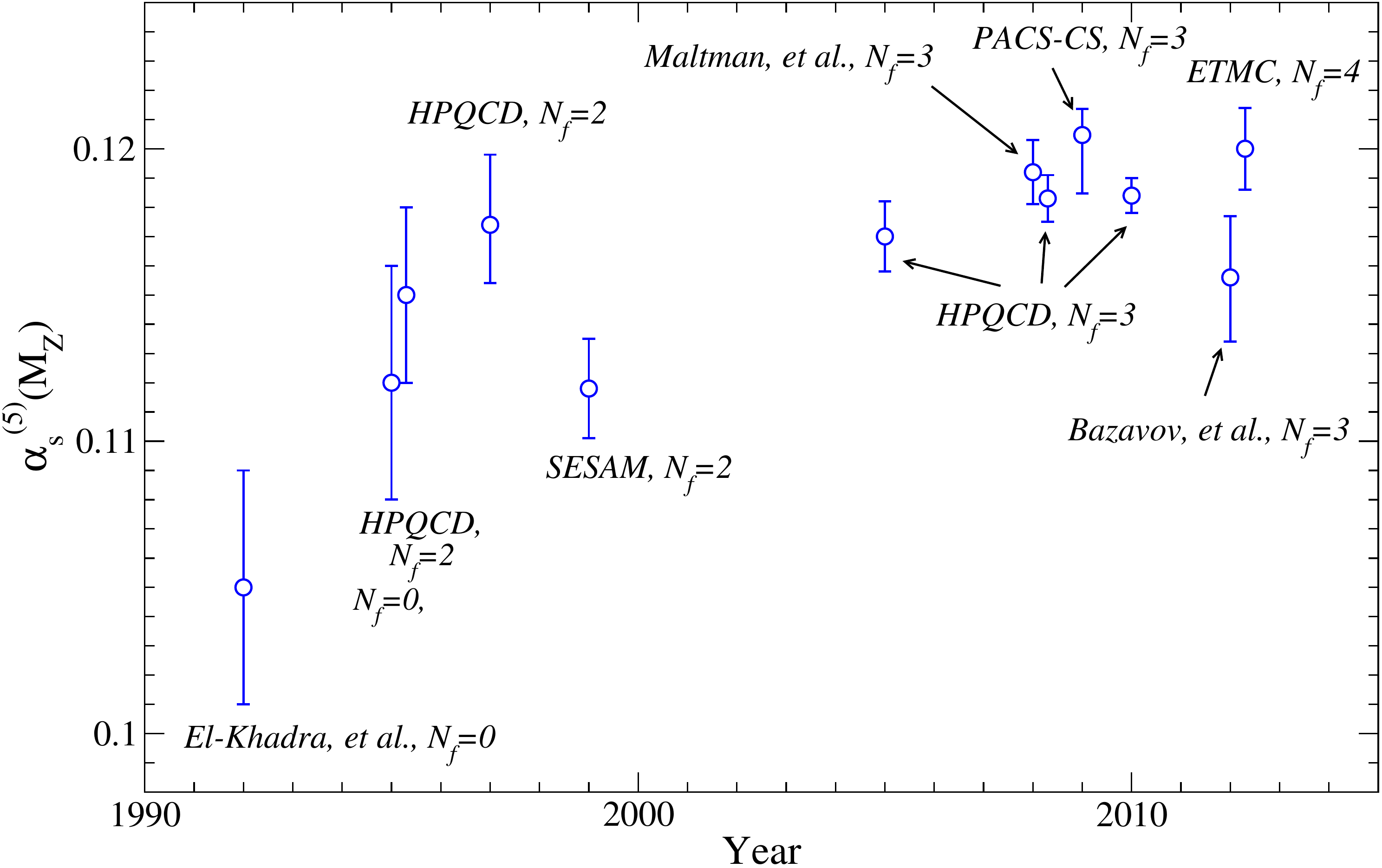}
\caption{History of lattice determination of $\alpha_s$.}\label{fig:history}
\end{center}
\end{figure}

As shown in Figure \ref{fig:history}, the precision of lattice  calculations of $\alpha_s$ 
increases, and currently its uncertainty is below the 1\% level.
In lattice QCD, we need to set a scheme to extract $\alpha_s$ from target quantities 
obtained in lattice calculation. 
From the point of view of convergence in perturbative expansion, 
the lattice perturbation is not an appropriate scheme as discussed in \cite{Lepage:1992xa}, 
and there have been several alternative schemes adopted in lattice studies. 
One idea is to directly extract the strong coupling constant by 
a method using the Adler function. 
Comparing the Adler function in the continuum theory with lattice VPF results, 
we directly evaluate the renormalized strong coupling constant $\alpha_s$
in the $\overline{\rm MS}$ scheme. 
As shown in Ref.\cite{Shintani:2008ga,Shintani:2010ph}, 
the precise value of $\alpha_s^{\rm\overline{MS}}$ is obtained 
from the pure perturbative formula at 3-loop and 
the contribution of quark and gluon condensate in OPE.

Here we show preliminary results of these observables 
using the all-mode-averaging (AMA) technique to drastically reduce the 
statistical error. 
In this study we use two cut-off ensembles, 
$24^3\times 64$ at $a^{-1}=1.73$ GeV 
and $32^3\times 64$ at $a^{-1}=2.23$ GeV, and 
estimate the systematic error due to lattice discretization. 
Table \ref{tab:param} shows the details of the simulation parameters.
We obtain the Lanczos method to compute the exact $N_\lambda$ low-lying modes 
of even-odd preconditioned Hermitian Dirac operator. 
In AMA, we employ relaxed CG solver with fixed CG iteration as approximation
\cite{Blum:2012uh,Blum:2012my}.

\begin{table}[b]
\begin{center}
\caption{Lattice parameters in the simulation.$N_\lambda$ means the number of 
low-lying mode we use in deflation method of CG process, and 
$N_{\rm CG}^{\rm AMA}$ means the stopping iteration of relaxed CG in AMA.
$N_G$ is the number of source locations of approximation.}
\label{tab:param}
\begin{tabular}{ccccccccc}
\hline\hline
Lattice size & $a^{-1}$ GeV & $m_s$ & $m_{ud}$ & $m_\pi$ (GeV) & $N_\lambda$
& $N_G$ & $N_{\rm CG}^{\rm AMA}$ & $N_{\rm conf}$\\
\hline
$24^3\times 64$ & 1.73 & 0.04 & 0.005 & 0.33 & 140 & 32 & 180 & 192\\
& & & 0.01 & 0.42 & 140 & 32 & 150 & 172 \\
& & & 0.02 & 0.54 & 140 & 32 & 150 & 105 \\
$32^3\times 64$ & 2.23 & 0.03 & 0.004 & 0.28 & 140 & 32 & 180 & 100\\
& & & 0.006 & 0.33 & 140 & 32 & 180 & 113 \\
& & & 0.008 & 0.38 & 140 & 32 & 180 & 122 \\
\hline\hline
\end{tabular}
\end{center}
\end{table}

\subsection*{Extraction of vacuum polarization function}

In the continuum theory, the VPF is given by the 
expansion of the vector two-point correlation function
in four-dimensional momentum representation, 
\begin{equation}
  i\int d^4x e^{iqx}\langle T\{V^a_\mu(x) V_\nu^{b\,\dag}(0)\}\rangle
  = -\big(q_\mu q_\nu - g_{\mu\nu}q^2\big)\delta^{ab}\Pi^{\rm cont}_V(q^2) 
  \equiv \delta^{ab}\Pi^{\rm cont}_{\mu\nu}(q).
  \label{eq:vpf_cont}
\end{equation}
The above equation is satisfied with charge conservation, 
$q^\mu\Pi^{\rm cont}_{\mu\nu}(q) =q^\nu\Pi^{\rm cont}_{\mu\nu}(q) = 0$
for vector current $ V^a_\mu = \bar q \gamma_\mu \tau^a q$. 
$\tau$ is the SU(3) isospin generator matrix.

In lattice QCD, we use a combination of the conserved and the local current
of the two-point correlation function also used in 
\cite{Shintani:2010ph}.
The local current $V^{\rm loc}_\mu$ has the same form as in the continuum theory, 
however, it is not satisfied with charge conservation. 
We multiply the renormalization constant $Z_V=0.7178$
which is computed non-perturbatively in Ref.\cite{Aoki:2010dy}.
The conserved current 
has the charge conservation $\sum_\mu\hat Q_\mu\Pi_{\mu\nu}=0$, 
where we define $\hat Q_\mu = e^{iQ_\mu}-1 = e^{iQ_\mu/2}2i\sin(Q_\mu/2)$.
To extract the VPF from the lattice vector two-point correlation function, 
we modify the expansion of Eq.(\ref{eq:vpf_cont}) as
\begin{eqnarray}
  \Pi^q_{\mu\nu}(Q) 
  &=& \Pi^q_V(Q)(\hat Q_\mu\hat Q_\nu -\delta_{\mu\nu}\hat Q^2)
       + \mathcal O((aQ)^6)\nonumber\\
  &=& \Pi^q_V(Q)\Big[ Q_\mu Q_\nu - \frac{Q_\mu^2Q_\nu^2}{16}
  - \delta_{\mu\nu}\Big( Q^2 - \frac{Q^4}{16}\Big)\Big] + \mathcal O((aQ)^6),
  \label{eq:pi_lat}
\end{eqnarray}
where, in the last equation, we also expand $\hat Q_\mu$ with 
$aQ_\mu = 2\pi n_\mu/L$.
Below the $(aQ)^2=0.7$ region, we ignore higher order contribution than $O((aQ)^6)$, 
which contains the lattice artifact due to using non-conserved current
in this formula. 
As shown in \cite{Shintani:2010ph}, 
this effect is negligibly small in this region. 
In practice, $\Pi_V(Q)$ is obtained by fitting lattice data 
for different combination of $Q_\mu$ at each $Q^2$
with the function of Eq.(\ref{eq:pi_lat}).

Furthermore we exclude the momentum which has only finite value
at one direction above $Q^2\simeq 0.2$ GeV$^2$ region, such as 
$n_\mu^{\rm exclude} 
  = \big\{ (n_i=1,n_t=0),\,(n_i=0,n_t=3),\,(n_i=0,n_t=4),\,(n_i=2,n_t=0)\big\}$,
except for $(n_i=0,n_t=1)$ and $(n_i=0,n_t=2)$
in the case of $24^3\times 64$ at $a^{-1}=1.73$ GeV, 
and $(n_i=0,n_t=1)$ at $a^{-1}=2.23$ GeV. 
As a consequence of the above subtraction, 
we can conservatively avoid the lattice artifacts effect.

\subsection*{Estimate of strong coupling constant}

The formula of the OPE of the vector VPF is expressed as  
\begin{eqnarray}
  \Pi_{V}(Q^2) &=& \frac{c}{\mu^2a^{2}} + C_{0}(l_\mu(Q^2),\alpha_s)
   + C^{V}_m(l_\mu(Q^2),\alpha_s)\frac{m_q^2}{Q^2}
   + C^{V}_{\bar qq}(l_\mu(Q^2),\alpha_s)
     \frac{\langle m\bar qq\rangle}{Q^4}\nonumber\\   
  &+& C_{\bar qq}^{\rm loop}(l_\mu(Q^2),\alpha_s)
      \frac{\sum_f \langle m_f\bar qq\rangle}{Q^4}
   + C_{GG}(l_\mu(Q^2),\alpha_s)\frac{\Big\langle\frac{\alpha_s}{\pi}GG\Big\rangle}{Q^4},
  \label{eq:ope}
\end{eqnarray}
where $\alpha_s=\alpha_s(\mu^2)$ and $m_q = m_q(\mu^2)$
at a renormalization scale $\mu=2$ GeV.
Note that we use the $\alpha_s$ and mass renormalization 
in the $\overline{\rm MS}$ scheme. 
The mass renormalization constant is determined non-perturbatively
as $Z_m^{\overline{MS}}(\mu=2{\rm GeV})=1.498$ in $24^3\times 64$ 
and $Z_m^{\overline{MS}}(\mu=2{\rm GeV})=1.527$ in $32^3\times 64$
in Ref.\cite{Aoki:2010dy}.  
We also define $l_\mu(Q^2)=\ln(Q^2/\mu^2)$.
The first term and second term 
are scheme dependence and pure QCD perturbation 
up to N$^3$LO $\mathcal O(\alpha_s^2)$
referring to \cite{Chetyrkin:1996cf,Braaten:1991qm}.
The third term is proportional to the quark mass, which 
is derived from the expansion with respect to $m_q^2$, and 
here we use 3-loop formula referring to \cite{Chetyrkin:1993hi}. 
From the forth term, since these are sub-leading contribution 
to the VPF, we use the leading order of Wilson coefficient, 
$C^V_{\bar qq} = 2$, $C^{\rm loop}_{\bar qq} = 0$ 
and $C_{GG}=1/12$ \cite{Chetyrkin:1985kn,Braaten:1991qm}. 

\begin{figure}
\begin{center}
  \includegraphics[width=72mm]{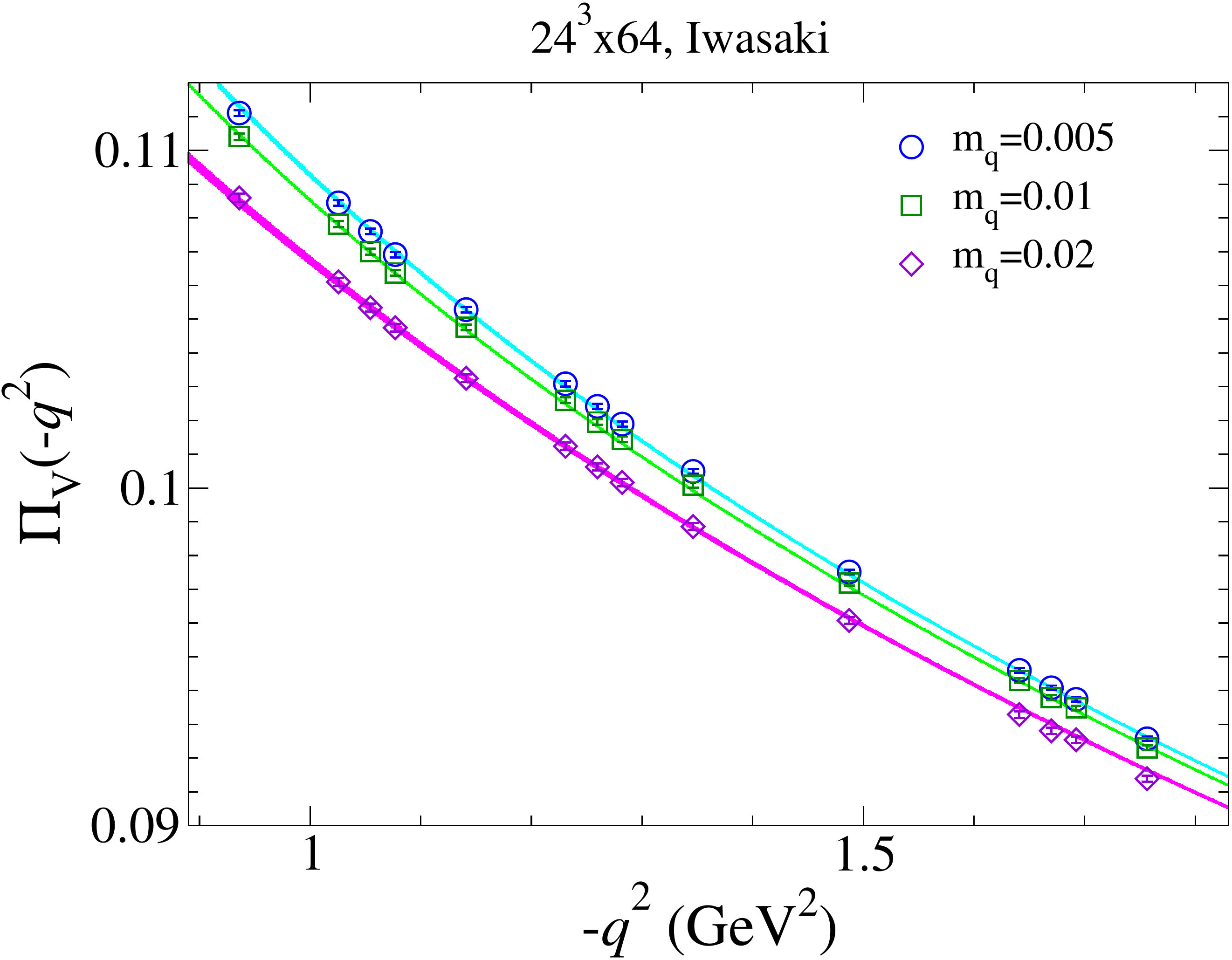} 
  \hspace{2mm}
  \includegraphics[width=72mm]{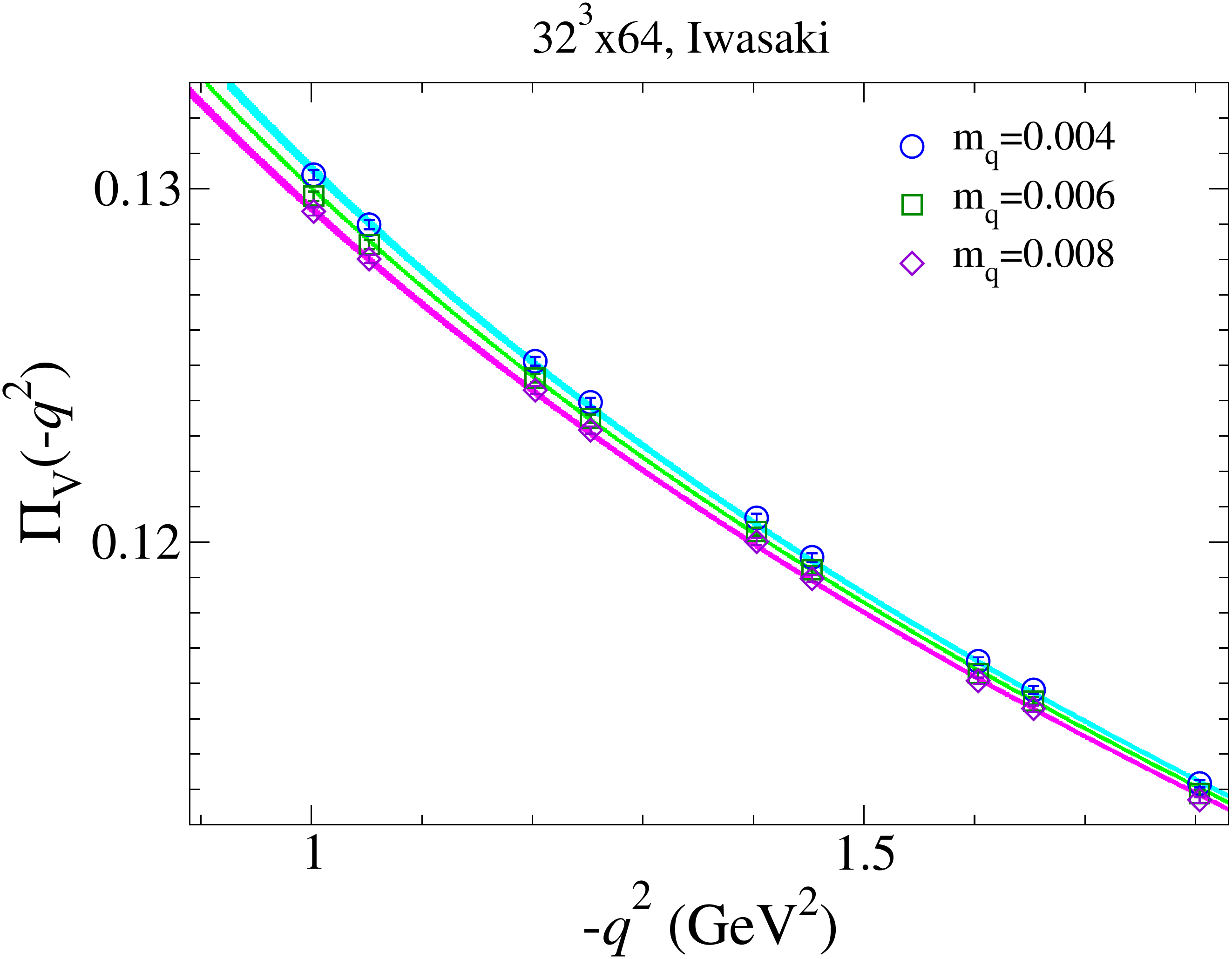}
  \caption{VPF in high-$Q^2$ in $24^3\times 64$ lattice (left)
  and $32^3\times 64$ lattice (right). The different symbols denote 
  the lattice data at different masses. The solid lines denote 
  the fitting function with OPE formula.}
  \label{fig:vpf_v_high}
\end{center}
\end{figure}

In this calculation, we use four free parameters in the
fitting of the VPF, which are 
$c$, $\alpha_s$, $\langle q\bar q\rangle_{\overline{MS}}$ 
and $\langle\frac{\alpha_s}{\pi}GG\rangle$.
We perform the simultaneous fitting for all quark masses
in each cut-off scale.
Figure \ref{fig:vpf_v_high} shows the VPF in our ensemble 
in the fitting $Q^2$ region, 0.93 GeV$^2$<$Q^2$<1.8 GeV$^2$.
The accuracy of the VPF at each momentum is almost subpercent in this fitting region, 
which is improved by a factor 10 and more compared to \cite{Shintani:2010ph}. 
One sees that the fitting function is in good agreement with 
our accurate lattice data within a 1$\sigma$ statistical error, and 
thus it turns out that the OPE formula up to N$^3$LO in Eq.(\ref{eq:ope}) is able to 
precisely describe the $q^2$ dependence of the lattice results above $Q^2=0.9$ GeV$^2$. 

\begin{figure}
\begin{center}
  \includegraphics[width=90mm]{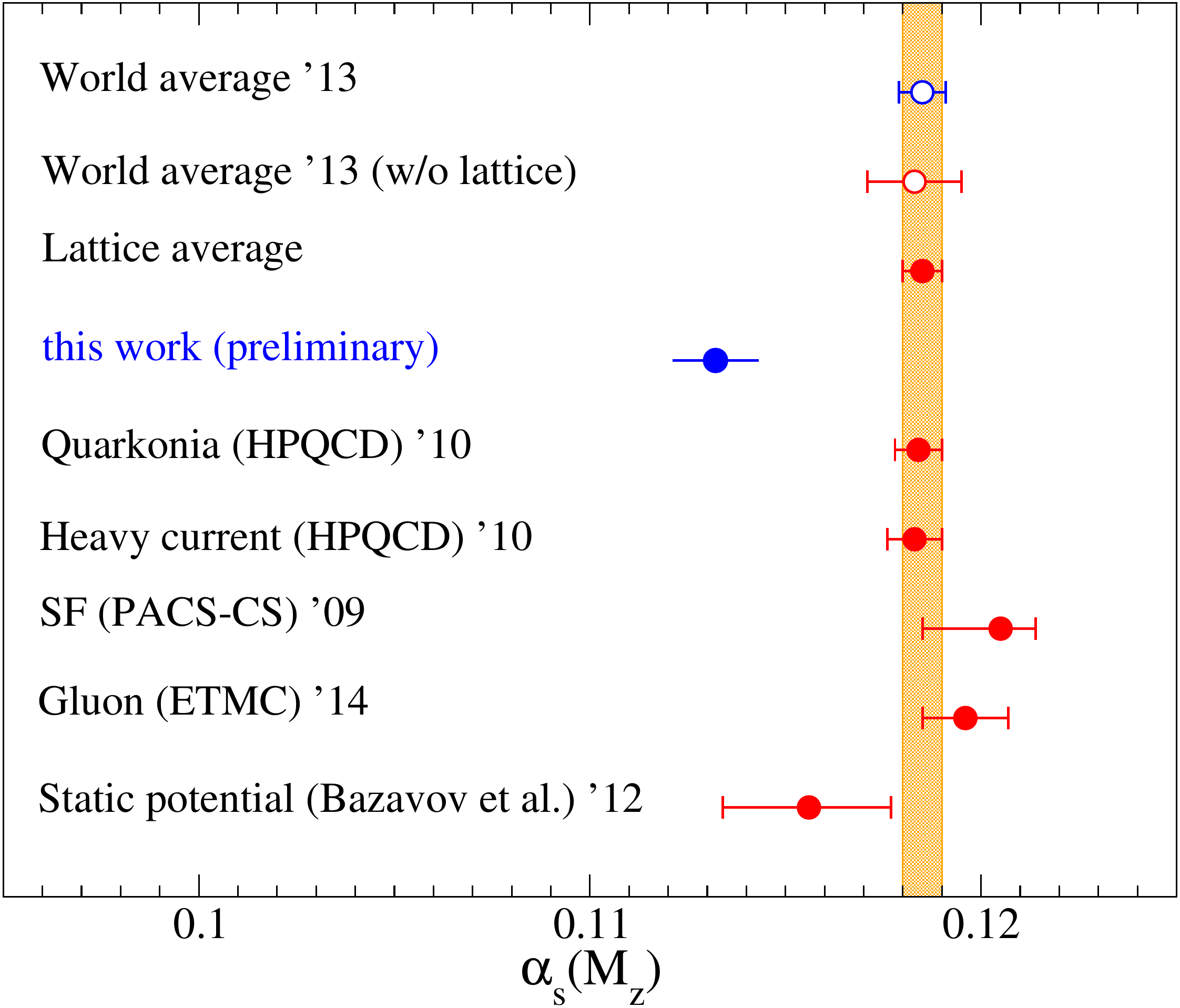}
  \caption{Summary of lattice calculation of $\alpha_s^{(5)}(M_Z)$.}
  \label{fig:summary}
\end{center}
\end{figure}

In Table \ref{tab:res}, we compare the lattice results obtained in 
this calculation and previous work by the JLQCD collaboration \cite{Shintani:2010ph}. 
In Ref.\cite{Shintani:2010ph}, they have used the overlap fermion 
at relatively small volume and one lattice spacing, and also 
the condensate is fixed in computed value in the others.
The strong coupling constant at $\mu=2$ GeV is consistent between 
our result and JLQCD, and our results are around factor 1.5 improved
for the $24^3$ lattice. 
The fitting result with a free chiral condensate also makes 
an improvement of $\chi^2$/dof, and furthermore 
which is compatible with $0.256(6)$ \cite{Aoki:2010dy} given from 
hadron spectroscopy. 
It is also compatible between different cut-off results, and 
taking continuum limit for $\alpha_s$ and $\langle q\bar q\rangle$
increases these values less than 5\% for $\alpha_s$
and 20--30\% for quark condensate. 
We notice that the chiral condensate obtained by analysis in the OPE formula is 
large over 1$\sigma$ error, although statistical and systematic errors are still large. 
It indicates that there is necessary to do more detailed analysis 
of systematic uncertainties for $\langle q\bar q\rangle$. 
Figure \ref{fig:summary} shows the comparison with other lattice 
calculations of the strong coupling constant at the $M_Z$ scale.

\begin{table}
\begin{center}
\caption{Result of fitting lattice VPF with OPE formula.}
\label{tab:res}
\begin{tabular}{cccccccc}
\hline\hline
Lattice & $a^{-1}$ GeV & $\alpha_s/\pi$ & $\Lambda^{(3)}_{\overline{MS}}$ (GeV) &
$-\langle q\bar q\rangle_{\overline{MS}}^{1/3}$ (GeV) &
$\langle\frac{\alpha_s}{\pi}GG\rangle$ (GeV$^4$) &
$\chi^2$/dof \\
\hline
$24^3\times 64$ & 1.73 & 0.08192(39) & 0.2486(27) & 0.276(11) & 0.237(18) & 1.8\\
                & & 0.08193(34) & 0.2486(24) & 0.256[fix] & 0.205(3) & 2.5 \\
$32^3\times 64$ & 2.23 & 0.08317(85) & 0.2572(58) & 0.325(29) & 0.741(138) & 1.3 \\
                & & 0.08196(28) & 0.2489(19) & 0.256[fix] & 0.475(10) & 2.7\\
$a\rightarrow 0$ & & 0.0844(20) & 0.265(14) & 0.390(68) & \\
\hline
JLQCD\cite{Shintani:2010ph} & 1.83 & 0.0817(6) & 0.247(5) & 0.242[fix] & -0.020(2) & 2.8\\
\hline\hline
\end{tabular}
\end{center}
\end{table}

\subsection*{Summary}

In these proceedings, we present the recent analysis of the vector vacuum polarization 
function (VPF) calculated in lattice QCD using domain-wall fermion. 
In this work, we adopt a new error reduction technique, so called as 
{\it All-mode-averaging (AMA)}, and thus the precise VPF is obtained. 
For the calculation of the strong coupling constant, 
$\alpha_s^{\rm \overline{MS}}$ is given from the OPE formula of the vector VPF
in two different cut-offs, 
and so that the lattice artifact effect is also taken into account.

\section*{7~~~~$\alpha_s$ from the QCD static energy~\footnote{X.~Garcia i Tormo; 
collaboration with Alexei Bazavov, Nora Brambilla, P\'eter
Petreczky, Joan Soto, and Antonio Vairo on the work
reported here is acknowledged.}}
\addcontentsline{toc}{section}{\protect\numberline{7}{$\alpha_s$ from the QCD static energy}}

The energy between two static sources in QCD as a function of the separation distance $r$, i.e. the QCD static energy $E_0(r)$, is a basic object to understand the dynamics of the theory. It can be computed in perturbation theory at short distances, and it encodes the confining dynamics of QCD at large distances. There has been a lot of progress, in the last few years, on its evaluation. On the one hand, lattice computations with 2+1 light-quark flavors have become available~\cite{Bazavov:2011nk} in a wide distance range, including the short-distance perturbative region; on the other hand, the perturbative evaluation of $E_0(r)$ has reached three-loop accuracy~\cite{Smirnov:2008pn,Anzai:2009tm,Smirnov:2009fh}, and logarithmically enhanced terms have been resummed to high accuracy~\cite{Brambilla:2009bi}. This progress opened the possibility to obtain a determination of $\alpha_s$ by comparing the short-distance part of $E_0(r)$ computed on the lattice with the perturbative predictions. 

\begin{figure}
\begin{center}
\includegraphics[width=0.75\textwidth]{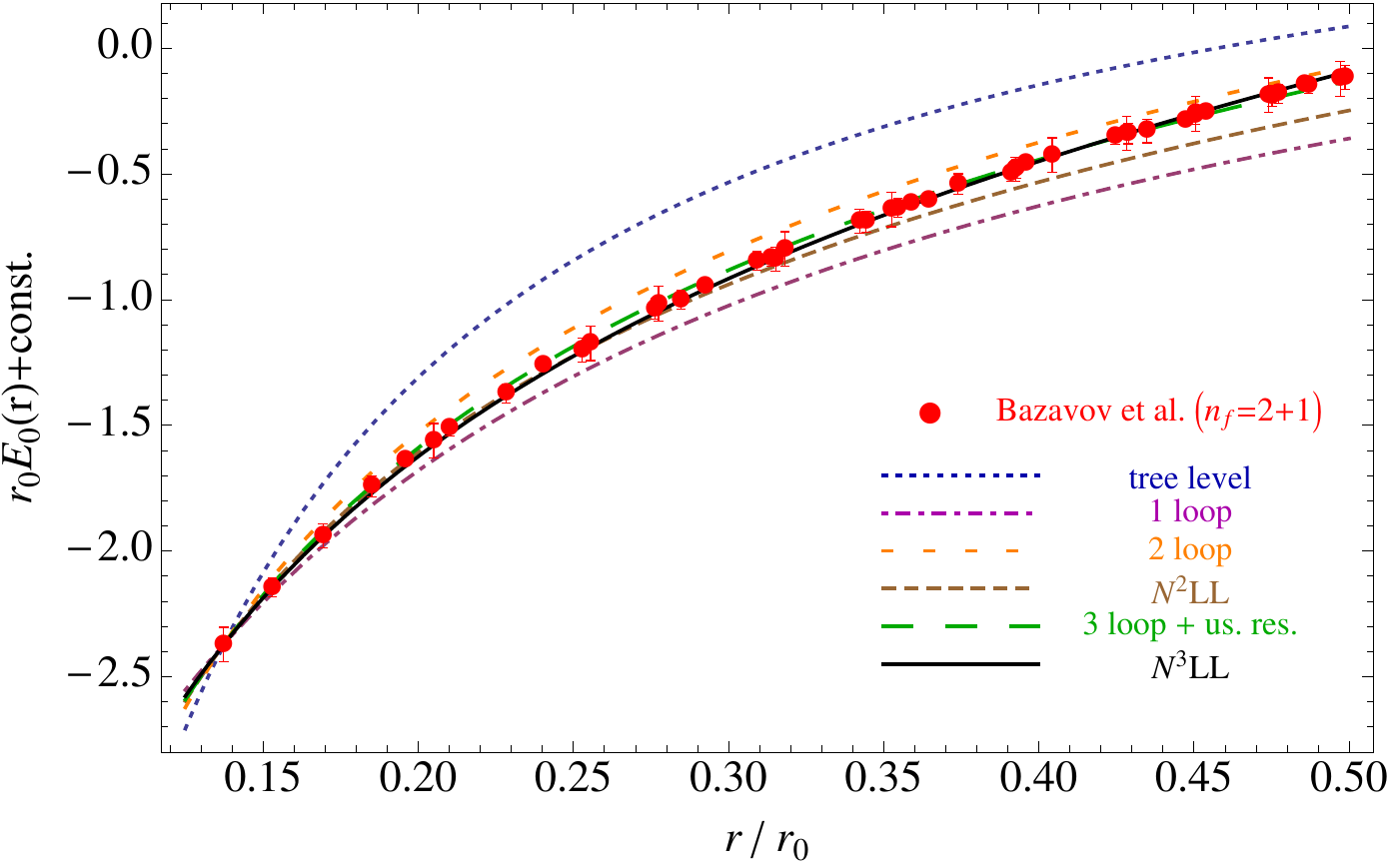}
 \caption{Comparison of the static energy with the lattice data of Ref.~\cite{Bazavov:2011nk}, red points, at different orders of perturbative accuracy (see Ref.~\cite{Tormo:2013tha} for a detailed description of which terms enter at each perturbative order). $\alpha_s(M_Z) = 0.1156$ is used for all the curves. All quantities are written in units of the scale parameter $r_0=0.468\pm0.004$~fm \cite{Bazavov:2011nk}.}
    \label{garcia:fig1}
\end{center}
\end{figure}

To achieve that, one goes to the short-distance region, where it is expected that perturbation theory is enough to describe the lattice results, and uses the comparison to find the values of $\alpha_s$ that are allowed by lattice data. This is done by demanding that the agreement of the theoretical predictions with the lattice data improves when the perturbative order of the computation is increased. In order to perform this analysis, it is important to remember that the normalization of the lattice result is not physical, but only its slope is. Therefore, one must either normalize the energy to a certain value at a given distance, or take a derivative and work directly with the force. This comparison and analyses were performed in Ref.~\cite{Bazavov:2012ka}, which is based on the lattice data of Ref.~\cite{Bazavov:2011nk}, see also Ref.~\cite{Tormo:2013tha} for a summary of all the perturbative expressions that enter in the analyses. The concrete implementation of the aforementioned guiding principle to extract $\alpha_s$, of finding values of $\alpha_s$ that make the agreement with lattice increase at higher perturbative orders, is based on the procedure devised in Ref.~\cite{Brambilla:2010pp}. Roughly speaking, it consists on letting the scale in the perturbative expansion vary around its natural value, then fit to the lattice data points at different perturbative orders, and finally select the values where the reduced $\chi^2$ of the fits decreases when increasing the perturbative order. To estimate the error of the result, which should reflect the uncertainties due to unknown higher perturbative orders, one considers the difference with the result for $\alpha_s$ obtained at the previous perturbative order, and the spread of values obtained within a given order. The final outcome of the analysis is
  \begin{equation}
    \label{Garcia:eq1}
    \alpha_s(M_Z) = 0.1156^{+0.0021}_{-0.0022},
  \end{equation}
i.e. we obtain an uncertainty of about $\pm2\%$ for $\alpha_s$ at the scale $M_Z$. The extraction is at three-loop accuracy, including resummation of logarithmically enhanced terms, and uses data in the energy range between 3~GeV and 0.8~GeV. In Fig.~\ref{garcia:fig1} we show, for illustration, a comparison of the static energy at different orders of accuracy with lattice data, using the value of $\alpha_s$ in Eq.~(\ref{Garcia:eq1}).

One difficulty in this analysis is to know whether or not the current lattice data has really reached the purely perturbative regime, and with enough precision to perform the extraction. To undoubtedly state this point is not an easy task, but in that regard, the facts that agreement with lattice is indeed found to improve when increasing the perturbative order, and that the perturbative curves can describe the data quite well, can be seen as positive evidence that this is the case. In addition, the result is not very sensitive to the precise upper limit in $r$ that we use for the fits, i.e. using a slightly larger or smaller distance range the result would be essentially the same. In any case, further studies to verify this point are certainly warranted. Another important issue is to have a reliable estimate of the systematic discretization errors for the shorter-distance lattice data points. 

An updated analysis, that will carefully and comprehensively address the points mentioned above, is ongoing and can be expected to appear shortly. In addition, it will also include new lattice data at shorter distances.

To summarize, the recent progress in the evaluation of the static energy opened the possibility to obtain a determination of $\alpha_s$ by comparing perturbative expressions with lattice data. It constitutes a novel determination of $\alpha_s$, complementary to other existing extractions. The current value from this extraction is given in Eq.~(\ref{Garcia:eq1}), and has a $\pm2\%$ error. An update of the analysis is ongoing and will appear in the near future.

\section*{8~~~~The program GAPP and $\alpha_s$~\footnote{J.~Erler; 
update of the contribution~\cite{Erler:2011ux} to the {\it Workshop on Precision Measurements of 
$\alpha_s$}~\cite{Bethke:2011tr}, Munich, Germany, Feb.~9--11, 2011.}}
\addcontentsline{toc}{section}{\protect\numberline{8}{The program GAPP and $\alpha_s$}}

The FORTRAN package GAPP~\cite{Erler:1999ug} (Global Analysis of Particle Properties) computes 
pseudo-observables and performs least-$\chi^2$ fits in the $\overline{\rm MS}$ scheme.
Fit parameters besides $\alpha_s$ and $M_H$ include the heavy quark masses which are
determined from QCD sum rule constraints thus affecting and being affected by $\alpha_s$.
This set can be extended to parameters describing physics beyond the Standard Model, 
such as new contributions to gauge boson self-energies (oblique parameters),
a 4th generation of fermions or extra neutral gauge bosons. 
When possible, analytical expressions (or expansions) are used to capture the full dependence on
$\alpha_s$ and the other fit parameters. 

$Z$-pole observables  from LEP~1 and SLC include the $Z$-width, $\Gamma_Z$, 
hadronic-to-leptonic partial  $Z$-width ratios, $R_\ell$, 
and the hadronic peak cross section, $\sigma_{\rm had}$.
These are most sensitive to $\alpha_s$ by far, 
but the weak mixing angle enters and needs to be known independently.  
Thus, the extracted $\alpha_s$ depends on the set of other, purely electroweak measurements employed in the fits,
such as various asymmetries and experiments exploiting parity violation. 
There is also explicit $\alpha_s$ dependence in the electroweak observables,
most notably in the mass and width of the $W$-boson, the Higgs boson branching ratios,
and the muon magnetic moment anomaly.
The statistical and systematic experimental correlations of $\Gamma_Z$, $\sigma_{\rm had}$ 
and the $R_\ell$ are known, small and included. 
The parametric uncertainties (such as from $M_H$) are non-Gaussian but treated exactly. 
The theoretical errors in $\Gamma_Z$, $\sigma_{\rm had}$, and the $R_\ell$ are identical
(except for those from unknown but very small higher order electroweak and mixed QCD/electoweak effects)
and induce a negligibly small uncertainty in $\Delta\alpha_s(M_Z) < 0.0001$, as we discuss below.

It is now dominated by the uncertainty induced by uncalculated terms in the massless non-singlet contribution,
which is known up to ${\cal O}(\alpha_s^4)$ and where the last known term~\cite{Baikov:2008jh}
produced an increase in the extracted $\alpha_s(M_Z)$ by $0.00047$.
As in the case of $\tau$ decays, one may opt for either fixed-order perturbation theory (FOPT) or
contour-improved perturbation theory (CIPT)~\cite{LeDiberder:1992te}, and one take the 
difference as the uncertainty ($\pm 0.00005$) associated with the non-singlet part.
(This difference has the opposite sign from $\tau$ decays indicating 
that the non-singlet expansions behave quite differently in higher orders and 
the theory errors can be taken as uncorrelated between $Z$ and $\tau$
decay observables.)

The singlet (QCD annihilation) contribution~\cite{Kniehl:1989bb,Larin:1994va} is numerically 
significant due to the $(t,b)$-doublet mass splitting which strongly affects the axial-vector correlator. 
Failure to include the singlet pieces would result in a value of $\alpha_s$ which be smaller by about 0.003.
The NNL order $(\alpha_s^4)$ terms have recently been obtained~\cite{Baikov:2012er},
except for those with additional mass suppressions like $M_Z^2/m_t^2$ or $m_b^2/M_Z^2$.
This resulted in a reduction of the extracted $\alpha_s$ by $6 \times 10^{-5}$,
and removed the previously~\cite{Erler:2011ux} dominant theory uncertainty.
The remaining unknown terms should not affect $\alpha_s$ by more than the $10^{-5}$ level. 

The global electroweak fit excluding $\tau$ decays yields $\alpha_s(M_Z) = 0.1192 \pm 0.0027$ 
(if one restricts oneself to the $Z$ factories, LEP and SLC, one finds $0.1198 \pm 0.0028$).
These results are expected to be stronger affected by physics beyond the Standard Model than other $\alpha_s$ determinations
which is the primary reason to include another $\alpha_s$ constraint in the fits as a control.
If the new physics affects only the gauge boson propagators (oblique corrections) the resulting 
$\alpha_s(M_Z) = 0.1199^{+0.0027}_{-0.0030}$ hardly changes,
while allowing new physics corrections to the $Zb\bar{b}$-vertex gives the lower $\alpha_s(M_Z) = 0.1167 \pm 0.0038$.        

As the aforementioned $\alpha_s$ control we choose the $\tau$ lifetime, $\tau_\tau$, 
not least because of its transparent (even if controversial) theory uncertainty. 
Moreover, $\tau$ decay kinematics provides a double zero near the branch cut of the two-point correlator, 
mitigating the impact of the region where the operator product expansion (OPE) is not strictly applicable.
And just as $Z$ decays, $\tau$ decays are also known~\cite{Baikov:2008jh} to NNNL order $(\alpha_s^4)$.
Our master formula~\cite{Erler:2002bu} reads,
\begin{eqnarray}
\label{tau}
\tau^{\rm expt}
\equiv \tau [{\cal B}_{e,\mu}^{\rm expt}, \tau_{\rm direct}^{\rm expt}]
=  \hbar\, \frac{1 - {\cal B}_s^{\rm expt}}{\Gamma_e^{\rm theo} + \Gamma_\mu^{\rm theo} + \Gamma_{ud}^{\rm theo}} 
= 291.31 \pm 0.45~{\rm fs} \,,
\end{eqnarray}
where $\tau_{\rm direct}^{\rm expt} = 290.6~(1.0)$~fs is the directly measured $\tau$ lifetime~\cite{Beringer:1900zz}.
$\tau [{\cal B}_{e,\mu}^{\rm expt}] = 291.49~(0.50)$~fs is the combination of indirect determinations,
using $\tau [{\cal B}_{e, \mu}] = \hbar\, {\cal B}_{e,\mu}^{\rm expt}/\Gamma_{e,\mu}^{\rm theo}$
and the experimental branching ratios, ${\cal B}_e^{\rm expt} = 0.1783~(4)$ and ${\cal B}_\mu^{\rm expt} = 0.1741~(4)$,
together with their 13\% correlation~\cite{Beringer:1900zz}.  
The constraint~(\ref{tau}) translates into a determination of the QCD correction (including non-perturbative corrections)
of $\delta_{\rm QCD} = 0.1948 \pm 0.0032$.
Decays into net strangeness, $S$, are plagued by the uncertainty in the  $\overline{\rm MS}$ strange mass, $\hat{m}_s (m_\tau)$,
and a poorly converging QCD series proportional to $\hat{m}_s^2$, so that in Eq.~(\ref{tau}) 
we employ the measured $\Delta S = -1$ branching ratio, ${\cal B}_s^{\rm expt} = 0.0287~(7)$~\cite{Beringer:1900zz}.

The partial $\tau$-width into light quarks contains logarithmically enhanced electroweak corrections, 
$S(m_\tau, M_Z) = 1.01907 \pm 0.0003$~\cite{Erler:2002mv}, and reads (employing FOPT as advocated in Ref.~\cite{Beneke:2008ad}),
\begin{eqnarray}
\label{Gammaud}
\Gamma_{ud}^{\rm theo}
&=& \frac{G_F^2 m_\tau^5 |V_{ud}|^2}{64\pi^3} S(m_\tau, M_Z) \left(1 + \frac{3}{5} \frac{m_\tau^2}{M_W^2} \right)  \times \\ \nonumber
&  & \left( 1 + \frac{\alpha_s(m_\tau)}{\pi} + 5.202 \frac{\alpha_s^2}{\pi^2} + 26.37 \frac{\alpha_s^3}{\pi^3} + 127.1 \frac{\alpha_s^4}{\pi^4} 
- 1.393 \frac{\alpha(m_\tau)}{\pi} + \delta_q \right) \,,
\end{eqnarray}
where $\delta_q$ collects quark condensate, $\delta_{\rm NP}$~\cite{Boito:2012cr}, as well as heavy and light quark mass effects.
Note, that $\delta_{\rm NP}$ is obtained from fits to $\tau$ decay spectral functions.
There is much activity targeted at improving such fits and the corresponding error can be expected to decrease in
the future due to the inclusion of more solid theoretical constraints, as well as more precise data
(for further reading, see~\cite{Boito:2012nt} and references therein).
The dominant experimental and theoretical errors are given in the following tables, respectively:
\begin{center}
\begin{tabular}{|c||c|r|}\hline
source                                       & uncertainty            & $\Delta\alpha_s(M_Z)$  \\ \hline\hline
$\Delta\tau^{\rm expt}$           & $\pm 0.45$~fs       & $\mp 0.00033$ \\ \hline 
$\Delta{\cal B}_s^{\rm expt}$ & $\pm 0.0007$       & $\mp 0.00015$ \\ \hline 
$\Delta V_{ud}$                       & $\pm 0.00015$     & $\mp 0.00004$ \\ \hline 
$\Delta m_\tau$                       & $\pm 0.16$~MeV & $\mp 0.00002$ \\ \hline \hline
total                                           &                                 & 0.00037             \\ \hline 
\end{tabular}
\hspace{40pt}
\begin{tabular}{|c||c|c|c|}\hline
source                                       & uncertainty                  &based on                      & $\Delta\alpha_s(M_Z)$  \\ \hline\hline
PQCD                                       & $\mp 0.0142$             & $\alpha_s^4$-term     & $^{+0.00180}_{-0.00145}$ \\ \hline 
RGE                                          & $\beta_4 = \pm 579$ & \cite{Erler:1999ug}     & $\mp 0.0004$ \\ \hline 
$\delta_{\rm NP}$                          & $\pm 0.012$             &\cite{Boito:2012cr} & $\mp 0.0013$ \\ \hline 
OPE \hspace{-27pt} ------  \hspace{27pt}& $\pm 0.0008$ & \cite{'tHooft:1976fv,Davier:2005xq} & $\mp 0.00012$ \\ \hline
$S(m_\tau, M_Z)$                   & $\pm 0.0003$             & \cite{Erler:2002mv}     & $\mp 0.00004$ \\ \hline\hline
total                                           &                                      &                                        & $^{+0.0022}_{-0.0020}$ \\ \hline 
\end{tabular}
\end{center}
The perturbative QCD (PQCD) error dominates and is estimated as the $\alpha_s^4$-term in Eq.~(\ref{Gammaud}).
It is re-calculated in each call in the fits to access its $\alpha_s$-dependence and features asymmetric.
It basically covers the range from the higher values favored by CITP down to the lower ones one obtains from 
assuming that the roughly geometric form of FOPT continues.
Note that if CIPT is used, the error from the renormalization group evolution (RGE) parametrized 
by the unknown 5-loop $\beta$-function coefficient, $\beta_4$, and part of the PQCD error are correlated.
Effects breaking the OPE, \mbox{OPE \hspace{-27pt} ------}, 
are estimated by assuming the instanton motivated functional form~\cite{'tHooft:1976fv}, 
$A\, \alpha_s^{-6} \exp[- 2\pi/\alpha_s(s_0)]$,
and adjusting $A$ to the difference between the OPE and data curves in Fig.~22 of Ref.~\cite{Davier:2005xq}.
Our result after evolution to the Z-scale is $\alpha_s[\tau_\tau] = 0.1192^{+0.0022}_{-0.0020}$.
Adding $\tau$ decays as a constraint to the global electroweak fit discussed before
results in the overall value, $\alpha_s (M_Z) = 0.1192 \pm 0.0016$.

In conclusion, due to their inclusive nature, both $Z$ decays and $\tau$ decays provide theoretically clean 
determinations of $\alpha_s$ at N$^3$LO, where the resulting central values are currently 
in perfect agreement with each other.
It should be mentioned, however, that one of the five chief individual $Z$ pole constraints on $\alpha_s$,
namely $\sigma_{\rm had}$, deviates by $1.7~\sigma$ from the SM best fit prediction.
By itself it would produce a very small value, $\alpha_s (M_Z) = 0.107 \pm 0.007$, 
thus dragging down the $Z$ pole average.

\section*{9~~~~Uncertainties in the renormalization group evolution of $\alpha_s$~\footnote{L.~de la Cruz, J.~Erler}}
\addcontentsline{toc}{section}{\protect\numberline{9}{Uncertainties in the renormalization group evolution of $\alpha_s$}}

Values of $\alpha_s$ which are determined at very low energy scales need to be
evolved to a common reference scale (typically taken as the $Z$ boson mass, $M_Z$) where they can be compared.
This introduces an additional (correlated) theoretical uncertainty.  
Here, we study this uncertainty for the case of $\tau$-decay extractions
and show that it is quite small, but it would be a non-negligible error component for ambitious
$\alpha_s$ determinations with errors targets at the few-$\permil$ level in $\alpha_s(M_Z)$.

The renormalization group evolution (RGE) of $\alpha_s$ is governed by the QCD beta function,
\begin{equation}
{d a \over d\ln \mu^2} = \beta(a) = - \sum_{k=0}^\infty {\beta_k}a^{k+2} \qquad\qquad 
\left[ \mbox{with } a \equiv {\alpha_s(\mu)\over \pi}\right].
\label{rge}
\end{equation}
In addition, if a quark threshold is crossed so that the number of active quark flavors, $n_f$, changes,
one needs to consider the associated threshold effects and their uncertainties.  
The beta function coefficients in the $\overline{\rm MS}$-scheme are known to four-loop precision 
for a more general gauge theory consisting of a single gauge group with arbitrary $n_f$, 
but restricted to only one {\em type\/} of fermion representation~\cite{vanRitbergen:1997va}.
Beyond that, partial results on the higher order coefficients are available from Ref.~\cite{Gracey:1996he}
where the leading terms in the large $N_F$ expansion have been obtained.
However, for the most relevant applications with $3\leq N_F \leq 6$, these terms are rather small.

One can also try to make educated guesses at the coefficients of the higher orders. 
A popular approach involves Pad\'e approximants, which has been quite successful~\cite{Ellis:1996zn} 
in predicting $\beta_3$ in QCD (before its calculation), but failed in other cases such as QED,
in part due to the difficulty to assess terms involving higher order Casimir invariants. 
There exists a number of variants, where the so-called weighted asymptotic Pad\'e approximant prediction
(WAPAP) gives rise to the prediction~\cite{Ellis:1997sb},
\begin{equation}
\beta_4^\text{WAPAP} \simeq 741 - 214\ n_f + 21\ n_f^2 - 0.0486\ n_f^3 - 0.0018\ n_f^4\ ,
\label{ldelacruz:eq1}
\end{equation}
(the last term is input to a fit~\cite{Gracey:1996he}) and thus 
$\beta_4^\text{WAPAP} \simeq 287\ (218, 190)$ for $n_f = 3\ (4,5)$, respectively.
We will not use these results, because it is difficult to quantify the uncertainty on such a prediction,
and even less is known on the factorially growing higher order $\beta_k$ for $k \geq 5$.
But they give a sense of the size of the truncation error, and agree roughly with what we find below.

To four-loop precision, the LO, NLO, NNLO, and N$^3$LO logarithms have to be summed correctly,
but beyond that both the truncation of the beta function and its solution are subject to ambiguities
which are formally of the same order as the unknown terms in Eq.~(\ref{rge}).
Specifically, without loss of precision (as it turns out, there is a gain in precision) Eq.~(\ref{rge})
can be rewritten (by dividing by $\beta(a)$ and re-expanding the power series) 
in a form that can be conveniently integrated, with the result,
\begin{equation}
\label{rge2}
{1\over \bar{a}} - {1\over \bar{a}_0} + \ln {\bar{a}\over \bar{a}_0} + \sum_{k=1}^\infty {q_{k+1} \over k} (\bar{a}^k - \bar{a}_0^k) = 
{\beta_0^2 \over \beta_1} \ln {\mu^2 \over\mu_0^2} \qquad\qquad
\left[ \bar{a}     \equiv {\beta_1 \over \beta_0} a\ , \quad
\bar{a}_0 \equiv {\beta_1 \over \beta_0} {\alpha_s(\mu_0)\over \pi}\right],
\end{equation}
where the lowest order $q_i$ are given by,
\begin{equation}
q_2 \equiv {\beta_0 \beta_2 \over \beta_1^2} - 1, \qquad
q_3 \equiv {\beta_0^2 \beta_3 \over \beta_1^3} - {2 \beta_0 \beta_2 \over \beta_1^2} + 1, \qquad
q_4 \equiv {\beta_0^3 \beta_4 \over \beta_1^4} - {2 \beta_0^2 \beta_3 \over \beta_1^3} - 
{\beta_0^2 \beta_2^2 \over \beta_1^4} + {3 \beta_0 \beta_2 \over \beta_1^2} - 1.
\end{equation}
Of course, nullifying all $\beta_k$ with $k \geq 4$ will differ formally from nullifying all $q_k$ with $k \geq 4$
only in the unknown orders, so that both truncations seem equally valid.
We note, however, that the known $\beta_k$ in the power series~(\ref{rge}) 
are all positive for $n_f < 6$ (that is, the cases at hand).
Upon truncation this produces $q_i$ of strong combinatorial growth and alternating signs.
On the other hand, while it is also true that $q_i > 0$ for $n_f < 5$,
the truncation, inversion and re-expansion of the series now produces strictly positive signs to all orders 
(also with combinatorial growth), thus introducing a bias.

Eq.~(\ref{rge2}) can be solved for $a$ iteratively, giving the compact closed-form analytical expression,
\begin{equation}
{1 \over \bar{a}} = {L \over \bar{a}_0} + \ln \left[ L + \bar{a}_0 \ln (L + \bar{a}_0 \ln L) + \bar{a}_0^2 q_2 {L - 1\over L} \right] + 
\bar{a}_0 q_2 \left[ 1 - {1 \over L + \bar{a}_0 \ln L} \right] + \bar{a}_0^2 q_3 {L^2 - 1\over 2 L^2}\ ,
\label{rge3}
\end{equation}
where we defined,
\begin{equation}
L \equiv 1 +         a_0 \beta_0 \ln {\mu^2 \over \mu_0^2} = 1 + \bar{a}_0 {\beta_0^2 \over \beta_1} \ln {\mu^2 \over \mu_0^2}\ .
\end{equation}
Eq.~(\ref{rge3}) may be useful for many practical applications (for example, if one wishes
to implement the contour improved perturbation theory of Ref.~\cite{LeDiberder:1992te}).
We now compare the results for $\alpha_s(M_Z)$ corresponding to Eq.~(\ref{rge2}) with $q_k = 0$ (for $k>3$)
with integrating Eq.~(\ref{rge}) with $\beta_k=0$ (for $k>3$), as well as with Eq.~(\ref{rge3})
and its lower order versions:
\begin{center}
\begin{tabular}{|c||c|c|c|}\hline
approximation & $q_k=0\ (k>3)$ & $\beta_k=0\ (k>3)$ & Equation~(\ref{rge3}) \\ \hline\hline
LO                    &     0.124 317     &     0.124 317           &     0.124 317              \\
NLO                  &     0.118 351     &     0.118 796           &     0.118 616              \\
NNLO               &     0.118 355     &     0.118 376           &     0.118 309               \\ 
NNNLO            &     0.118 201     &     0.118 216           &     0.118 155               \\ \hline\hline
\end{tabular}
\end{center}
For this comparison we ignored threshold effects (see below), but evolved $\alpha_s$ from the assumed initial value 
$a_0 = 0.1$ at $\mu_0 = m_\tau$ in the 3-flavor theory {\em down} to $m_c(m_c) = 1.27$~GeV 
and then in the 4-flavor theory up to $m_b(m_b) = 4.2$~GeV and from there with $n_f = 5$ to $\mu = M_Z = 91.1876$~GeV.
Somewhat coincidentally, the N$^3$LO effect is very similar in all cases, and one may take it ($\pm 1.54 \times 10^{-4}$) 
as the uncertainty in $\alpha_s(M_Z)$ corresponding to the $\beta$-function truncation.
Note, that this effect could be mimicked by $\beta_4 \approx 248$ 
(assuming it was $n_f$-independent and $\beta_k=0$ for $k>4$) 
which is of similar size as $\beta_4^\text{WAPAP}$ in Eq.~(\ref{ldelacruz:eq1}).

At the quark thresholds, which we identify with the $\overline{\text{MS}}$ masses $m_c(m_c)$ and $m_b(m_b)$
quoted above, one needs to include matching effects, relating $a$ in the theory with $n_f$ and $n_f + 1$ active quarks,
respectively. These are known to four-loop order~\cite{Kniehl:2006bg},
\begin{equation}
\label{matching}
a_{n_f + 1} = a_{n_f} \left[ 1 - {11 \over 72}\, a_{n_f}^2 - (0.9721 - 0.0847\, n_f)\, a_{n_f}^3 - 
(5.100 -1.010\, n_f - 0.022\, n_f^2)\, a_{n_f}^4 \right],
\end{equation}
which is one loop-order higher than what would be necessary to correctly sum all N$^3$LO logarithms.
Estimating the error associated with the truncation of Eq.~(\ref{matching}) again as the last calculated term,
yields $1.8 \times 10^{-4}$ and $4.2 \times 10^{-6}$ for $\alpha_s(m_c)$ and $\alpha_s(m_b)$, respectively.
The error budget for $\alpha_s(M_Z)$ is then,
\begin{equation}
\alpha_s(M_Z) = 0.117\, 967\ (154)_{\beta}\ (14)_{c-\text{threshold}}\ (1)_{b-\text{threshold}}\ 
(65)_{\Delta m_c= \pm 35~{\rm MeV}}\ (12)_{\Delta m_b= \pm 24~{\rm MeV}}\ ,
\label{ldelacruz:eq7}
\end{equation}
where we also indicated the additional error introduced by the imperfect knowledge of the quark masses, 
even though these are input and thus not purely theoretical.
In total we find an uncertainty in the scale evolution between $m_\tau$ (with $n_f = 3$)
and $M_Z$ (with $n_f = 5$) of about 
\begin{equation}
\delta_{\rm RGE}\ \alpha(M_Z) = \pm 0.000\, 17
\end{equation}

\section*{10~~~~{\boldmath Electroweak corrections in the determination of
    $\alpha_\mathrm{s}$}~\footnote{S.~Dittmaier, A.~Huss, K.~Rabbertz}}
\addcontentsline{toc}{section}{\protect\numberline{10}{\boldmath Electroweak corrections in the determination of
    $\alpha_\mathrm{s}$}}

In the following we briefly summarize the salient features of electroweak (EW) radiative corrections
to processes at $\mathrm{e}^+\mathrm{e}^-$ and hadron colliders that are used in precision
determinations of the strong coupling constant $\alpha_\mathrm{s}$.
In the past, those corrections played a minor role in those processes, but EW corrections
are known to grow large at TeV energies due to enhancements by EW Sudakov and other high-energy 
logarithms, so that EW higher-order effects have to be carefully inspected in view of 
measurements at the LHC and at future $\mathrm{e}^+\mathrm{e}^-$ colliders. 

\subsection*{\boldmath Jet event-shape observables in $\mathrm{e}^+\mathrm{e}^-$ colliders}

At $\mathrm{e}^+\mathrm{e}^-$ colliders the strong coupling constant can be determined
upon measuring the 3-jet production cross section and related event-shape observables.  
The QCD corrections, which are known to next-to-next-to-leading-order 
including dedicated resummations, comprise the major part of radiative corrections,
but at this level of accuracy EW corrections are relevant as well. 
In Refs.~\cite{Denner:2009gx,Denner:2010ia} the
next-to-leading order (NLO) EW corrections to 3-jet
observables in $\mathrm{e}^+\mathrm{e}^-$ collisions including the
quark--antiquark--photon ($q\bar{q}\gamma$) final states were calculated and discussed.
The QCD corrections to these final states are of the same perturbative
order as the genuine EW corrections to
quark--antiquark--gluon ($q\bar{q}\mathrm{g}$) final states.  Since
photons produced in association with hadrons can never be fully
isolated, both types of corrections have to be taken into account.

\begin{figure}
\centerline{
\includegraphics[width=.46\textwidth,bb=150 490 380 710, clip]{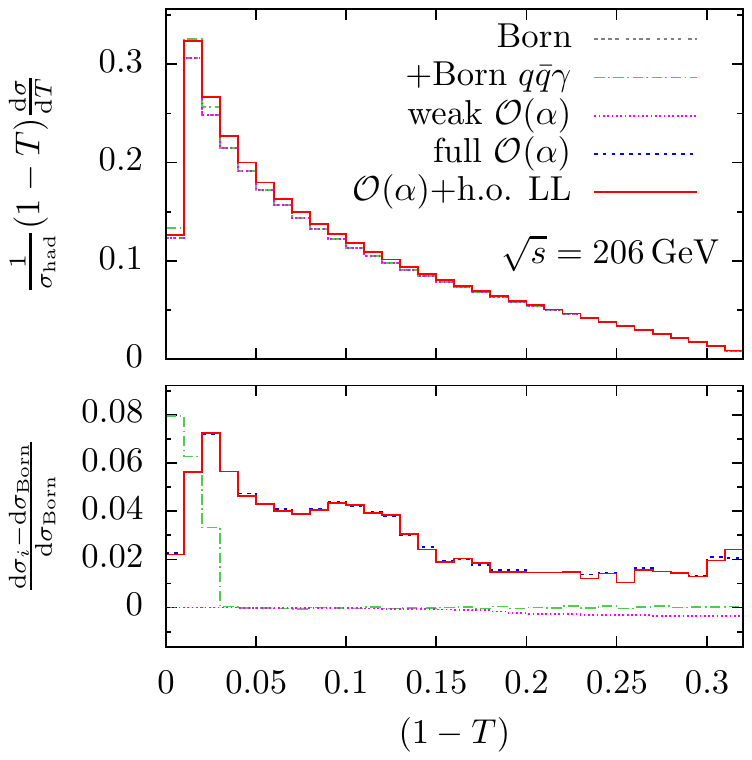} 
\hspace*{1em}
\includegraphics[width=.46\textwidth,bb=150 490 380 710, clip]{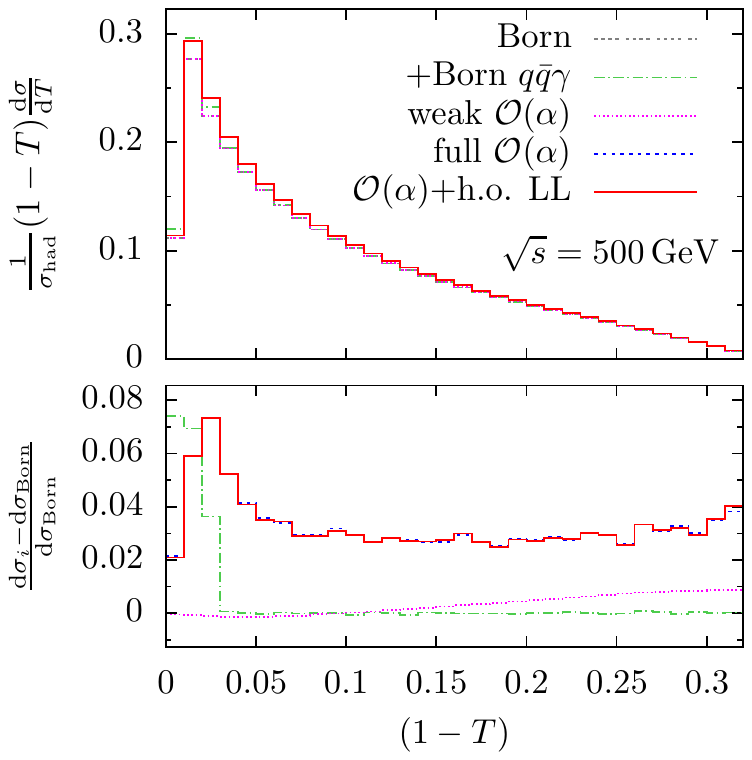}
}
\vspace*{-1em}
\caption{Differential thrust distribution
normalised to $\sigma_{\mathrm{had}}$ at different CM energies
$\sqrt{s}$ (upper plots) and corresponding relative corrections (lower plots).}
\label{dittmaier-fig:sigthrustnorm}
\end{figure}

Exemplarily, in Fig.~\ref{dittmaier-fig:sigthrustnorm} we show the differential thrust
distribution normalised to the total hadronic cross section
including NLO EW corrections 
for a typical LEP2 energy and for an energy of a future linear collider.
The distributions are weighted by $(1-T)$, evaluated at each bin centre.
The full ${\cal O}(\alpha)$ corrections
contain the  tree-level $q\bar q \gamma$ contribution
$\delta_\gamma$ and the NLO EW contribution
$\delta_{{\rm EW}}$, as defined in Eq.~(3) of Ref.~\cite{Denner:2009gx}. 
With ``weak ${\cal O}(\alpha)$'' we
denote the EW NLO corrections without purely photonic
corrections, and ``h.o.~LL'' indicates the inclusion of the
higher-order effects from photonic initial-state radiation via 
leading-logarithmic structure functions.

As discussed in Ref.~\cite{Denner:2009gx} for the Z~pole in more detail,
large ISR corrections cancel upon normalizing the event-shape distribution
to the hadronic cross section, resulting in EW
corrections of a few per cent. Moreover, effects from ISR
resummation are largely reduced as well.
Note that the photonic corrections develop a distinctive peak structure
of up to $9\%$ in size inside the thrust distribution, an effect that is
a remnant of the radiative return to the Z~pole,
which is suppressed, but not fully excluded, by the event-selection cuts.
The purely weak corrections are at the per-mille level at LEP energies
and only grow to the per-cent level for linear-collider energies.
Results on other event-shape variables, which are discussed in 
Ref.~\cite{Denner:2010ia} in detail, show corrections of the same generic size.

At LEP, data on event-shape distributions and jet cross sections have been corrected
for photonic radiation effects modelled
by standard leading-logarithmic parton-shower Monte Carlo
programs. Afterwards, it was not possible to reanalyze data again including the full
NLO EW corrections.
At a future linear collider, where the corrections become larger and, in particular,
the weak corrections reach some per cent, the full NLO EW corrections plus leading
higher-order effects from initial-satte radiation should be taken into account in predictions.

\subsection*{Jet production at hadron colliders}


The NLO results on hadronic jet production
shown here are based on the calculation of the weak radiative correction to 
dijet production presented in Ref.~\cite{Dittmaier:2012kx} 
and its extension to inclusive-jet production~\cite{Butterworth:2014efa}.
Corrections at this order have been previously calculated for the inclusive-jet case in Ref.~\cite{Moretti:2006ea},
and preliminary results to dijet production were shown in Ref.~\cite{Scharf:2009sp}.
The EW corrections to the production of three jets at hadron colliders are currently unknown, however,
significant cancellations of the EW corrections are expected in the cross section ratio 
$\sigma^\mathrm{pp\to3jets}/\sigma^\mathrm{pp\to2jets}$, 
which is important in the $\alpha_\mathrm{s}$ determination.

The EW interaction not only affects the jet-production cross section through radiative corrections,
but also via the exchange of EW gauge bosons already at leading order (LO). 
To this end, we define the correction factor $K_\mathrm{EW}^\mathrm{tree}$, which is given by the ratio of the
LO contributions of $\mathcal{O}(\alpha_\mathrm{s}\alpha,\,\alpha^2)$ 
with respect to the LO QCD prediction of order $\alpha_\mathrm{s}^2$.
Motivated by the known enhancement of EW corrections in the high-energy domain,
we have restricted our investigation 
at NLO to the purely weak corrections, which will be denoted by
$\alpha_\mathrm{s}^2\alpha_\mathrm{w}$ in the following with the corresponding 
correction factor $K_\mathrm{weak}^\mathrm{1-loop}$.
The remaining photonic corrections were separated in a gauge-invariant manner and are expected to 
give contributions at or below the per-cent level.

\begin{figure}
\begin{center}
  \parbox[c]{.5\textwidth}{\centering\hspace*{-3em}
    \includegraphics[width=\linewidth,trim=0 330 225 25, clip]
    {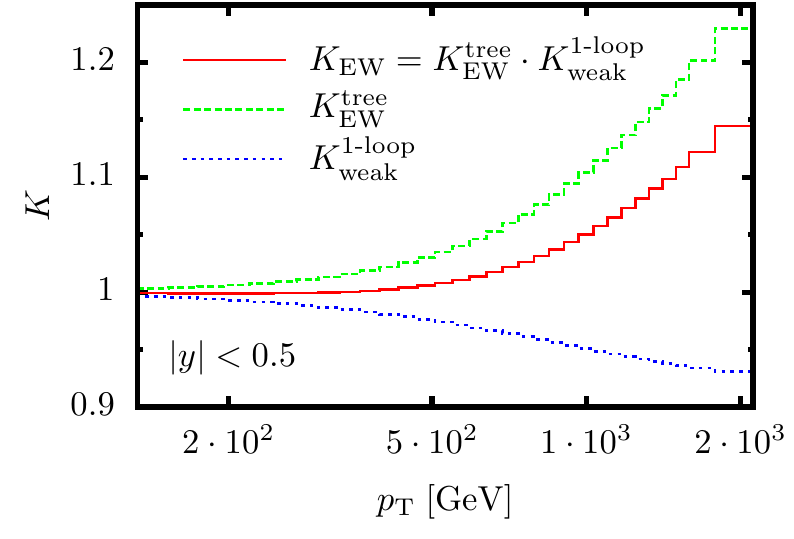} }
  \parbox[c]{.45\textwidth}{\centering
    \includegraphics[width=\linewidth]
    {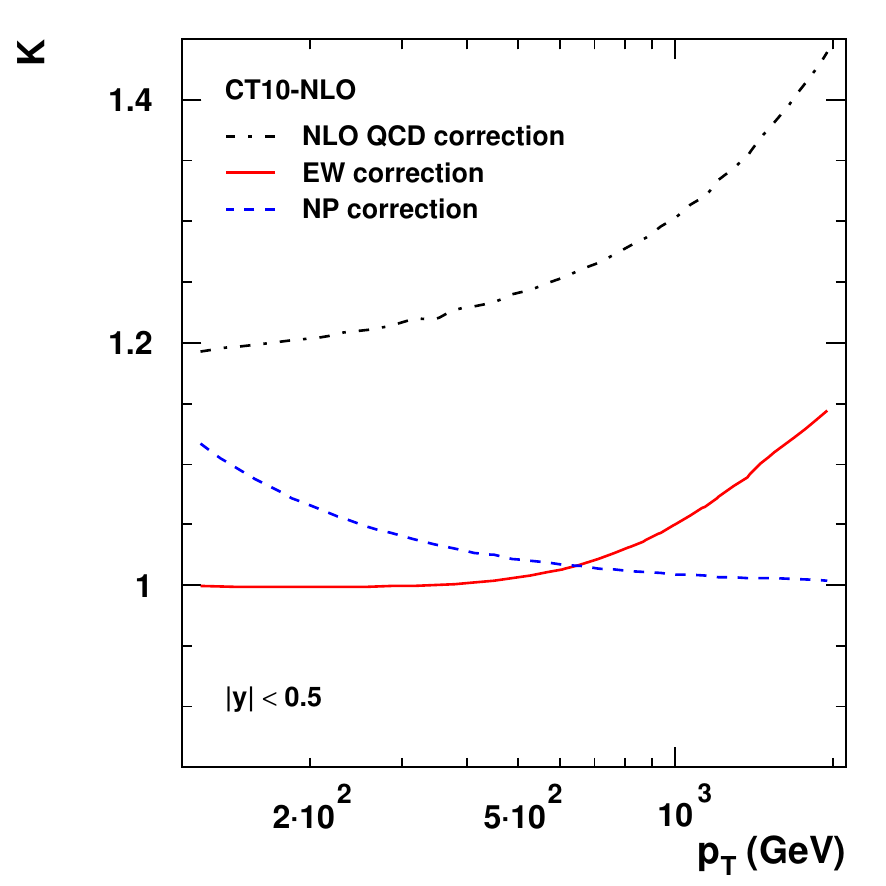} }\\[.5em]
  \parbox[c]{.45\textwidth}{\footnotesize
    (a)~The EW corrections~(solid red) broken up into 
    weak one-loop corrections of
    $\mathcal{O}(\alpha_\mathrm{s}^2\alpha_\mathrm{w})$ (dotted blue)
    and tree-level contributions of 
    $\mathcal{O}(\alpha_\mathrm{s}\alpha,\,\alpha^2)$ (dashed green).
  }\hspace{.05\textwidth}
  \parbox[c]{.45\textwidth}{\footnotesize
    (b)~The three correction factors entering the theory--data
    comparison: NLO QCD corrections (dash-dotted
    black), EW corrections (solid red), and NP
    effects (dashed blue).
  }
  \caption{The correction factors to the transverse-momentum distribution
    for inclusive jet production at the LHC with a centre-of-mass energy of $\sqrt{s}=7~\mathrm{TeV}$
    and for the lowest rapidity bin $\vert y\vert<0.5$
    (taken from Ref.~\cite{Butterworth:2014efa}).}
  \label{dittmaier:corr}
\end{center}
\end{figure}

\begin{figure}
\begin{center}
  \includegraphics[scale=0.75, trim=0 0 15 0]{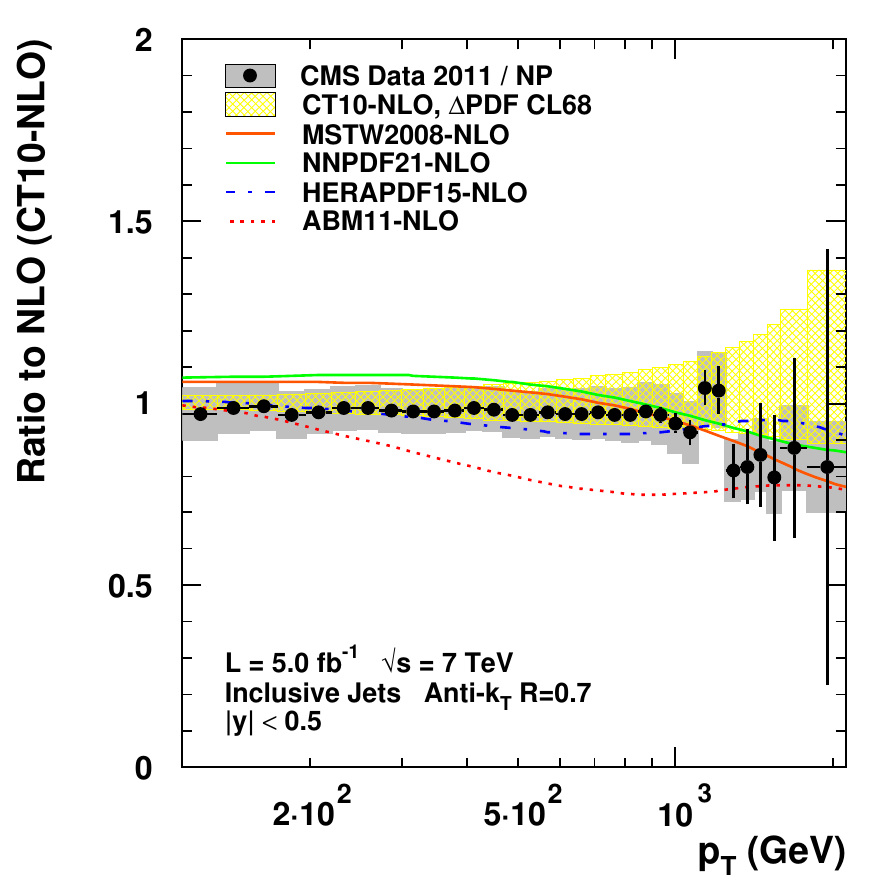}
  \includegraphics[scale=0.75, trim=52 0 0 0, clip=true]{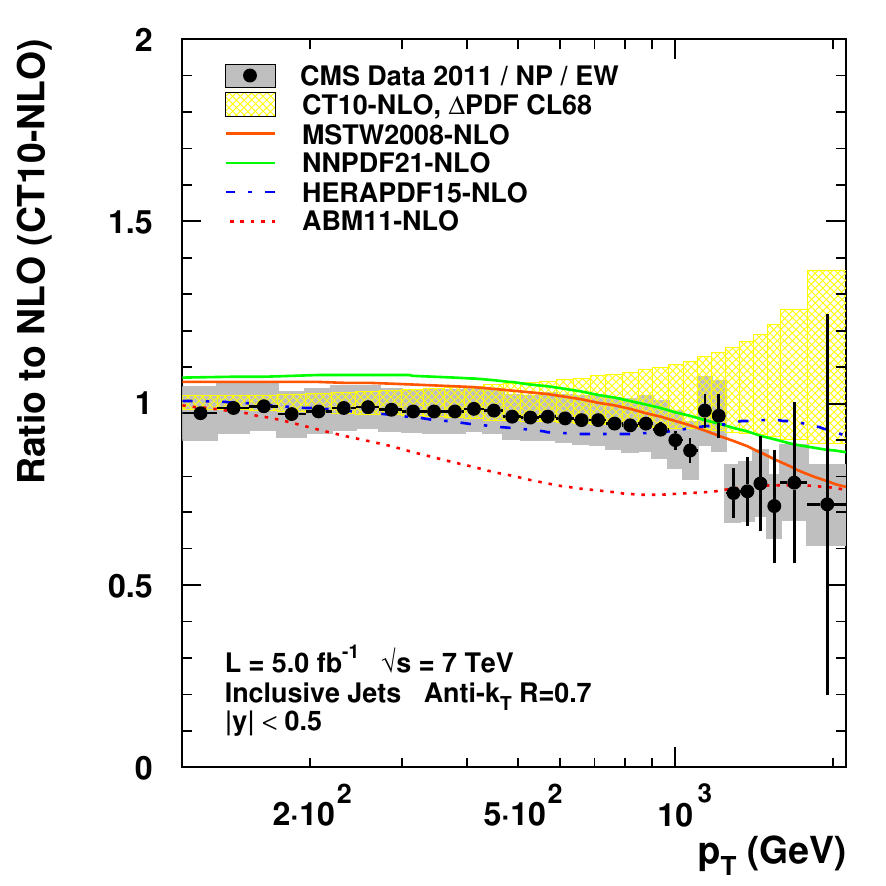}
\vspace*{-1em}
  \caption{The theory--data comparison for the rapidity bin 
    $\vert y\vert < 0.5$ 
    illustrated in terms of ratios with respect to the NLO QCD
    prediction obtained with the \textsc{CT10-NLO} PDF set. On the
    left, only NP correction factors are taken into account in this
    ratio, and on the right also the EW corrections
    (taken from Ref.~\cite{Butterworth:2014efa}).}
  \label{dittmaier:theory-data}
\end{center}
\end{figure}

Although the weak radiative corrections are small in inclusive quantities, 
their contribution to kinematic distributions grows towards higher scales and can reach 
$\sim10\%$ in the $\mathrm{TeV}$ range for transverse jet momenta $p_\mathrm{T}$ (see Fig.~\ref{dittmaier:corr}(a)).
The tree-level EW contributions display corrections of the same generic size but with the opposite sign at the LHC,
leading to significant cancellations between 
$K_\mathrm{EW}^\mathrm{tree}$ and $K_\mathrm{weak}^\mathrm{1-loop}$.
However, aside from being cut-sensitive, this cancellation is purely accidental and should not be mistaken 
to be a general property of the EW corrections.
In Fig.~\ref{dittmaier:corr}(b) the EW corrections 
$K_\mathrm{EW}=K_\mathrm{EW}^\mathrm{tree}\cdot K_\mathrm{weak}^\mathrm{1-loop}$ are compared to the different correction factors that enter
the theory--data comparison shown in Fig.~\ref{dittmaier:theory-data}.
Although the non-perturbative~(NP) corrections become negligible in the tails of the distributions, where the EW corrections are most pronounced,
the dominant contributions are still given by the NLO QCD corrections.
Figure~\ref{dittmaier:theory-data} illustrates the impact of the EW corrections to the theory--data comparison 
in terms of ratios with respect to the NLO QCD prediction:
In the left plot the data points were only corrected for NP effects, 
whereas in the right plot the EW corrections are also accounted for.
A small shift due to the EW corrections is noticeable, but their impact is below the current experimental and theoretical uncertainties.
However, these corrections will certainly become more relevant once the collision energy of the LHC is increased to
$\sqrt{s}=13{-}14~\mathrm{TeV}$ and with higher luminosity.
A more detailed discussion of the theory--data comparison for the inclusive-jet distributions
and the results for the other rapidity bins ($0.5\leq\vert y\vert<2.5$) are given in Ref.~\cite{Butterworth:2014efa}.

Unlike the $p_\mathrm{T}$ spectra,
the tails of the dijet invariant mass $M_{12}$ distributions do not probe the Sudakov regime, 
where all the scales are simultaneously much larger than the gauge-boson masses, 
but are dominated by the Regge-like regime of forward scattering.
As a consequence, the corrections to the $M_{12}$ distributions turn out to be generally smaller by at least a factor of two 
compared to the corresponding reach in the $p_\mathrm{T}$ distributions.
A detailed discussion of this difference together with results on the dijet invariant mass distributions can be found in Ref.~\cite{Dittmaier:2012kx}.

\cleardoublepage
\setcounter{section}{0}
\chapter{The top-quark mass}

\section{Summary on $m_t$}

The top-quark is the heaviest elementary particle currently known and its mass 
is a fundamental parameter of quantum chromodynamics.
Its numerical value not only affects theory predictions of collider cross sections 
required in the endeavor of exploring the Higgs boson properties or in
searching for new physics phenomena. 
Owing to its large value of the order of the electroweak scale the top quark mass 
has also direct impact on the Higgs sector of the Standard Model, and on extrapolations 
of the model up to high scales. 
The stability of the electroweak vacuum, i.e., requiring a well-defined minimum 
of the scalar potential that breaks the electroweak symmetry, depends 
crucially on the precise numerical value of $m_t$.
Thus, the top-quark mass provides input to an important test of the
self-consistency of the Standard Model.

\noindent During the past almost two decades since the discovery of the top-quark, 
the mass parameter $m_t$ has been measured in a variety of ways.
The most precise determinations of $m_t$ have been achieved experimentally 
from kinematical reconstruction of the measured top-quark decay products. 
As a caveat, this measurement has not been linked so far in an 
unambiguous manner to a Lagrangian 
top-quark mass in a specific renormalization scheme employed in the Standard Model.
In order to achieve measurements of the Lagrangian top-quark mass, 
several suggestions for appropriate observables together with their dependence 
on the mass parameter in a theoretically well-defined renormalization scheme
have been made in the recent past.
Most of these methods have been discussed at the workshop 
and it has become apparent that this task has to address
both QCD corrections in perturbation theory as well as the influence of non-perturbative effects.
The latter are related partly to the fact that the top is a colored unstable object such
that that its identification after the decay through leptons, jets and missing
$E_T$ involves low-energy color exchange, and partly to the complicated
soft environment due to the hadronic initial state.
Some non-perturbative effects are also specific to the observable under
consideration, e.g., power corrections in the definition of particular
jet event shape variables, the impact of parton distributions or of $\alpha_s$, 
and so on.

\noindent
In summary, it has become clear during the workshop, that coherent action 
regarding the precision and the interpretation of $m_t$ measurements 
is needed. Two such topics were identified during the discussion. 
The first issue concerns the 
definition of the $m_t$ parameter in the Monte-Carlo generators used in the 
Tevatron and the LHC analyses  
and their interpretation in theory, i.e., the relation of those $m_t$ measurements
from kinematically reconstructed top-quark decays to a Lagrangian mass in a given renormalization scheme.
The second topic concerns the validation of Monte Carlo tools, in particular a
comparison of the MC mass definition with well-defined theoretical calculations 
for events in differential kinematics.

\subsection*{Top mass definition of Tevatron and LHC results}

The latest experimental combinations of top mass measurements at the LHC and the
Tevatron~\cite{LHC2013,ATLAS:2014wva} include a statement like the following,
regarding the mass definition that is used for the input measurements:  
{\it 
``In all measurements considered in the present combination, the analyses are calibrated to the Monte Carlo (MC)
top-quark mass definition. It is expected that the difference between the MC mass definition and the formal
pole mass of the top quark is up to the order of 1 GeV (see
Refs.~\cite{Buckley:2011ms,Juste:2013dsa} and references therein).''}

\noindent 
Both references~\cite{Buckley:2011ms,Juste:2013dsa} 
refer to the arguments and conclusions made by  A. Hoang and
I. Stewart~\cite{Hoang:2008xm} in proceedings to workshops in 2008. 
However, in our discussion at the workshop it became clear that the message of
the original work was partly lost in translation.  
A more appropriate description of the content of Ref.~\cite{Hoang:2008xm} is as follows:

\noindent
{\it ``The uncertainty on the translation from the MC mass definition to a
theoretically well defined short-distance mass definition at a low scale is
currently estimated to be of the order of 1 GeV~\cite{Hoang:2008xm}.''}

\noindent 
We recommend that future experimental publications use a description that is
consistent with this sentence. In the following we describe the line of
reasoning of Ref.~\cite{Hoang:2008xm}, clarify the statements made on the 
uncertainty being {\it ``of the order of 1 GeV''} 
and quantify the additional theoretical uncertainty originating from the conversion 
of the short-distance mass to pole mass.

\noindent
The arguments made by Ref.~\cite{Hoang:2008xm} are based on earlier
work~\cite{Fleming:2007qr,Fleming:2007xt} on the invariant mass distribution of
reconstructed boosted top quarks in $e^+e^-$ collisions (at NLO+NLL order). In
Ref.~\cite{Fleming:2007qr,Fleming:2007xt} it was shown that the distribution in the
peak region which is highly sensitive to the top mass can be factorized into a hard function, a universal
jet function and a soft non-perturbative function. The jet and soft functions
affect the shape of the distribution including the peak position while the hard function affects the norm. 
The factorization entails a full treatment of the color flow (such that 
the reconstructed top object is fully color neutral), of hadronization effects and a systematic summation
of perturbative collinear and soft large logarithms involving the hard interaction scale, the
top mass and the hadronization scale.
The merit of this factorization is that in each factor separately the top quark
mass dependence can be analytically controlled in a fully coherent and
transparent way. This allows one to systematically implement proper short-distance
mass schemes that avoid the spurious higher-order renormalon effects of the
pole-mass. It was found in particular that the soft function is top mass 
independent (such that information on it can be obtained from massless jet
distributions), that the hard function is governed by a top mass at the hard
scale and that the jet function is governed by a top mass at the scale of the
top quark width $\Gamma_t$. The latter low-scale top mass definition
is tied to virtual as well as to real radiation corrections related to the color
neutralization of the reconstructed top mass system.
Because measurements of the top invariant mass
distribution mostly provide information on the shape and the peak position of
the distribution and not on its norm, it is the low-scale top mass contained in the
jet function that can be measured from data on the reconstructed top invariant
mass distribution.   
This means that the ${\overline {\rm MS}}$
mass cannot be directly determined from a kinematic reconstruction observable.
However, it is possible to define a different short-distance mass that is more
suitable for this type of observable.

\noindent
In Ref.~\cite{Hoang:2008xm} four arguments were made:
\begin{itemize}
\item[(1)] The most precise top mass measurements at the LHC and the Tevatron
  are and were essentially fits based on multi-observable analyses. In these analyses
  effectively a high weight is assigned to the top invariant mass distribution
  because it is the most sensitive and direct way to access precise quantitative
  information on the mass of the top system. 
\item[(2)] In proton-proton collisions for reconstructed high-$p_T$ top quarks
  the invariant mass distribution with properly defined top jet cones can be
  theoretically factorized in a way having features common to the $e^+e^-$
  case. The factorization is more complicated due to the initial state partons
  and because one has to impose a veto on additional hard non-top jets. 
  Moreover the distribution is further affected by hadron collider specifics such as the
  underlying event or pile up issues. However, the shape of the invariant mass
  distribution in the top mass sensitive peak region is described by the 
  jet function from the $e^+e^-$ case. So differences to the $e^+e^-$ distribution in the
  peak region concern the normalization and the effects of soft top mass
  independent non-perturbative effects. Assuming that the non-perturbative
  effects can be quantified, measurements of the reconstructed top invariant
  mass distribution allow to measure the low-scale top quark mass known from the $e^+e^-$
  factorization~\cite{Fleming:2007qr,Fleming:2007xt}. 
\item[(3)] There is an analogy of the factorization  supplemented by the 
  summation of collinear and soft logarithms for the top quark invariant mass 
  distribution and the components for hard interaction, partons shower and
  hadronization model contained in the MC description. It is known that the invariant mass
  distribution in the peak region is very sensitive to all these
  components. Moreover the energy scales involved in the MC description (hard 
  interaction scale, shower cutoff $\Lambda_c$ and non-perturbative scale) are 
  numerically compatible with the corresponding scales in the factorization for
  the invariant mass distribution in the peak region of
  Refs.~\cite{Fleming:2007qr,Fleming:2007xt}. 
\item[(4)]  Real emission and virtual collinear effects
  play an  important role for the definition of the mass parameter
  within the factorization. Likewise real emission effects, details of the
  parton shower implementation and the extent to which virtual effects are
  accounted for determine the physical meaning of the  mass parameter in the
  MC.  Because the shower
  cut provides a strict IR cutoff for soft low-momentum effects, the MC top mass
  is related to a low-scale short-distance mass and not the pole mass. 
  It was therefore concluded that the MC top quark mass parameter is closely 
  related to the low-scale top quark mass in the jet function, i.e.\ to the mass
  parameter that can be measured from the factorization for the 
  invariant mass distribution in the peak region
\end{itemize}

\begin{table}[t!]
  \centering
\begin{tabular}{|c c c | c | c c c | c c c |}
\hline
$m^{\rm MSR}(1)$ & $m^{\rm MSR}(3)$ & $m^{\rm MSR}(9)$ & 
$\overline m(\overline m)$ &
$m^{\rm pl}_{\rm 1lp}$ & $m^{\rm pl}_{\rm 2lp}$ & $m^{\rm pl}_{\rm 3lp}$ &
$m^{\rm pl}_{\rm 1lp}$ & $m^{\rm pl}_{\rm 2lp}$ & $m^{\rm pl}_{\rm 3lp}$ \\
\hline
172.52&172.20&171.58&162.62&170.14&171.75&172.25&172.52&172.67&172.78\\
172.72&172.40&171.78&162.81&170.34&171.95&172.45&172.72&172.87&172.98\\
172.92&172.60&171.98&163.00&170.54&172.15&172.65&172.92&173.07&173.18\\
173.12&172.80&172.18&163.19&170.73&172.35&172.85&173.12&173.27&173.38\\
173.32&173.00&172.38&163.38&170.93&172.55&173.05&173.32&173.47&173.58\\
173.52&173.20&172.58&163.57&171.13&172.75&173.25&173.52&173.67&173.78\\
173.72&173.40&172.78&163.76&171.33&172.95&173.45&173.72&173.87&173.98\\
173.92&173.60&172.98&163.95&171.53&173.15&173.65&173.92&174.07&174.18\\
174.12&173.80&173.18&164.14&171.72&173.35&173.85&174.12&174.27&174.38\\
174.32&174.00&173.38&164.33&171.92&173.55&174.05&174.32&174.47&174.58\\
174.52&174.20&173.58&164.52&172.12&173.74&174.25&174.52&174.67&174.78\\
\hline
\end{tabular}
\caption{
\label{tab:MSRMSbar}
Top quark MSR and $\overline{\mbox{MS}}$ masses at different scales converted at
  ${\cal O}(\alpha_s^3)$ for $\alpha_s(M_Z)=0.1185$. 
Columns 5-7 show the 1, 2 and 3 loop pole masses converted from the
$\overline{\mbox{MS}}$ mass $\overline m(\overline m)$. 
Columns 8-10 show the 1, 2 and 3 loop pole masses converted from the
MSR mass $m^{\rm MRS}(3~\mbox{GeV})$. All numbers are given in GeV units.}
\end{table}

The statements of item~(2) are supported by upcoming results reported on in
Refs.~\cite{pptop:2014,ppISRUE:2014}. The statements of item~(4) still need to be
quantitatively checked by dedicated comparisons of MC and factorization
descriptions to be conducted in the future for observables {\it at the hadron
  level}. Comparisons at the hadron level are essential in this context since
the separation of partonic and hadronic effects in MC generators (related to
shower cut and hadronization model) is not compatible to the one usually used in analytic calculations
(related to perturbative regularization schemes). The fact that the many different
MC based top quark mass analyses, even if they are not based on the top
invariant mass distribution, lead to compatible MC top mass measurements,
support the view that a quantitative relation between the MC top mass and a
low-scale short-distance top mass might be established in a reliable way.

\noindent
Assuming that the line of arguments made above is complete and that the MC mass
can indeed be associated to a well defined mass scheme, Hoang and Stewart
related the Pythia MC top quark mass parameter $m_t^{\rm MC}$ to the MSR
mass $m_t^{\rm MSR}(R)$ (see~\cite{Hoang:2008yj}), which is a scale-dependent short-distance mass that 
provides a valid mass definition at low-scales much below the mass and
smoothly interpolates to the $\overline{\mbox{MS}}$ mass for scales larger than the mass:
\begin{equation}
\label{eq:MCMSR}
m_t^{\rm MC} = m_t^{\rm MRS}(3^{+6}_{-2}~\mbox{GeV})\,.
\end{equation}
For each choice of $R$ the MSR mass $m_t^{\rm MSR}(R)$ represents a different
mass definition, and the variable $R$ allows to smoothly vary between them.
The quoted scale uncertainty is an estimate of the conceptual uncertainty
that is currently contained in this relation. This uncertainty might also be thought
of to be associated to unknown higher order corrections in the relation
Eq.~(\ref{eq:MCMSR}), but it 
should be emphasized that these corrections are likely not calculable since a
complete analytic control of the MC machinery is out of question. Moreover, the
exact physical definition of the MC mass can depend on details of how the parton
shower, the shower cut and the hadronization model are implemented (see, e.g.~\cite{Skands:2007zg}).

\noindent
Numerically the conceptual scale uncertainty quoted on the right hand side of
Eq.~(\ref{eq:MCMSR}) refers to a numerical uncertainty in the MRS top mass
{\it ``of the order of 1 GeV''} as can be seen from Tab.~\ref{tab:MSRMSbar}
and as was summarized in the quoted statement above. 
Future concrete analyses should aim to test this relation, 
to quantify better the size of the conceptual uncertainty and to make it more precise.  
From the Tevatron top mass average $m_t^{\rm tev}=172.6\pm 0.8\pm 1.1$~GeV 
from Ref.~\cite{Group:2008nq} 
Hoang and Stewart then calculated the top $\overline{\mbox{MS}}$ mass using
${\cal{O}}(\alpha_s^3)$ QCD perturbation theory and resummation as 
$\overline m_t(\overline m_t)=163.0\pm 1.3^{+0.6}_{-0.3}$~GeV, where the asymmetric
errors were related to the conceptual uncertainty given in Eq.~(\ref{eq:MCMSR}).  
In Tab.~\ref{tab:MSRMSbar} in each line numerical values for the conversion at
${\cal{O}}(\alpha_s^3)$ are shown 
for the top MSR mass defined at the scales $R=1$, $3$ and $9$~GeV and the
$\overline{\mbox{MS}}$ mass $\overline m_t(\overline m_t)$ for
$\alpha_s(M_Z)=0.1185$ with $n_f=5$ flavors.  
For the MSR mass, the spread in values coming from the range of scales assumed in
Eq.~(\ref{eq:MCMSR}) can be quantified (Tab.~\ref{tab:MSRMSbar}, column 1-3)
and the numerical value for the $\overline{\mbox{MS}}$ mass corresponding to
$m_t^{\rm MRS}(3~\mbox{GeV})$ is quoted (Tab.~\ref{tab:MSRMSbar}, column 4).
Also displayed are pole mass values at 1-, 2- and 3-loop
order for fixed-order conversions.
As can be seen,  the numerical values of the pole mass are strongly
dependent on the order of perturbation theory (a well-known fact, indeed), but also
carry a significant dependence on the scale, where this conversion is
performed. Shown are conversions 
$\overline m_t(\overline m_t) \to m^{\rm pl}$ with using $\mu=\overline m_t(\overline m_t)$ (Tab.~\ref{tab:MSRMSbar}, column 5-7)  
and 
$m_t^{\rm MRS}(3~\mbox{GeV}) \to m^{\rm pl}$ using $\mu = 3~\mbox{GeV}$ 
(Tab.~\ref{tab:MSRMSbar}, column 8-10).
Those differences can be sizeable and have not been addressed at all yet in the
interpretation that the experimental measurements of the top mass in 
Refs.~\cite{LHC2013,ATLAS:2014wva} refer to the pole mass.

\subsection*{Comparing MC mass definition with well-defined theoretical calculations}

\begin{figure}[t!]
\begin{center}
\includegraphics[width=0.49\textwidth]{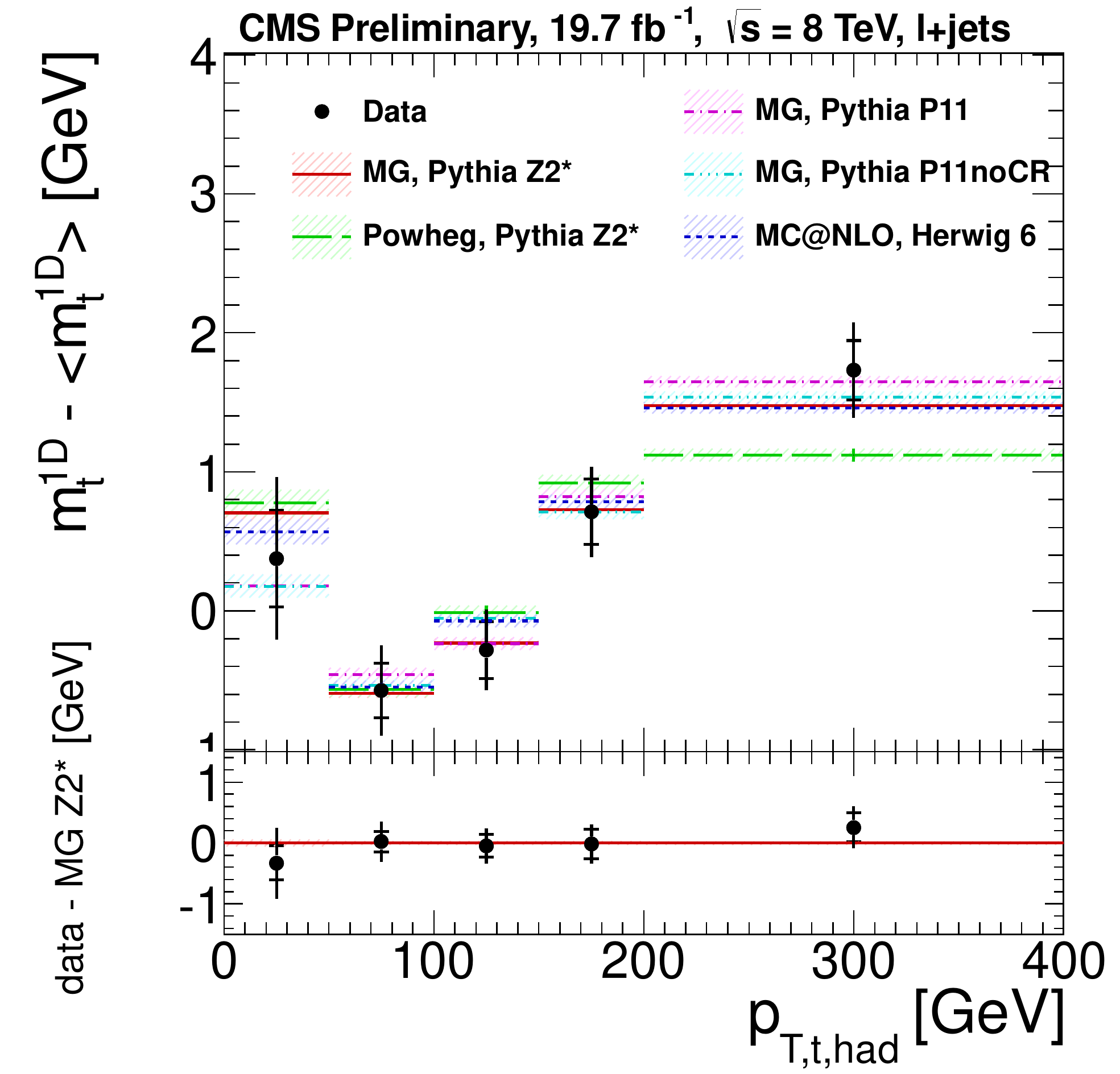}
\includegraphics[width=0.49\textwidth]{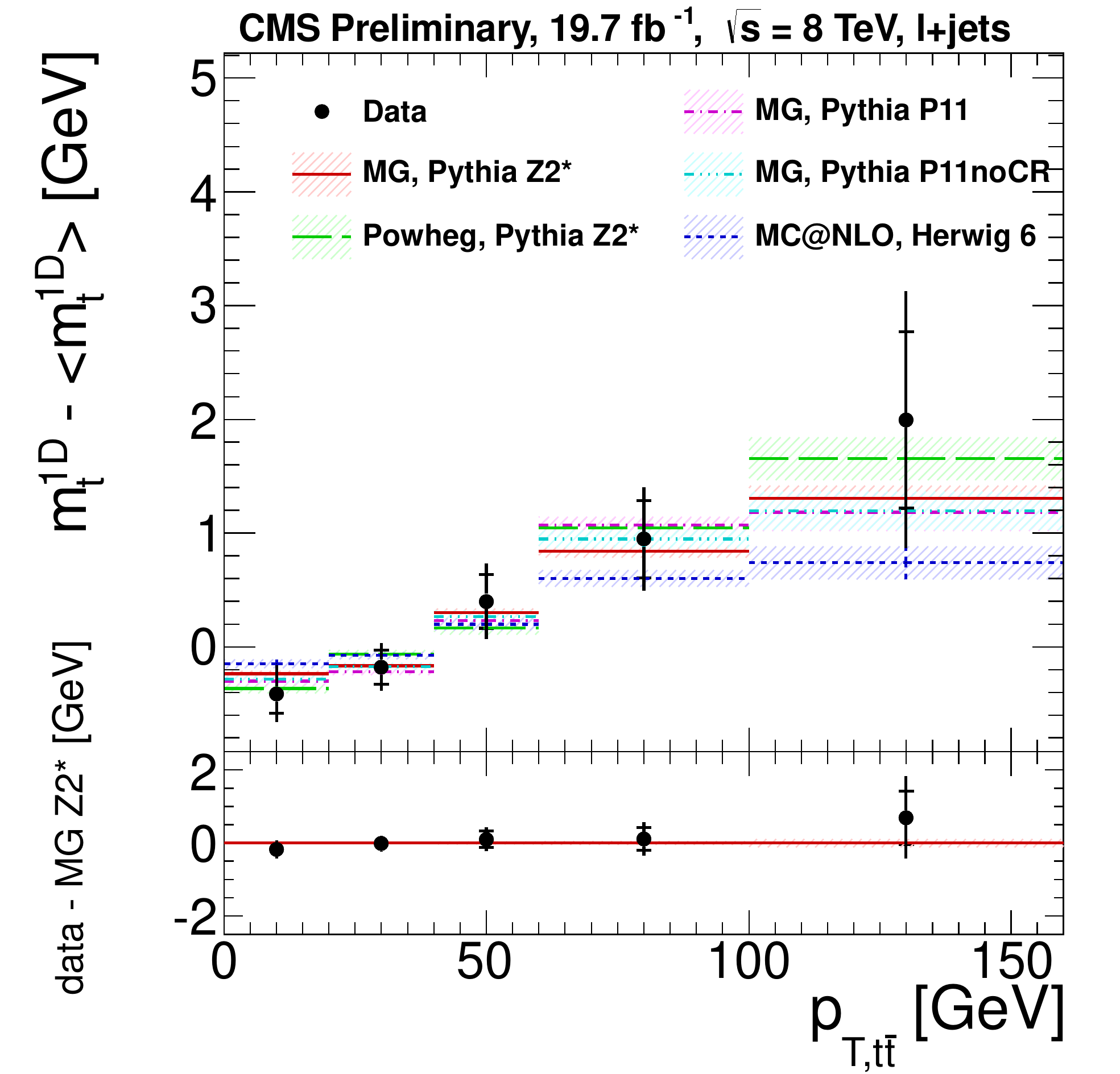}
 \caption{Left: Dependence of the measured top mass observable on the transverse momentum of the hadronically decaying top~\cite{CMS-PAS-TOP-14-001}. Right: Dependence of the measured top mass observable on the transverse momentum of the $t \bar{t}$ pair~\cite{CMS-PAS-TOP-14-001}.}
    \label{mulders:fig5}
\end{center}
\end{figure}

At the workshop we agreed that the approach of studying top-mass related
observables as a function of kinematic event
variables~\cite{CMS-PAS-TOP-12-029,CMS-PAS-TOP-14-001} promises to be a useful
method to improve the understanding of top mass interpretation.  
An example from CMS~\cite{CMS-PAS-TOP-14-001} for two such variables is shown in Fig.~\ref{mulders:fig5}.

\noindent
Even without looking at the data, a comparison between the prediction by the
default MC tool used in the experiment and a well-defined higher order QCD calculation as a
function of several event variables would allow to quantify possible differences
in mass definition. If there are significant differences due to QCD effects,
these differences are not expected to be constant in all corners of phase
space. A good agreement as a function of several relevant variables would be a
strong indication that effects are well modeled.  
Again, it is important to stress the comparison be carried out at the hadron-level. 
For observables with small hadronization corrections, hadron-level and
parton-level predictions are expected to be very similar, but given the 
anticipated precision in the top-quark mass of less than $1$ GeV, full control 
over the hadronization corrections will unavoidably become important.

\noindent
In summary, testing the MC mass definition with well-defined theoretical calculations 
will become more precise as more MC and data statistics become available. 

\section*{2~~~~Top mass measurements at the LHC~\footnote{M.~Mulders}}
\addcontentsline{toc}{section}{\protect\numberline{2}{Top mass measurements at the LHC}}

\noindent Thanks to the copious production of top quarks at the LHC~\cite{LHC}, the 
excellent performance of the LHC detectors, and the availability of advanced theoretical calculations and Monte Carlo tools, the ATLAS~\cite{ATLAS} and CMS~\cite{CMS} collaborations have been able to study top quark properties and the mechanism of their production and decay with unprecedented detail and precision. Already with the 7 TeV dataset, LHC measurements of the top quark mass have reached a precision comparable to the latest Tevatron results.

\subsection*{LHC combination and World Average}

\begin{figure}[htb]
\begin{center}
\includegraphics[width=0.54\textwidth]{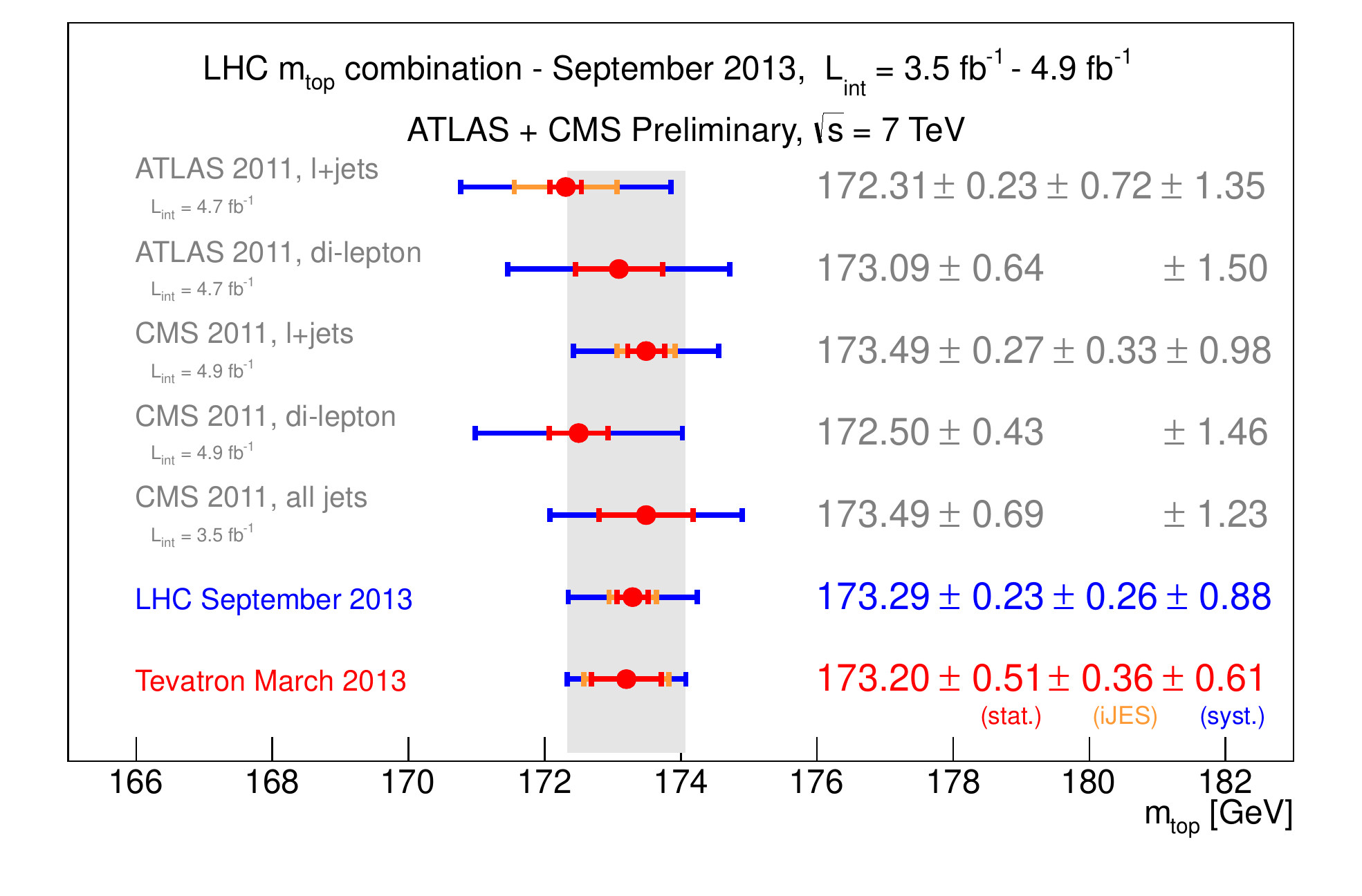}
\includegraphics[width=0.45\textwidth]{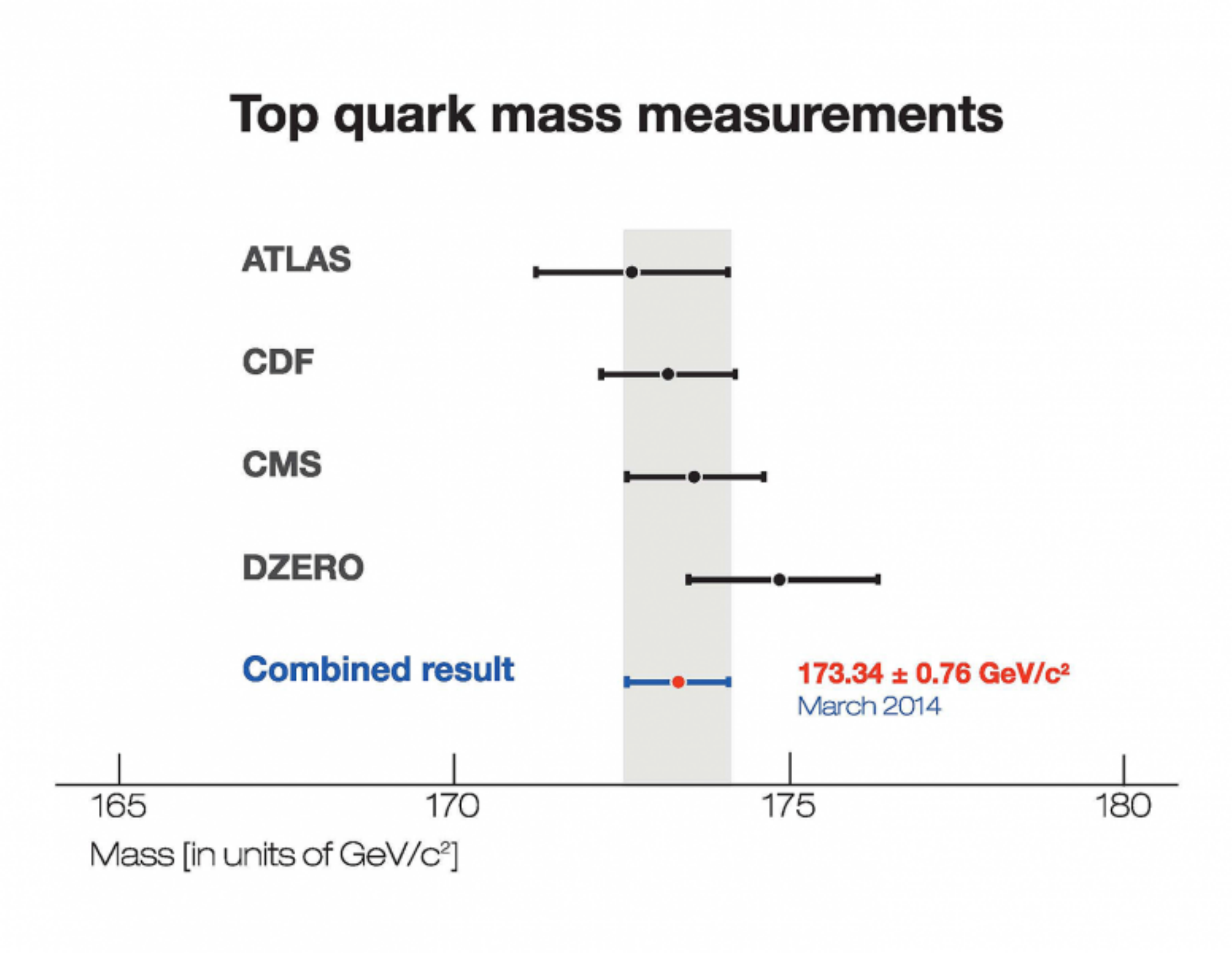}
 \caption{Left: Summary of input mass measurements of the latest LHC top mass average and the combined result. The LHC combination with the separated iJES uncertainty is shown in blue and compared to the latest Tevatron result, which is shown in red~\cite{LHC2013}. Right: Summary of the top mass results of the four experimental collaborations ATLAS, CDF, CMS and DZERO in the first ever joint measurement involving experiments from Fermilab and CERN~\cite{ATLAS:2014wva}.}
    \label{mulders:fig1}
\end{center}
\end{figure}

\noindent The ATLAS and CMS collaborations provided an updated combination of their most precise top mass measurements available as of September 2013~\cite{LHC2013}. This combination relies on a proper understanding of the systematic uncertainties assigned to the measurements by both experiments, and correlations between them. These uncertainties and their correclations were studied and discussed in detail in the framework of the TOPLHC Working Group (TOPLHCWG), involving experts in experimental and theoretical aspects of the top mass determination. Only the most precise LHC results were considered for the combination, and they are summarized together with the combined result in Fig.~\ref{mulders:fig1}(left).

During the MITP workshop, CERN and Fermilab announced the first joint measurement of the top quark mass, adding the most precise measurements of the CDF and DZERO collaborations obtaining the first 4-experiment top mass measurement, illustrated in Fig.~\ref{mulders:fig1}(right) and yielding $m_{\rm top} = 173.34 \pm 0.76$~GeV~\cite{ATLAS:2014wva}.

Please note that the input measurements are conventional top mass measurements, using full reconstruction of the $t\bar{t}$ events to extract the top mass from the invariant mass of its decay products. They rely on the use of MC simulation and are designed to yield the value of the input MC top mass parameter that gives the best agreement between the experimental observables in data and MC simulation.

Shortly after the MITP workshop, the CMS collaboration released the world's most precise single measurement of the top-quark mass in the semileptonic decay channel, using the experiment's full sample of data at 8 TeV. Combined with the previous CMS results, this gives a mass of $172.22 \pm 0.73$ GeV~\cite{CMS-PAS-TOP-14-001}.

\subsection*{New approaches and alternative techniques}

\begin{figure}[htb]
\begin{center}
\includegraphics[width=0.49\textwidth]{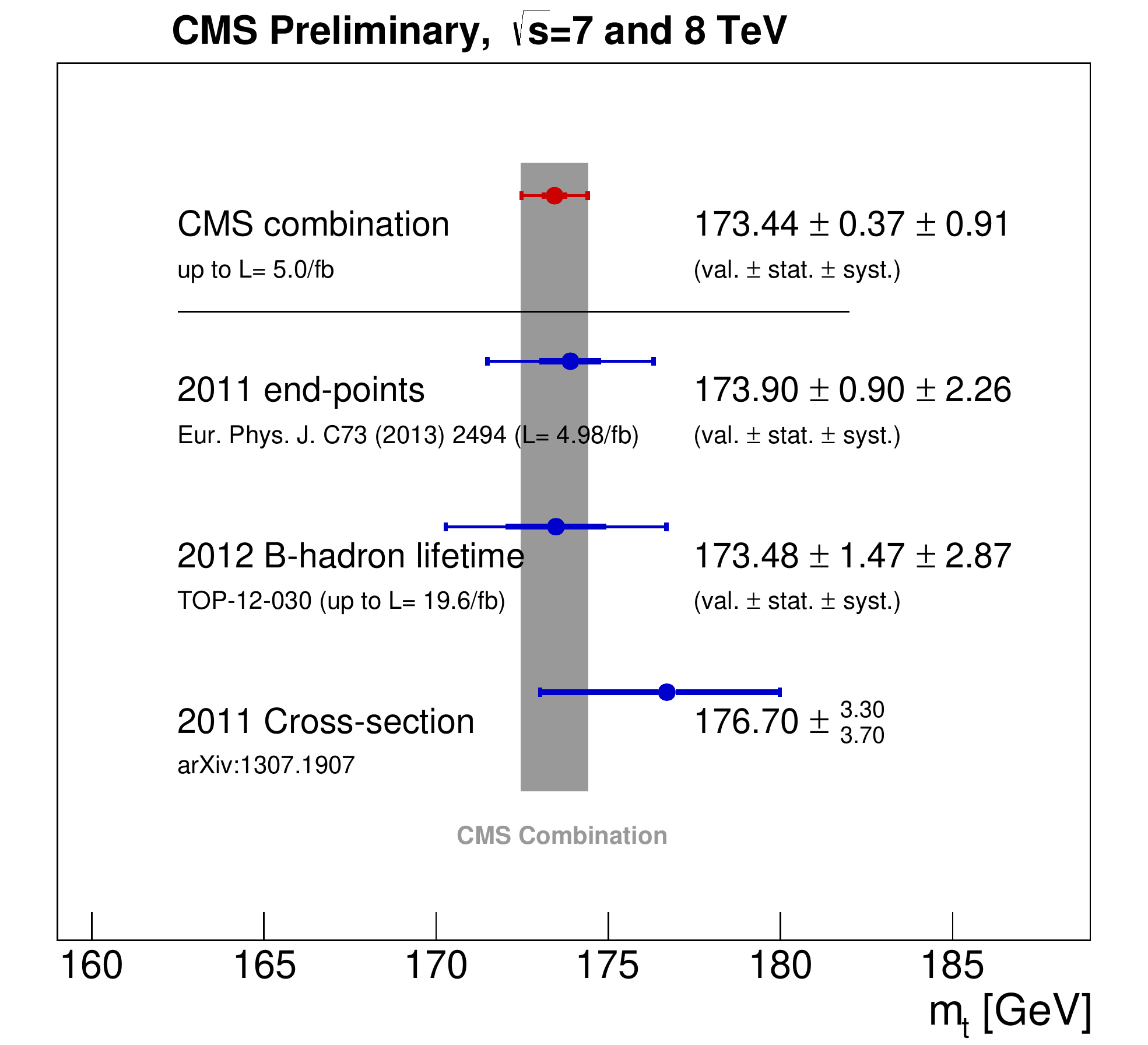}
 \caption{Summary of mass measurements by CMS using alternative measurement techniques, compared to the combination of standard measurements~\cite{CMS-PAS-TOP-13-002}. }
    \label{mulders:fig2}
\end{center}
\end{figure}

\noindent Several alternative techniques to extract the top quark mass are also being explored at the LHC. They include extraction of the top ``pole'' mass from the inclusive  $t\bar{t}$ cross-section~\cite{Chatrchyan:2013haa}, which is predicted at NNLO precision~\cite{Czakon:2013goa}. An alternative approach that has been proposed and may lead to superior ultimate precision is the extraction from the  $t\bar{t}$+jet cross-section, discussed elsewhere in these proceedings. No results using this method have been released at the LHC thus far.

Another technique pursued is to use the transverse decay length ($L_{\rm xy}$) of B hadrons in the b-jets from top quark decay. Pioneered by the CDF collaboration, this method can give relevant statistical precision at the LHC. However, the transverse boost of the b jets relies strongly on the top production modeling, in particular modeling of the top transverse momentum spectrum. This is the limiting uncertainty, of order 2-3 GeV, in the latest CMS measurement~\cite{CMSLxy2012}. The same top modeling uncertainty has a far smaller effect on a Lorentz-invariant quantity such as the invariant mass of the top decay products, where the effect on the measured top mass is estimated to be of the order of 200 MeV~\cite{CMS-PAS-TOP-14-001}.

A third alternative technique presented by the CMS collaboration uses the measured end points of various observables, including the invariant mass of the lepton and b-jet $m_{lb}$ from top decay~\cite{CMS-PAS-TOP-12-027}. By using only the endpoint of the distributions, the sensitivity to certain systematic uncertainties is different than for standard techniques. Also, the endpoint can be calculated analytically (using a narrow-width picture) allowing this analysis to report a top mass that does not rely on MC simulation and is therefore not using the MC mass definition. This study also includes a MC calibration for illustration, showing that the difference between the purely analytical mass used in the endpoint calculation and the full MC mass is of the order of 300$\pm$300 (stat) MeV. 

A summary of the measurements by CMS using alternative techniques is shown in Fig.~\ref{mulders:fig2}. The alternative measurements are in good agreement with the combined result of the standard CMS measurements, within measurement precision. 

\begin{figure}[htb]
\begin{center}
\includegraphics[width=0.49\textwidth]{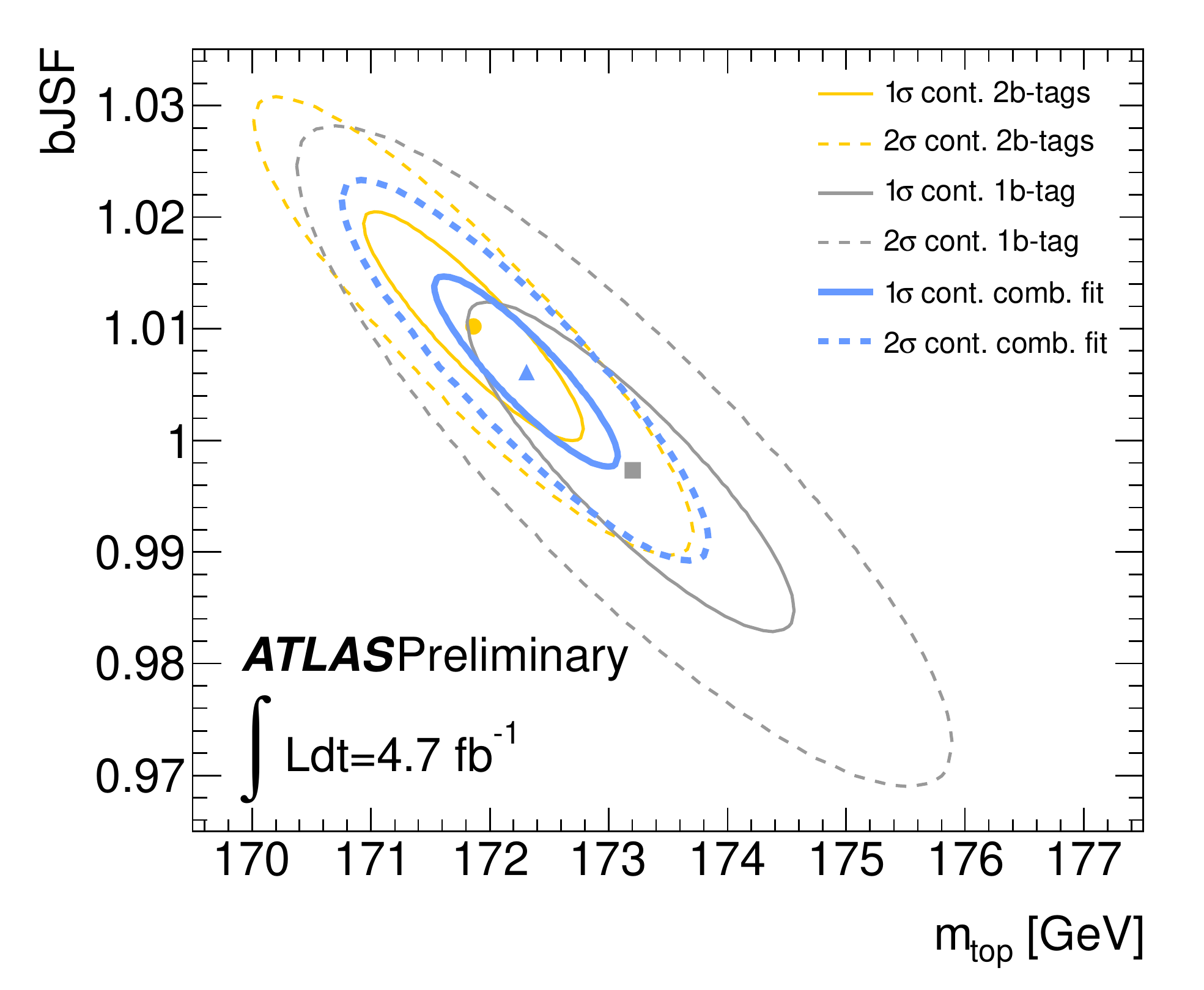}
\includegraphics[width=0.49\textwidth]{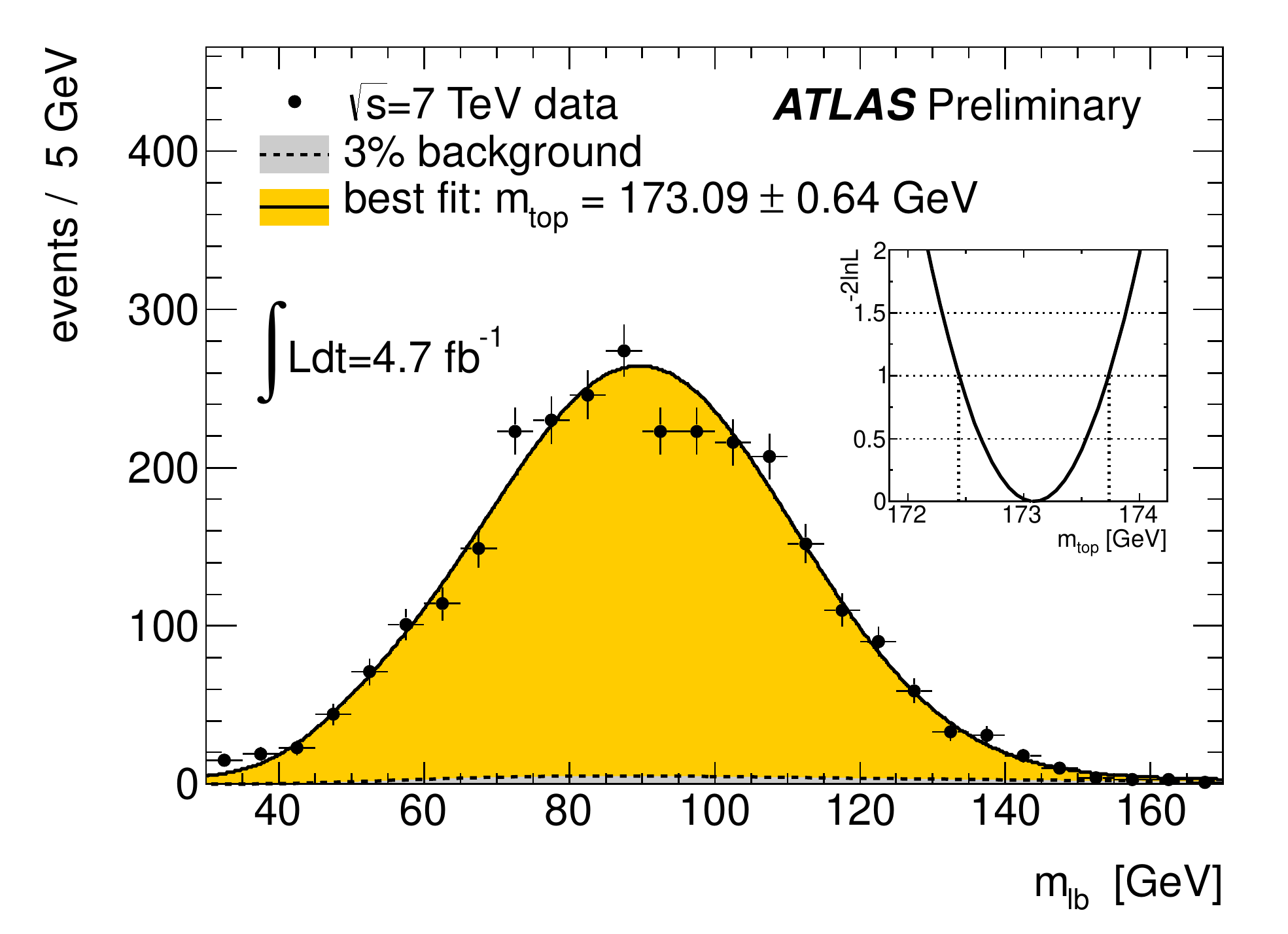}
 \caption{Left: Correlation between the fitted b-jet energy scale factor and the extracted top mass, in the ATLAS 3D template fit~\cite{ATLlj2011}. Right: Fitted mlb distribution in data, in the di-lepton decay channel.  The inset shows a -2 log likelihood profile as a function of the fitted top quark mass.~\cite{ATLdil2011}}
    \label{mulders:fig3}
\end{center}
\end{figure}

At the same time the ATLAS collaboration achieved a major milestone with a conventional top mass measurement in the lepton+jets channel, demonstrating for the first time that it is possible to fit both the light-quark jet energy scale and the b-jet energy scale {\em in-situ} in $t\bar{t}$ events using a 3-dimensional template fit~\cite{ATLlj2011}. The fit uses the invariant mass of the hadronically decaying W in the events, and the transverse momentum balance between b-tagged and non-b-tagged jets. Fig.~\ref{mulders:fig3}(left) shows the strong correlation between the fitted relative b-jet energy scale factor, and the extracted top mass. With further increase in statistics, this method promises to provide a strong handle on the jet energy scale calibration, still one of the limiting experimental uncertainties. 

Finally, another recent result by ATLAS shows that with the large top quark samples at the LHC, the statistical uncertainty is no longer an important limitation even in the di-leptonic decay channel, and a basic but robust observable such as $m_{lb}$ the invariant mass of the lepton from W decay and the corresponding b-jet, as shown in Fig~\ref{mulders:fig3}(right) can yield a precise and competitive top mass measurement~\cite{ATLdil2011}.

\subsection*{Prospects for top mass measurements at the LHC}

\begin{figure}[htb]
\begin{center}
\includegraphics[width=0.49\textwidth]{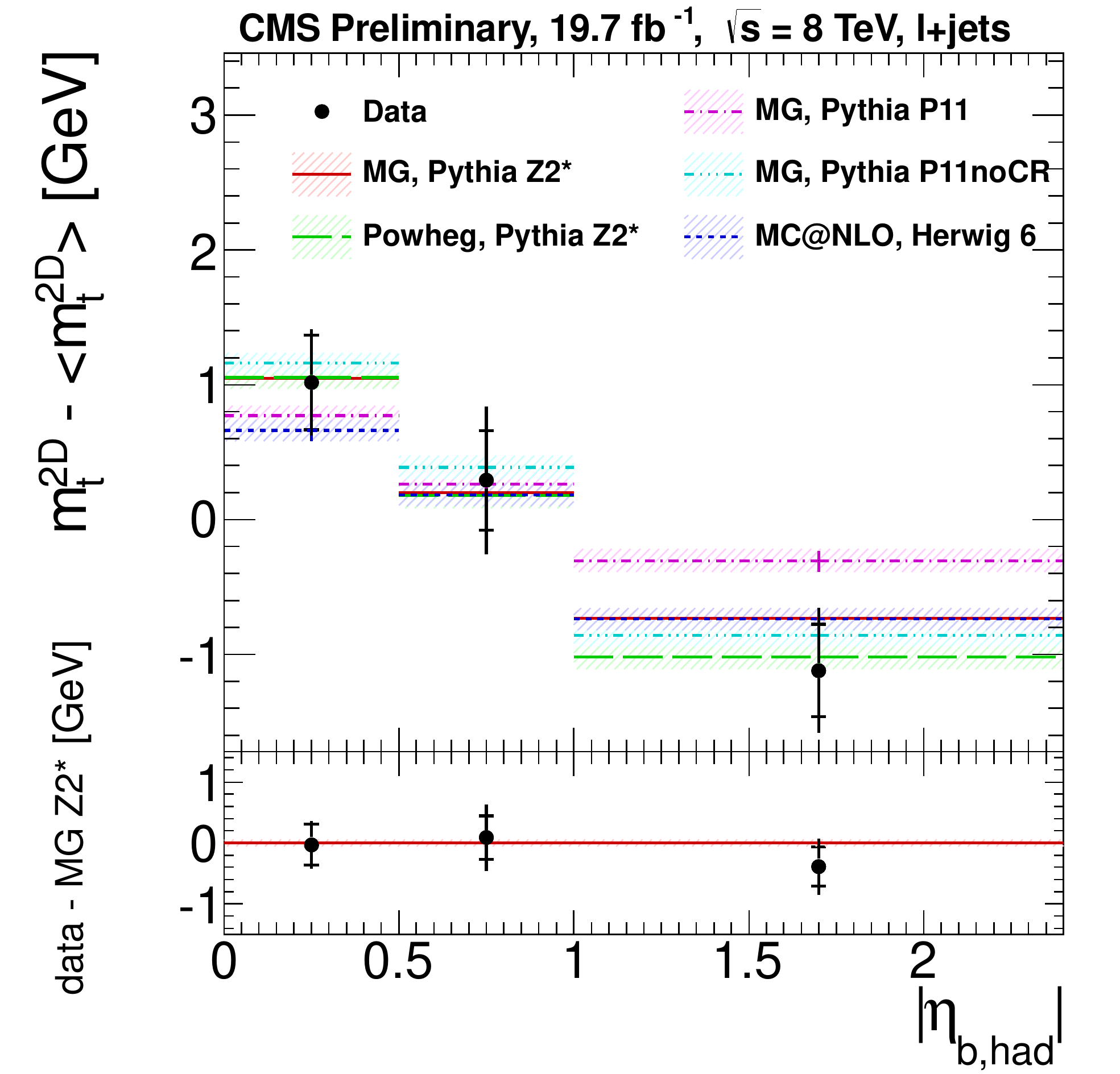}
\includegraphics[width=0.49\textwidth]{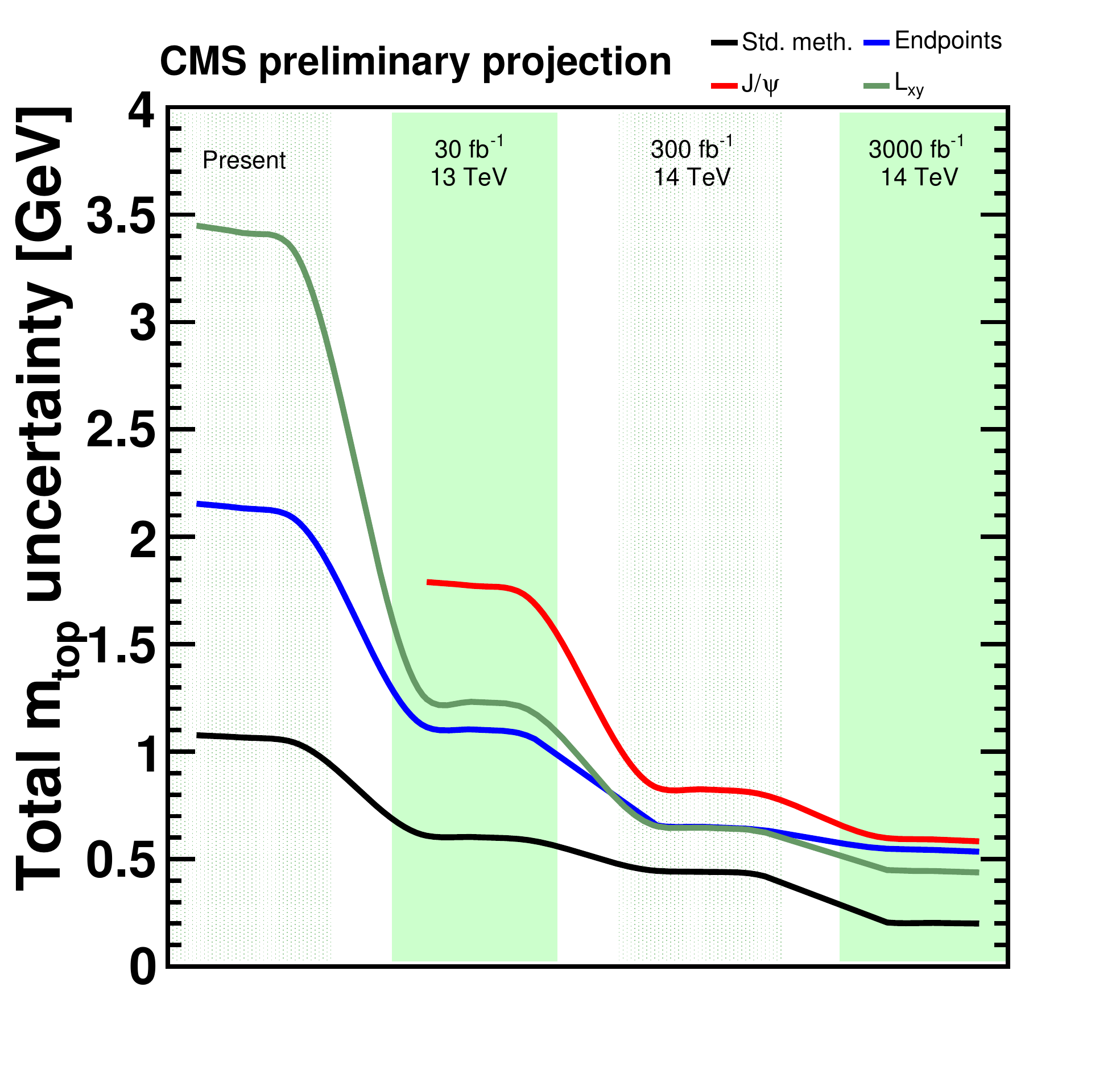}
 \caption{Left:  Dependence of the measured top mass observable on event variables, here the pseudo-rapidity of the b jet from the hadronically decaying top~\cite{CMS-PAS-TOP-14-001}. Right: Projection of the top-quark mass precision obtained with different methods, for increasing integrated LHC luminosity~\cite{CMS-PAS-FTR-13-017}.}
    \label{mulders:fig4}
\end{center}
\end{figure}

\noindent The large samples of top quark events at the LHC allow to study the dependence of the top mass observable as a function of event kinematics. In Fig.~\ref{mulders:fig2}(right) one example is shown of the measured top mass observable as a function of the pseudo-rapidity of the b-jet from the hadronically decaying top quark. A comprehensive study was performed by CMS, monitoring the behaviour of the top mass observable as a function of 12 different variables, looking for any corner in phase space where the MC models might disagree with the data~\cite{CMS-PAS-TOP-12-029,CMS-PAS-TOP-14-001}. No disagreements were found with the current statistical precision. 

With further increase in LHC integrated luminosity, methods like the 3D template fit can be used to further reduce experimental uncertainties on jet reconstruction while studies of the dependence of the top mass observable on other variables are expected to allow to get a handle on other theoretical uncertainties such as the modeling of radiation and non-perturbative QCD effects. 

Figure~\ref{mulders:fig4}(right) shows projections for possible improvements in experimental precision in the measurements of the top mass, in several steps of additional integrated luminosity of LHC data. With the methods outlined above, it may be possible to reach an experimental precision of about 0.2 GeV~\cite{CMS-PAS-FTR-13-017}. These estimates do not include an uncertainty for the translation of the measured MC mass to a well-defined theoretical mass definition.

\section*{3~~~~Determinations of the top-quark pole mass from measurements of the total cross section~\footnote{C.~Schwinn}}
\addcontentsline{toc}{section}{\protect\numberline{3}{Determinations of the top-quark pole mass from measurements of the total cross section}}
A determination of the top-quark mass in a known and well-defined mass
scheme can be obtained from a comparison of a measurement
of the total top-pair production cross section to a theoretical
calculation using a given mass scheme such as the pole mass
or the $\overline{\mathrm{MS}}$ scheme~\cite{Langenfeld:2009wd}.  
Such measurements have been performed recently by the experimental collaborations at the Tevatron and the LHC~\cite{Abazov:2011pta,ATLAS:2011qga,Aldaya:2012tv,Chatrchyan:2013haa} as well as by theorists~\cite{Beneke:2011mq,Alekhin:2012py}.
In this contribution, the extraction of the pole mass from~\cite{Beneke:2011mq} is updated by including the complete next-to-next-to-leading order~(NNLO) corrections~\cite{Baernreuther:2012ws,Czakon:2012zr,Czakon:2013goa} and taking recent experimental results into account. The prospects and limits of the method are also briefly discussed. 
In addition to the top-quark mass, the theoretical prediction of the total cross section within perturbative QCD depends also on the strong coupling constant $\alpha_a$ and the parton distribution functions~(PDFs). Here several PDF sets with their default $\alpha_s$ values will be used and the resulting differences are treated as theoretical uncertainty. A report on the determination of the  $\overline{\mathrm{MS}}$-mass and the inclusion of the top-quark mass in an PDF fit~\cite{Alekhin:2013nda} is given elsewhere in these proceedings.

 In practice, the top-quark 
extraction is complicated by the fact that the measurement of the
total cross section relies on Monte-Carlo simulations using an input
top mass $m_t^{\mathrm{MC}}$, resulting in a dependence of the measured total cross section on the Monte-Carlo input mass. 
We employ the 
method described in~\cite{ATLAS:2011qga} where the top-quark mass is obtained by maximizing the joined likelihood function 
\begin{equation} \label{schwinn:eq1}
f(m_t) = \int f_{\mathrm{th}}(\sigma|m_t) 
\cdot f_{\mathrm{exp}}(\sigma|m_t) d \sigma \, ,
\end{equation} 
where $f_{\mathrm{th}}$ is a normalized Gaussian 
distribution centered on the theoretical prediction of the cross section,
\begin{equation}
f_{\mathrm{th}}(\sigma|m_t) = \frac{1}{\sqrt{2 \pi} \Delta 
\sigma^\mathrm{th}_{t \bar{t}}(m_t)} \exp \left[-\frac{\left( 
\sigma-\sigma^\mathrm{th}_{t \bar{t}}(m_t)\right)^2}{2 (\Delta
\sigma^\mathrm{th}_{t \bar{t}}(m_t))^2} \right] \, .
\end{equation}
Following~\cite{ATLAS:2011qga}, the theoretical uncertainty  $\Delta 
\sigma^\mathrm{th}_{t \bar{t}}$ is obtained by
adding the error estimate of the perturbative calculation linearly to
the $68\%$ CL PDF$+\alpha_s$ uncertainty. (The effect of including the perturbative uncertainty  with a rectangular instead of a Gaussian distribution has found to be small, see e.g.~\cite{Chatrchyan:2013haa}.) 
The distribution $f_{\mathrm{exp}}$ is defined in a similar way, with the central value and width of the Gaussian given by the measured  cross section, 
$\sigma^\mathrm{exp}_{t \bar{t}}(m_t)$, and the
total experimental error, $\Delta \sigma^\mathrm{exp}_{t \bar{t}}(m_t)$. 
Here it is assumed that the Monte-Carlo input mass $m_t^{\mathrm{MC}}$
is equal to the pole mass $m_t$, the resulting systematic uncertainty
is discussed below.
The top-quark mass is then extracted from the maximum of the likelihood function (\ref{schwinn:eq1}), 
with the error obtained from the $68\%$ area around the maximum. 
The theoretical cross section $\sigma_{t \bar{t}}^{\mathrm{th}}(m_t)$  will be fit by a function of the form
\begin{equation} \label{schwinn:eq2}
\sigma_{t \bar{t}}^{\mathrm{th}}(m_t) = \left(\frac{172.5}{m_t}\right)^4 \left(c_0+c_1 (m_t-172.5)+c_2 (m_t-172.5)^2+c_3 (m_t-172.5)^3 \right) \, \mathrm{pb}\, ,
\end{equation}
where all the masses are given in GeV.
As a default, the full NNLO prediction~\cite{Baernreuther:2012ws,Czakon:2012zr,Czakon:2013goa} is used that is
incorporated in the public programs {\sc top++
  v2.0}~\cite{Czakon:2011xx}, {\sc HATHOR v1.5}~\cite{Aliev:2010zk}
and {\sc Topixs v2.0}~\cite{Beneke:2012wb}, where {\sc top++} adds
soft-gluon resummation at next-to-next-to-leading logarithmic~(NNLL)
accuracy and {\sc Topixs} further includes combined soft and
Coulomb-gluon NNLL resummation.  Resummation has a minor impact on the
central value but further reduces the theoretical uncertainty of the
perturbative cross-section prediction to the $4\%$-level which is of
the same order as the PDF+$\alpha_s$ uncertainty.  The numerical
results used here have been obtained with {\sc Topixs}. For the LHC at
$\sqrt s=8\,\mathrm{TeV}$ the coefficients in the
fit~(\ref{schwinn:eq2}) for the MSTW08 PDFs~\cite{Martin:2009bu} and
the NNLO approximation are obtained as $c_0=244.81\pm 15.21\pm 10.61$,
$c_1=-(1.5565\pm 0.1017\pm 0.0503)$, $c_2=(5.7081\pm 0.3852\pm
0.1411)\times 10^{-3}$, $c_3=-(1.4646\pm 0.1118\pm 0.0392))\times
10^{-5}$ where the first error denotes the total theory error and the
second the $68\%$ CL PDF$+\alpha_s$ error.

Table~\ref{schwinn:tab1} shows a collection of recent measurements
of the total cross section providing a parameterization of the
dependence of the measured cross section on the Monte-Carlo input
mass. The slope of the extracted cross section at $m_t^{\mathrm{MC}}=172.5$ is
quoted in the third column of the table. The table shows the pole mass
extracted using the procedure discussed above for a selection of PDF
sets. For definiteness the
fixed-order NNLO prediction has been used for the theoretical predictions.
The results for the pole mass obtained with the MSTW08,
CT10~\cite{Gao:2013xoa}, NNPDF2.3~\cite{Ball:2012cx} agree with each other within
uncertainties, and are also in good agreement with the world average
$\QTY{173.34\pm 0.76}{GeV}$ from direct measurements~\cite{ATLAS:2014wva}. The ABM11 results at the Tevatron are
consistent with the other PDF sets while
there is a larger difference at the LHC as also noted in~\cite{Alekhin:2012py}.
\begin{table}[h]
  \centering
  \caption{Experimental cross-section measurements providing the dependence on $m_t^{\mathrm{MC}}$ and the resulting values for the pole mass obtained using the NNLO prediction for several PDF sets. The result for the mass is given in GeV.}
  \label{schwinn:tab1}
{\renewcommand{\tabcolsep}{.1cm}
    \begin{tabular}{c|c|c||c|c|c|c}
     Ref. &$\sigma_{t\bar
          t}\,{\scriptstyle (172.5)}/\mathrm{pb}$ &
        $\frac{1}{\sigma_{t\bar t}}
        \frac{d\sigma_{t\bar t}(172.5)}{dm_t^{\mathrm{MC}}}$ &
        $m^{\mathrm{MSTW}}_t$ & $m_t^{\mathrm{CT10}}$ &
        $m_t^{\mathrm{NNPDF}}$ & $m_t^{\mathrm{ABM}}$\\\hline
D0~\cite{Abazov:2011cq}&  $7.56^{+0.63}_{-0.56}$&$-
      1.1\%\, \mathrm{GeV}^{-1}$ & $170.7^{+ 5.9}_{ -6.8}$ 
      & $172.5^ {+6.2}_{-7.8}$& $171.8^{+5.4}_{-5.8}$& $168.2^{ +5.7}_{ -6.2}$ \\\hline
      ATLAS~\cite{ATLAS:2011xha}$(\QTY{7}{TeV})$ &$179.0^{+11.8}_{-11.7}$&$-
      0.75\%\, \mathrm{GeV}^{-1}$ &  $171.3^{ +5.5}_{-5.6}$ & $171.7^{+6.7}_{-6.8}$ & $172.0^{ +5.5}_{-5.5}$&
      $161.8^{+4.8}_{-4.9}$    \\
      CMS~\cite{Chatrchyan:2012bra}$(\QTY{7}{TeV})$ &$161.9^{+6.7}_{-6.7}$ &$-
      0.80\%\, \mathrm{GeV}^{-1}$   &$175.9^{ +6.5}_ {-5.5}$&$176.5^{ +8.5}_ {-6.8}$ &$176.7^{ +6.7}_ {-5.4}$
      & $165.8^{+4.3}_{-4.3}$
      \\\hline
      ATLAS~\cite{TheATLAScollaboration:2013dja}$(\QTY{8}{TeV})$ &
      $237.7^{+11.3}_{-11.3}$&
      $-0.26\%\, \mathrm{GeV}^{-1}$&  $174.0^{+4.1}_{-4.5}$ &$174.3^{+4.9}_{-5.4}$  &$174.8^{+4.1}_{-4.4}$ & $166.4^{+3.7}_{-4.0}$ \\ 
       CMS~\cite{Chatrchyan:2013faa} $(\QTY{8}{TeV})$& $239^{+13.1}_{-13.1}$    &
         $-0.90\%\, \mathrm{GeV}^{-1}$ &  $174.8^{+7.0}_{-5.7}$& $175.8^{ +7.5}_{-5.7}$&
         $175.8^{+ 7.3}_{ -5.8}$& $165.0^{ +4.8}_{ -4.2}$\\
   \end{tabular}
}
\label{tab:mt-pdfs}
\end{table}
It is seen that the recent ATLAS
measurement~\cite{TheATLAScollaboration:2013dja} has the weakest
dependence on the input Monte-Carlo mass which translates into the
smallest uncertainty of the top mass. We therefore use this measurement
to investigate further current and potential future improvements of the method. 
First, replacing the NNLO prediction by the resummed NNLL+NNLO
calculation from~\cite{Beneke:2012wb} results in the
value
\begin{equation}
 m_t=\QTY{174.2^{+3.6}_{-3.9}}{GeV}  
\end{equation}
where the MSTW08 PDF set has been used. This result and the
uncertainty are consistent with the value
$m_t=\QTY{177.9^{+4.1}_{-3.6}}{GeV}$ obtained
in~\cite{Chatrchyan:2013haa}.  This shows that the current precision
of the top-mass measurement from the total cross section is limited to
the $\pm 2\%$ level. This uncertainty as well as that quoted in
Table~\ref{tab:mt-pdfs} does not include the systematic uncertainty
from identifying $m_{t}^{\mathrm{MC}}$ with the pole mass in the
experimental measurement. Due to the weak
$m_{t}^{\mathrm{MC}}$-dependence of the measurement
from~\cite{TheATLAScollaboration:2013dja}, a difference
$m_t-m_t^{\mathrm{MC}}=\QTY{\pm 1}{GeV}$ translates into a shift of
only $\Delta m_t=\QTY{\pm 0.1}{GeV}$.  

To investigate potential
further improvements, consider first the case of a hypothetical
measurement with the same central value and error
as~\cite{TheATLAScollaboration:2013dja} but without any
$m_t^{\mathrm{MC}}$ dependence. This would only lead to a moderate
reduction of the uncertainty to $\QTY{+ 3.3-3.6}{GeV}$. Similar values
are obtained assuming a future reduction of the experimental
uncertainty by one-half.  To obtain a significantly improved
sensitivity therefore requires further improvements in the theoretical
predictions and PDF$+\alpha_s$ uncertainty. With a very optimistic
assumption of a future reduction of both experimental and theoretical
uncertainties by one-half one finds a reduction of the uncertainty to
the $1\%$, level, $\QTY{\Delta m_t\sim \pm 1.9 }{GeV}$. An
experimental study of future prospects~\cite{CMS:2013wfa} found a
potential improvement to the $\QTY{\Delta m_t\sim \pm 1.0 }{GeV}$
level assuming similar improvements on the theoretical and
experimental uncertainties and no dependence of the measured cross
section on $m_t^{\mathrm{MC}}$. Therefore the top-mass determination
from the total cross section measurement will not be able to reach the
sensitivity of direct methods in the future, but will be
useful for consistency checks with more precise but theoretically less
clean measurements.

\section*{4~~~~Top quark mass determination from kinematic distributions
~\footnote{M.~Schulze, fruitful discussions with Jan Winter on the template method are acknowledged.}}
\addcontentsline{toc}{section}{\protect\numberline{4}{Top quark mass determination from kinematic distributions}}

The study of kinematic distributions to measure the top quark mass is another promising opportunity which has to be explored at the upcoming LHC run.
It allows to contrast uncertainties obtained from other methods and helps revealing possible missing systematics in the $m_t$ extraction. 
Evidently, those kinematic distributions have to fulfill requirements which apply to other precision observables:
They have to be calculable to higher orders in perturbation theory, their experimental signature should be as clean as possible, 
and their sensitivity to $m_t$ has to be sufficiently strong.
Other important requirements are weak dependence on non-perturbative effects such as hadronization and underlying events.
A few suitable observables have been identified in the literature 
(see e.g. \cite{Beneke:2000hk,Chakraborty:2003iw,Wagner:2005jh,Quadt:2007jk,Bernreuther:2008ju}) 
and their use for a $m_t$ determination has been investigated. 
Depending on the details of these studies projected uncertainties in the range of $\pm 1..2\,\mathrm{GeV}$ are found,
values which are highly competitive with other extraction methods.
Their power typically draws from sensitivity to kinematic boundaries and from being relatively insensitive to uncertainties of the production dynamics 
such as pdfs, threshold corrections or underlying event.  
However, it has to be pointed out that none of these observables were used in analyses of Tevatron data, and LHC experiments
are only beginning to use kinematic distributions to measure the top quark mass. 
Hence, a test under fully realistic conditions and a confirmation of the promised low uncertainties is still outstanding. 
This emphasizes the importance of studies which are precise and realistic,
and which allow quoting reliable uncertainties.
\\
At present, any infrared-save kinematic distribution of the top quark decay products in $t\bar{t}$ production can be predicted 
at next-to-leading order accuracy.
Important effects such as spin correlations, corrections to the decay dynamics and finite width contributions are under good 
theoretical control. 
Also, hadronization of $b$-jets can be modeled by $B$-meson fragmentation functions or through a subsequent parton shower approximation.
In the near future one can even expect next-to-next-to-leading order QCD predictions for differential distributions, as all basic ingredients
for such a calculation are available \cite{Czakon:2013goa,Gao:2012ja,Brucherseifer:2013iv,Melnikov:2011qx}.
This will further reduce the theoretical scale uncertainties from about $\pm 10\%$ at NLO QCD to approximately $\pm 5\%$ at NNLO QCD.
In the following sections we review a few proposals for measuring the top quark mass from kinematic distributions at the LHC.

\subsection*{$J/\psi$ method}
In Ref.~\cite{Kharchilava:1999yj} Kharchilava pointed out that the top quark mass may be measured to high precision in 
the exclusive final state $pp \to t\bar{t} \to b(\to J/\psi)\; W^+(\to \ell^+ \nu)  \; \bar{b}(\to \ell^-\!+\! X) \; W^-(\to jj)$
with the additional requirement of a leptonic decay of the $J/\psi$ meson.
In this case the invariant mass of the reconstructed $J/\psi$ and the lepton from the $W$ boson of the same top quark can be measured to high precision
and with good sensitivity to $m_t$.
Furthermore, the combinatorial background is very low and the observable is independent of jet energy scale uncertainties.
The only drawback of this method is the small cross section due to the required leptonic branching fractions. 
It is expected that an integrated luminosity of at least $100~\mathrm{fb}^{-1}$ needs to be accumulated for a precise measurement \cite{Kharchilava:1999yj}. 
Dropping the requirement of the leptonic decay of the $\bar{b}$-quark increases the combinatorial background 
and might allow for a less precise determination with only $20~\mathrm{fb}^{-1}$ \cite{Chierici:951386}.
\\
More detailed studies of this method have been performed in Refs.~\cite{Corcella:2009rs,Corcella:2000wq} 
using the parton showers PYTHIA \cite{Sjostrand:2006za} and HERWIG \cite{Marchesini:1991ch} 
for the top quark decay process $t \to W b$.
For simplicity only the invariant mass of the lepton and the $B$-meson (decaying into a $J/\psi$) has been considered.
This approximation is expected to be good since the decay of $B$-mesons into $J/\psi$-mesons is well understood 
from experiments at $B$-factories.
The top quark mass is then extracted from a linear fit of $m_t$ to the first moment of the invariant mass
of lepton and $B$-meson, $\langle m_{\ell B} \rangle$. 
Ref.~\cite{Corcella:2009rs} finds uncertainties on an $m_t$ extraction of about $1.5-2$~GeV.
These errors have been obtained by comparing results of the two parton showers PYTHIA and HERWIG.
A complementary analysis which avoids the use of parton showers and evaluates uncertainties based on a 
NLO QCD calculation was performed in Ref.~\cite{Biswas:2010sa}.
The transition of the $b$-jet into a $B$-meson was implemented through a fragmentation function at NLO QCD 
with two-loop running from $m_b$ to $\mu_\mathrm{frag}$ and an initial condition fixed by a fit to data from 
$e^+ e^- \to b\bar{b}$ \cite{Heister:2001jg,Abe:1999ki}.
The non-perturbative contribution can be parameterized through a fit to the same data as shown in Refs.~\cite{Mele:1990yq,Corcella:2001hz}.
Variations of scale uncertainties and the parameters of the non-perturbative part reveal an uncertainty of 
$\pm 1$~GeV on a top quark mass determination in the process $t \to W b$.
\\
The framework of Ref.~\cite{Biswas:2010sa} was also used to consider the full production and decay process 
$pp \to t\bar{t} \to b(\to B)\; W^+(\to \ell^+ \nu)  \; \bar{b} \; W^-(\to jj)$ at NLO QCD 
in the narrow width approximation, including acceptance cuts on the top quark decay products.
This allows to account for additional uncertainties from e.g. pdfs, jet recombination with 
initial state radiation, as well as NLO spin correlations between production and decay dynamics.
From a linear fit of $m_t$ to $\langle m_{\ell B} \rangle$ an uncertainty of 
$\pm 1.5$~GeV on the top quark mass is found. 
These uncertainties are obtained by assuming an uncertainty of $\pm 0.4$~GeV on the $\langle m_{\ell B} \rangle$ extraction 
and by varying renormalization, factorization and fragmentation scales independently as well as by
using two different parameterization of the fragmentation function.

\begin{figure}
    \begin{center}
      \includegraphics[width=0.43\textwidth]{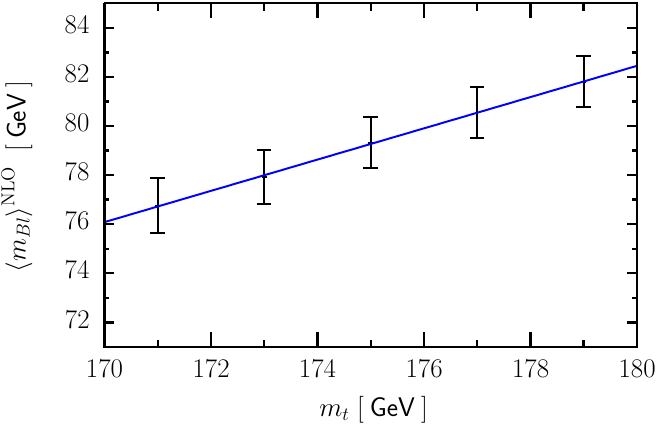}
      \hfill
      \includegraphics[width=0.495\textwidth]{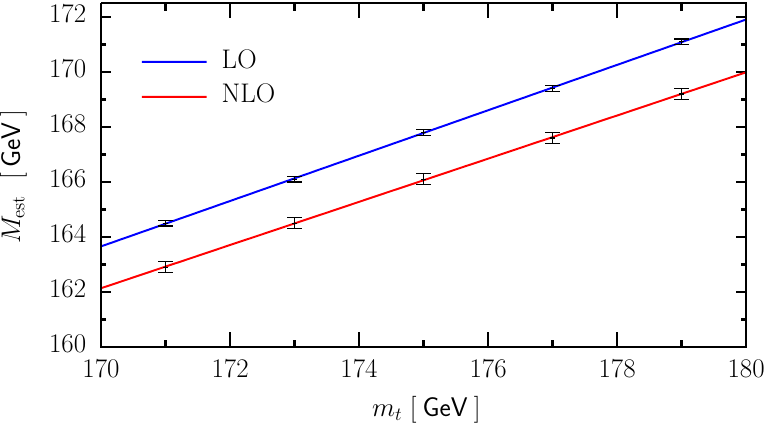}      
      \caption{Linear fit of $m_t$ to $\langle m_{\ell B} \rangle$ in the $J/\psi$ method (left) and 
               $M_\mathrm{est}$ in the $m_{\ell b}$ method (right)
               at the 14~TeV LHC ~\cite{Biswas:2010sa}.
               Error bars are obtained from scale variations, different parton distribution sets and variation of the 
               $b$-quark fragmentation function} \label{schulze:fig1}
    \end{center}
\end{figure}

\subsection*{$m_{\ell b}$ method}
A similar but more inclusive extraction of $m_t$ from the invariant mass of a lepton and a $b$-jet was suggested in Ref.~\cite{Beneke:2000hk}.
Not requiring leptonic decays of the $b$-quark decay products increases the rate such that one can study the di-leptonic decay mode of the two $W$ bosons.
In such case one can construct an estimator for $m_t$ given by 
\begin{equation} \label{schulze:Mest}
    M_\mathrm{est}^2 = M_W^2 + \frac{2 \langle m_{\ell b}^2 \rangle}{1-\langle \cos\theta_{\ell b} \rangle},
\end{equation}
where $\theta_{\ell b}$ is the angle between the lepton and the $b$-jet from the same top quark. 
The power of this estimator becomes evident when considering the LO values of $\langle m_{\ell b}^2 \rangle$ and $\langle \cos\theta_{\ell b} \rangle$ 
which imply $M_\mathrm{est}^2 = m_t^2$, if no constraints on the final state are applied.
It is obvious that in practice Eq.~(\ref{schulze:Mest}) is violated by several effects:
(a) acceptance cuts on leptons, jets and missing energy to account for detector effects, 
(b) higher order corrections which include additional radiation off the top quark decay products,
(c) imperfect pairing of lepton and $b$-jet,
(d) experimental issues such as particle misidentification.
However, it should be noted that points (a)-(c) can be addressed exactly in perturbative QCD. 
This was archived in Ref.~\cite{Biswas:2010sa} to NLO accuracy using the narrow width approximation for top quarks.
Similar to the $J/\psi$ method, the value and uncertainty of $m_t$ can be obtained from a linear fit to $M_\mathrm{est}$.
It is found that for realistic acceptance cuts at the 14~TeV LHC and assuming infinite precision on $M_\mathrm{est}$, 
the top quark mass can be extracted to $\pm 0.25$~GeV. 
Those uncertainties are derived from independent variations of factorization and renormalization scales as well as
from the evaluation of two different parton distribution sets of MSTW~\cite{Martin:2002aw} and CTEQ~\cite{Pumplin:2002vw,Nadolsky:2008zw}.
Additional experimental systematics on $m_t$ from $b$-quark fragmentation and jet energy scale are expected to be of the order of 
$\pm 0.7$~GeV and $\pm 0.6$~GeV \cite{Beneke:2000hk}, respectively.
\\
On the theoretical side there are a few more systematic effects on $m_{\ell b}$ which should be discussed. 
The effect of NLO QCD corrections to the top quark decay were studied in Refs.~\cite{Bernreuther:2004jv,Melnikov:2009dn,Campbell:2012uf}.
Refs.~\cite{Melnikov:2009dn,Campbell:2012uf} find that those correction lead to shape changes of the $m_{\ell b}$ distribution 
when compared to a calculation which includes NLO QCD correction to the $t\bar{t}$ production followed by a LO decay $t \to W b$.
Also, finite width effects in $t\bar{t}$ production have been studied thanks to the NLO QCD calculations of $WW b \bar{b}$ production 
\cite{Denner:2010jp,Bevilacqua:2010qb,Schlenk:2013gza,Frederix:2013gra}.
A tuned comparison with the results of a narrow width approximation in Ref.~\cite{AlcarazMaestre:2012vp} reveals that 
finite width effects have negligible effects on the total cross section but 
can lead to $\mathcal{O}(10\,\%)$  effects in the tail of certain kinematic distributions. 
In the case of $m_{\ell b}$, finite width effects turn out to be small (at the level of 1\%) over the entire kinematic range
until the mass drop at $m_{\ell b} \approx \sqrt{m_t^2-M_W^2}$.
Beyond that point finite width effects become large. 
However also the cross section drops by one order of magnitude and unknown non-perturbative effects might be sizable.
We conclude that the $m_{\ell b}$ distribution is under good theoretical control and can be predicted reliably to high precision
over the entire phenomenologically relevant range.
\\
\begin{figure}
    \begin{center}
      \includegraphics[width=0.5\textwidth]{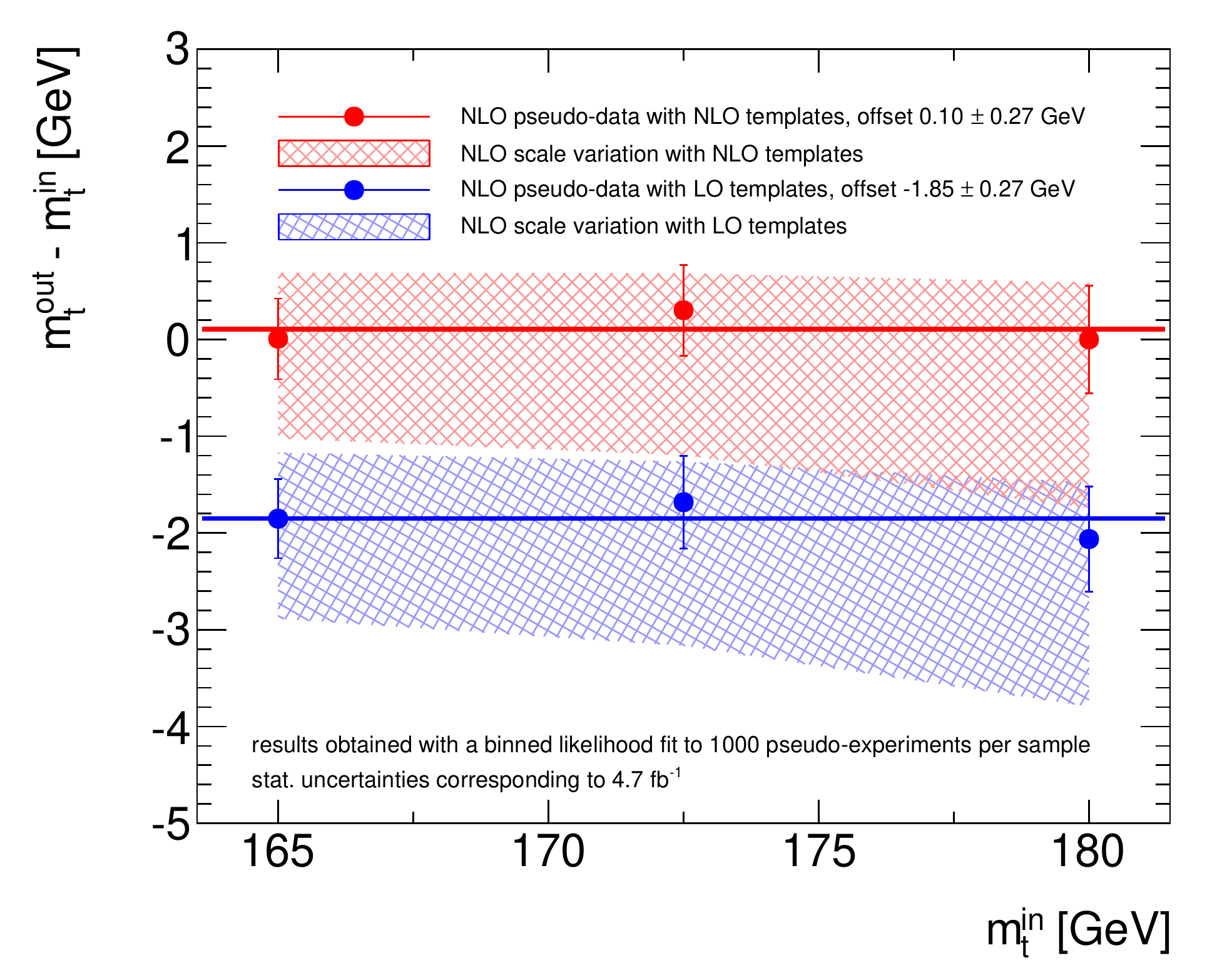}    
      \caption{ Taken from Ref.~\cite{Heinrich:2013qaa}, Fig.~8b.
      The points show the observed mean differences in $m_t^\mathrm{out}-m_t^\mathrm{in}$ from the template fit to pseudo data together with
their statistical uncertainty corresponding to a luminosity of 4.7~fb. The bands
indicate uncertainties when replacing the NLO pseudo-data by the ones obtained from NLO scale variation. } \label{schulze:fig2}
    \end{center}
\end{figure}
Finally, we would like to point out an interesting result of Ref.~\cite{Heinrich:2013qaa}.
The authors study the top quark mass determination from a $m_{\ell b}$ template fit 
and closely follow a procedure used by ATLAS~\cite{ATLdil2011}.
Templates for various (input) values of $m_t$ are generated from LO and NLO QCD predictions.
These templates are then used to fix a multi-dimensional parameterization   
leaving only $m_t$ as a free variable~\cite{ATLdil2011}.
The obtained parameterization can be employed to obtain a fit value of the top quark mass from data.
For the theoretical study of Ref.~\cite{Heinrich:2013qaa}, pseudo-data 
are generated from the NLO QCD predictions for a given value of $m_t$ assuming they
represent real data.
Theoretical uncertainties are estimated from scale variations
by generating three sets of pseudo-data for the central scale choice and its variations by a factor of two.
The three sets are then used for individually fitting the template parametrization of $m_t$.
The results are presented in Fig.~\ref{schulze:fig2} for two scenarios (hypotheses): LO templates (blue) and NLO templates (red).
The width of the uncertainty bands shown in Fig.~\ref{schulze:fig2} arises from 
the variations of the scales.
It is interesting to note that this uncertainty of about $\pm 0.8$~GeV is significantly larger than has been reported in previous experimental studies.
These studies typically rely on a factorized calculation of $t \bar{t}$ production and decay where
decay dynamics are treated at LO or in the parton shower approximation.
It is therefore important to investigate whether the differences in the uncertainty estimates arise
from NLO corrections in the decay or from finite width effects,
both of which are taken into account in the full calculation of $pp \to WWb \bar{b}$.
Another interesting feature in Fig.~\ref{schulze:fig2} is the gap between the blue and the red band.
This effect can be traced back to shape differences between the LO and NLO description of the $m_{\ell b}$ distribution. 
The mismatch between NLO pseudo-data and LO templates is compensated for by a value of $m_t$ which has an offset of almost 2~GeV.
In summary, the results of Ref.~\cite{Heinrich:2013qaa} are intriguing because they draw attention to the details of template methods which might be contaminated with
systematics that are not accounted for in a typical error budget.

\section*{5~~~~Determination of the running top quark mass~\footnote{S.~Moch, M.~Dowling}}
\addcontentsline{toc}{section}{\protect\numberline{5}{Determination of the running top quark mass}}

\noindent 
Quark masses are simply parameters of the QCD Lagrangian and, as such, they are subject to renormalization.
The most commonly used mass renormalization scheme defines the pole mass as
the location of the single pole in the two-point function at each order in perturbation theory.
Although being inspired by QED and the definition of the electron mass for an asymptotic state in the $S$-matrix, 
the notion of a pole mass for quarks is not really practical, because quarks 
do not appear asymptotically as free particles. 
Therefore, in the full theory including non-perturbative corrections,
the quark two-point function does not exhibit any pole.
Moreover, the value of the pole mass is strongly dependent on the order of
perturbation theory and, at high energies, large logarithms spoil the perturbative convergence.
This calls for alternative quark mass definitions and, particularly, in high
energy reactions for the use of a so-called short-distance mass, like the
running mass $m_t(\mu)$ in the $\overline{\rm MS}$ scheme, which is defined at
the scale appropriate to the hard scattering.
Much like the strong coupling constant $\alpha_s(\mu)$ in the $\overline{\rm MS}$ scheme, 
also $m_t(\mu)$ runs with $\mu$.

A determination of the top quark mass has to identify an appropriate
observable together with the dependence on the mass parameter in a chosen renormalization scheme.
For the running top quark mass therefore, this requires to perform a scheme
transformation from the standard perturbative results at NLO and NNLO, which
commonly employ the pole mass scheme.
For the inclusive cross section this has been achieved in \cite{Langenfeld:2009wd,Aliev:2010zk}
and for differential distributions in \cite{Dowling:2013baa}. 
In both cases, it has been found that the perturbative expansion converges faster 
and the scale dependence improves by using the mass 
in the $\overline{\rm MS}$ scheme as opposed to the on-shell scheme. 
\begin{figure}[t!]
\centerline{
  \includegraphics[width=6.5cm]{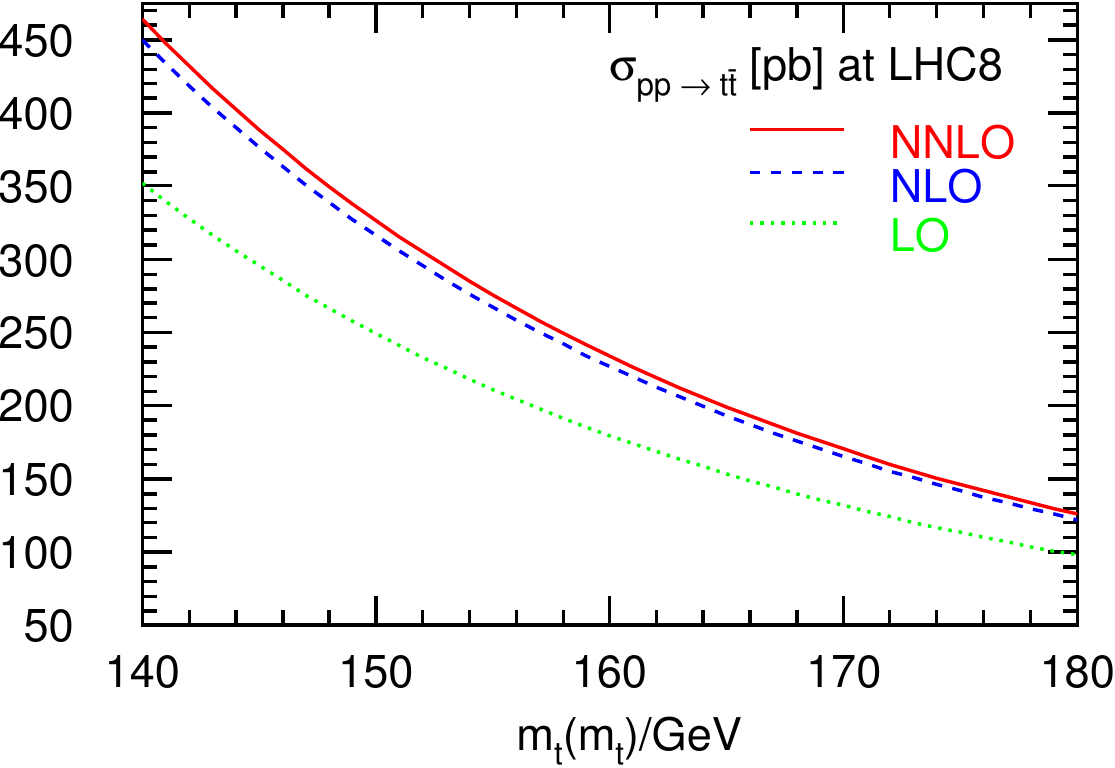}
  \hspace*{5mm}
  \includegraphics[width=6.5cm]{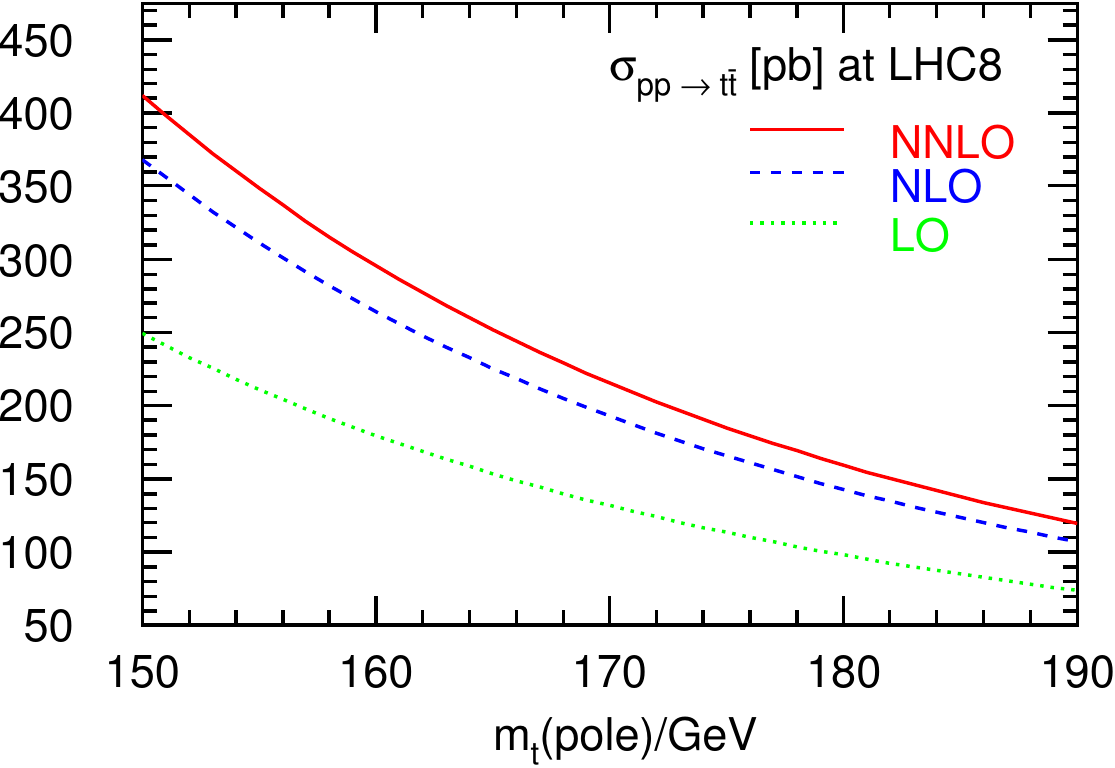}}
  \caption{\small
    \label{fig:moch-ttbar-incl}
    The LO, NLO and NNLO QCD predictions for the 
    $t{\bar t}$ total cross section at the LHC ($\sqrt{s} = 8$~TeV) 
    versus the top-quark mass 
    in the $\overline{\rm MS}$ scheme $m_t(m_t)$ at the scale $\mu = m_t(m_t)$ (left) 
    and in the on-shell scheme $m_t({\rm pole})$ at the scale $\mu = m_t({\rm pole})$ (right) 
    with the ABM12 PDFs. (Figure from Ref.~\cite{Alekhin:2013nda}).
  }
%
\vspace*{5mm}
\centerline{
\includegraphics[width=6.5cm]{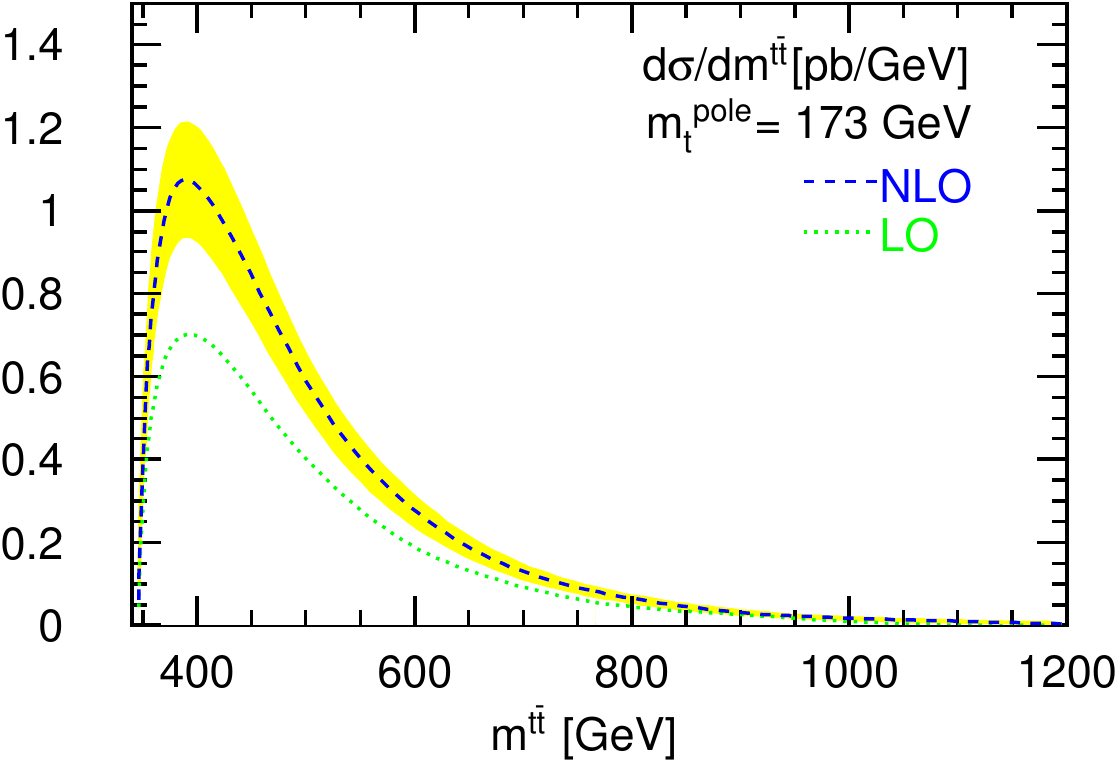}
  \hspace*{5mm}
\includegraphics[width=6.5cm]{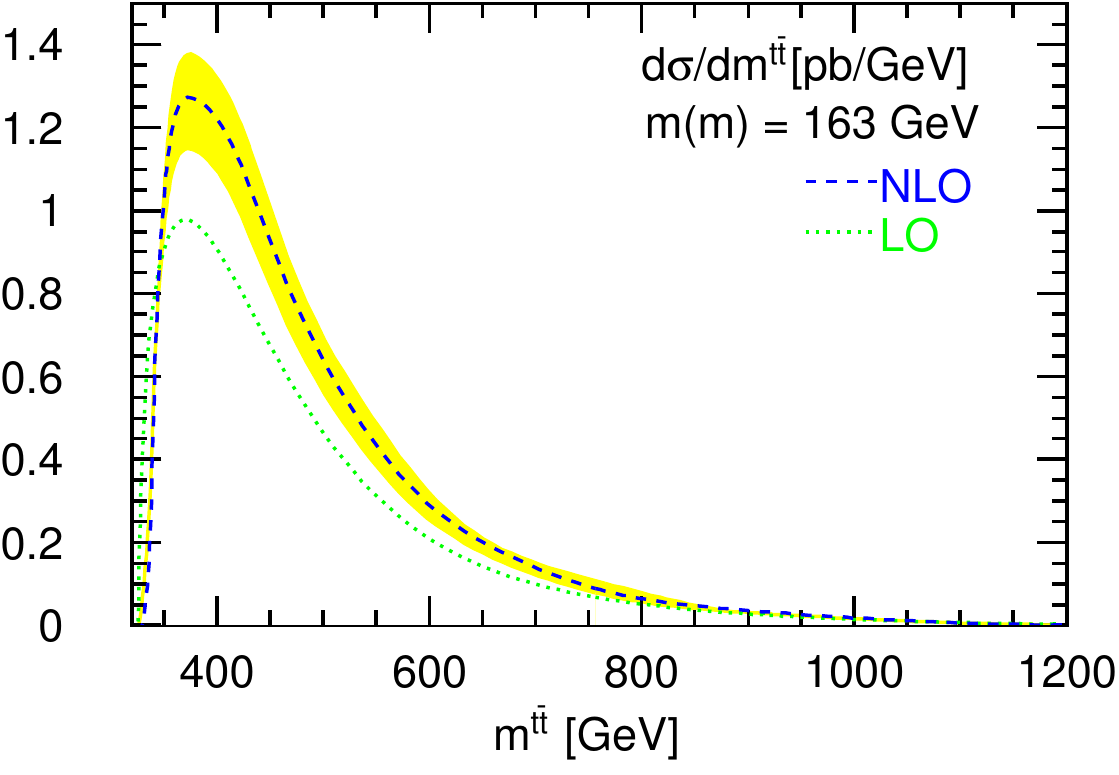}}
\caption{\small \label{fig:moch-mtt}
The differential cross section versus the invariant mass $m^{t{\bar t}}$ of the top quark pair
in the pole (left) and the $\overline{\rm MS}$  (right) mass scheme at the LHC
with $\sqrt{S}=8$~TeV.
The dotted (green) curves are the LO contributions, the dashed (blue) 
curves include NLO corrections and are obtained with the PDF set CT10~\cite{Gao:2013xoa}.
The scale dependence in the ranges $\mu/m_t^{\rm pole}$ 
or $\mu/m(m) \in [1/2,2]$ is shown as a band around the NLO curve.
(Figure from Ref.~\cite{Dowling:2013baa}).
} 
\end{figure}

Fig.~\ref{fig:moch-ttbar-incl} displays the theory predictions for inclusive $t{\bar t}$-pair production 
in the $\overline{\rm MS}$ scheme as well as for the pole mass $m_t({\rm pole})$ based on 
the available complete NNLO QCD corrections~\cite{Baernreuther:2012ws,Czakon:2012zr,Czakon:2012pz,Czakon:2013goa}.
The former results, i.e., the predictions as a function of the running mass $m_t(\mu_r)$
display much improved convergence and better scale stability 
of the perturbative expansion~\cite{Langenfeld:2009wd,Aliev:2010zk}.
Fig.~\ref{fig:moch-mtt} illustrates the same findings for the distribution in the invariant mass 
$m^{t{\bar t}}$ of the top quark pair, 
though the QCD corrections are available only to NLO in this case~\cite{Campbell:2010ff,Campbell:2012uf}. 
In addition to the improvement of the convergence, when moving from the pole mass 
to the $\overline{\rm MS}$ scheme, the overall shape of the distribution changes, 
the peak becomes more pronounced, while the peak position remains stable against radiative corrections.
The latter feature is essential for precision determinations of $m_t(\mu)$ from
LHC data in the forthcoming high-energy runs.

\begin{figure}[th!]
\centerline{
  \includegraphics[width=9.25cm]{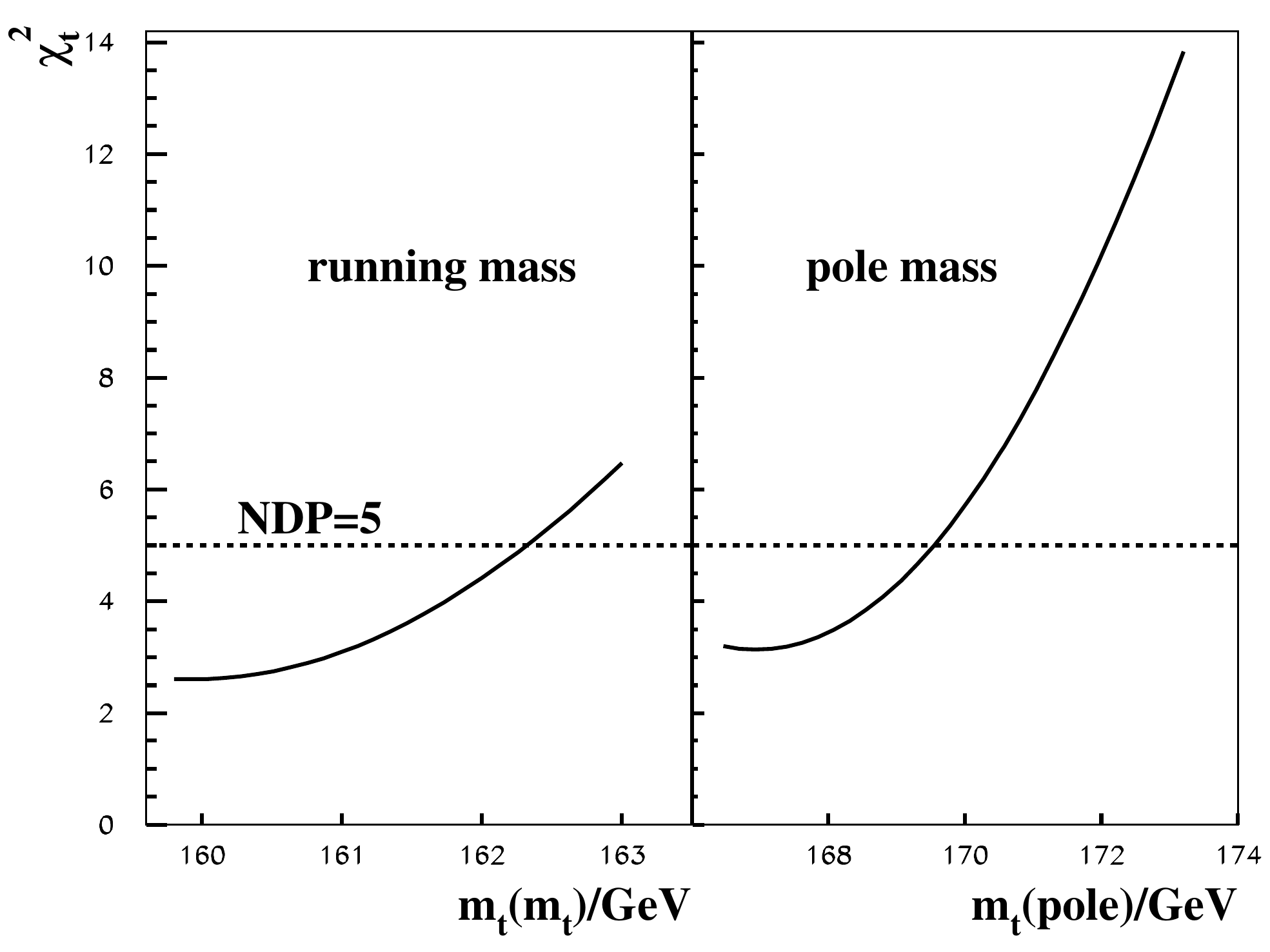}}
  \caption{\small
    \label{fig:moch-chi2t}
      The $\chi_t^2$ profile versus the $t$-quark mass for the $t\bar{t}$ cross section data subset
      in the variants of ABM12 fit with the $t\bar{t}$ cross section data included
      and the different $t$-quark mass definitions: 
      running mass (left) and pole mass (right). 
      The $NDP=5$ for this subset is displayed by the dashed line. 
      (Figure from Ref.~\cite{Alekhin:2013nda}).
}
\end{figure}

Using the NNLO theory predictions for the total top-quark pair-production cross section 
as displayed in Fig.~\ref{fig:moch-ttbar-incl}, Ref.~\cite{Alekhin:2013nda} has 
performed a consistent determination of the top-quark mass in a well-defined 
renormalization scheme with full account of the correlations 
with the gluon distribution and the strong coupling $\alpha_s$ based on the available data from 
LHC~\cite{ATLAS:2012fja,Chatrchyan:2012bra,ATLAS:2012jyc,CMS:2012dya,CMS:2012gza} 
and the Tevatron combination~\cite{tevewwg:2012}. 
Fig.~\ref{fig:moch-chi2t} shows the $\chi^2$ profile versus the top-quark mass, 
i.e., for the $\overline{\rm MS}$ mass $m_t(m_t)$ and the pole mass $m_t({\rm pole})$, 
in two variants of the ABM12 fit~\cite{Alekhin:2013nda} of parton distributions 
with the $t\bar{t}$ cross section data included.
From Fig.~\ref{fig:moch-chi2t} it is apparent, that the requirement of 
$\chi_t^2/NDP = 1$ leads to the value for the $\overline{\rm MS}$ mass at NNLO
\begin{eqnarray}
\label{eq:moch-mt}
m_t(m_t) = 162.3 \pm 2.3~{\rm GeV}\, ,
\end{eqnarray}
with an error in $m_t(m_t)$ due the experimental data, the PDFs and the value of $\alpha_s(M_Z)$
as the difference between the value for $m_t(m_t)$ at $\chi_t^2/NDP = 1$ 
and the minimum of the $\chi^2$-profile in Fig.~\ref{fig:moch-chi2t}.
An additional theoretical uncertainty from the variation of the factorization
and renormalization scales in the usual range is small, $\Delta m_t(m_t) = \pm
0.7~{\rm GeV}$ and Eq.~(\ref{eq:moch-mt}) is equivalent to the pole mass value of
\begin{eqnarray}
\label{eq:moch-mtpole}
m_t({\rm pole}) \,=\, 171.2 \pm 2.4~{\rm GeV}
\, ,
\end{eqnarray}
using the known perturbative conversion relations~\cite{Gray:1990yh,Chetyrkin:1999qi,Melnikov:2000qh}.
Within the uncertainty, this indicates good consistency of the procedure and also with the top-quark mass values obtained from other determinations.


\section*{6~~~~An Effective Theory approach to finite-width effects with an application to mass-scheme ambiguities
\footnote{A.S.~Papanastasiou, based on work with P.~Falgari (Utrecht) and A.~Signer (PSI) 
published in \cite{Falgari:2013gwa}. } }
\addcontentsline{toc}{section}{\protect\numberline{6}{An Effective Theory approach to finite-width effects with an application to mass-scheme ambiguities}}

Currently, the most precise hadron-collider experimental extractions of the top quark mass parameter come about through 
analyses based on kinematically reconstructing top quarks from their decay products, namely, leptons, jets and missing energy. 
Extractions of the mass often depend on fitting templates produced by Monte Carlo (MC) generators to data.  
The issue discussed in this section is that when NLO MCs are used to create such templates 
(and it is only when going beyond LO that the top-mass parameter becomes well-defined in perturbation theory), 
the NLO hard matrix elements \emph{only} describe the production of
on-shell stable tops. This in turn means that predictions for observables that are sensitive to the offshellness of top quarks (most observables
used for extraction fall in this category) are not actually NLO accurate. 
Given that NLO hard matrix elements with full off-shell effects matched to parton showers are not yet available, the systematic errors on 
top-signal modelling as well as mass-extraction, introduced by not including such effects is thus far unknown.     

The first step in addressing this issue is the systematic description at NLO of the parton-level process including finite-width effects. 
For the top pair-production process, one successful approach has been to compute the NLO corrections to the fully decayed final state, $W^+W^-b\bar{b}$ 
\cite{Bevilacqua:2010qb,Denner:2010jp,Denner:2012yc,Frederix:2013gra,Cascioli:2013wga,Heinrich:2013qaa}   
via the use of the complex-mass scheme \cite{Denner:1999gp,Denner:2005fg}. The alternative framework outlined here is one that describes the dominant 
off-shell effects of the \emph{resonant} production and decay of unstable top quarks. In detail, this approach captures the physics when the 
invariant mass of the system of top decay products is close to the top mass, $m_t$, i.e. when
\begin{align}
p^2_t = \left( \sum_{i \in \text{decay}} p_i \right)^2 \sim m^2_t.  
\end{align}  
In the resonant region we have
\begin{align}
\frac{p^2_t-\mu^2_t}{m^2_t} \sim \frac{m_t \, \Gamma_t}{m^2_t} \sim \frac{\Gamma_t}{m_t} \ll 1,
\end{align}
with $\mu_t$ being the complex pole of the full top quark propagator, indicating that the typical virtuality is controlled by the top width, $\Gamma_t$. 
Given this, the idea is to expand the full matrix elements in the new small parameter $\Delta_t := (p^2_t-\mu^2_t)/m^2_t$, in addition to expanding 
in the couplings $\alpha_s$ and $\alpha_{ew}$. The starting point for expansion is the introduction of a power-counting 
\begin{align}
\alpha^2_s \sim \alpha_{ew} \sim \Delta_t \sim \delta \ll 1,
\end{align} 
which allows one to assign a parametric scaling in $\delta$ to any part of a Feynman diagram before any computation. In turn, this
means that it is possible to efficiently identify the set of contributions that need to be computed in order to achieve a given accuracy in $\delta$.  
The expansion in the generic small parameter, $\delta$, can be viewed as a systematization of the pole expansion method \cite{Aeppli:1993rs,Stuart:1991cc}. 
Such an expansion has been employed in the computation of inclusive quantities in \cite{Beneke:2004km,Beneke:2007zg} and recently adapted for the fully
differential description of offshell single-top and $t\bar{t}$ production \cite{Falgari:2010sf,Falgari:2011qa,Falgari:2013gwa}. Details of precisely 
how the expansion is carried out in the fully-differential case can be found in the latter references. 

A pleasing aspect to the expansion in $\delta$ is that it reshuffles the contributions to the full $W^+W^-b\bar{b}$ process into a structure 
resembling an Effective Theory (ET) calculation. That is to say, the expansion
re-arranges parts of amplitudes into contributions which in a formal ET approach would be obtained by Wilson coefficients multiplying 
operators describing the production, propagation and decay of top quarks in addition to contributions from dynamical, soft gluons. The former and latter 
contributions are usually referred to as factorizable and non-factorizable corrections respectively.
A comparison of this ET-approach and a computation using the complex-mass scheme for $t$-channel single top \cite{Papanastasiou:2013dta} confirmed that 
the ET-expansion indeed captures all the relevant physics in its region of validity.  

A relevant detail to highlight here is that the ET-like expansion naturally lends itself to a class of suitable mass-schemes. This is desirable since  
it is well known that the pole scheme suffers from QCD renormalon effects \cite{Bigi:1994em, Beneke:1994sw,Smith:1996xz} which render the determination 
of the pole mass to an accuracy better than $\Lambda_{\text{QCD}}$ impossible. In order for the 
power-counting to remain valid order by order in $\delta$, the renormalized mass, $m_{r,t}$, must satisfy the condition 
$m^2_{r,t} - m^2_{\text{pole},t} \sim m_{\text{pole},t} \delta$. This excludes the use of, for example, the $\overline{\text{MS}}$-mass 
as a suitable mass in the expansion in $\delta$, since  
$m^2_{\overline{\text{MS}},t}-m^2_{\text{pole},t} \sim m_{\text{pole},t} \alpha_s \sim m_{\text{pole},t} \delta^{1/2}$.
However, there are other short-distance masses (threshold masses \cite{Bigi:1994em,Beneke:1998rk,Hoang:1999zc,Pineda:2001zq} 
and jet-masses \cite{Fleming:2007qr,Fleming:2007xt}), which, like $m_{\overline{\text{MS}},t}$, don't feature the problems inherent to 
$m_{\text{pole},t}$ but that do adhere to the power-counting. Here we choose to examine the use of the PS-mass \cite{Beneke:1998rk}, which, 
up to $\mathcal{O}(\alpha^2_s)$, is defined as 
\begin{align}
m_t=m_{\text{PS},t}+\mu_{\text{PS}} \left[
\frac{\alpha_s}{2 \pi} \delta_1^{\text{PS}}
+\left(\frac{\alpha_s}{2 \pi} \right)^2 \delta_2^{\text{PS}}\right]
\end{align}  
where $\delta_{1,2}^{\text{PS}}$ are terms that scale as $\delta_{1,2}^{\text{PS}} \sim 1$ in the power-counting.
In order to preserve the ET-counting, 
we must choose the PS-scale so as to satisfy $\mu_{\text{PS}} \sim m_t \alpha_s$ and doing so means that the use of the PS-mass instead of the pole mass
in the calculation is perfectly acceptable. At NLO in the expansion in $\delta$, the introduction of the PS-mass modifies the structure of the renormalized 
top-quark propagator \cite{Falgari:2013gwa} but leaves the factorizable corrections to the production and decay subprocesses as well as the 
non-factorizable contributions unchanged.  

Given that in the ET-approach the option to use alternative (suitable) mass-schemes is straightforward, it is possible to study the effect of 
using a scheme other than the pole scheme on the extraction of the top mass. In what follows we work only with the $q\bar{q}$-initiated channel.  
We first generate a pseudo-data set for a $(e^+,\nu_e,\text{b-jet})$-invariant mass distribution 
using the pole scheme with $m_{\text{pole},t}=172.9$ GeV.
Invariant mass distributions are then generated in the PS-scheme with $\mu_{\text{PS}} \in \{0,10,20\}$ GeV, where $\mu_{\text{PS}}=0$ GeV clearly corresponds to the 
pole scheme. The top mass in each scheme is extracted by adjusting its value, such that the predicted invariant mass distribution is in optimal agreement with 
the pseudo-data. The extracted masses are then converted to the $\overline{\text{MS}}$ and pole masses using a three-loop conversion \cite{Melnikov:2000qh} accompanied 
by a Pad\'e  estimate for higher-order corrections (the precise details of the latter do not affect our findings). The results of this toy study are shown in 
Table \ref{papanastasiou:tablemass-extraction}, where the left and right panels 
display the results of extractions using LO and NLO predictions respectively. Importantly, the NLO predictions here not only include corrections to the production 
and decay of the tops, but also non-factorizable corrections and corrections to the propagation. Only including corrections to the production and decay does not improve 
the LO extraction.      

\begin{table}
\centering
\begin{tabular}{c||ccc||ccc}
\hline
& & & & & &  \\[-9pt]
 & \multicolumn{3}{c||}{ LO}  & \multicolumn{3}{c}{NLO} \\
$\mu_{\text{PS}}$ & $m_{\text ext.}$ & $m_{\overline{\text{MS}},t}$ & $m_{\text{pole},t}$ \qquad
& $m_{\text ext.}$ & $m_{\overline{\text{MS}},t}$ & $m_{\text{pole},t}$ \\[3pt]
\hline \hline
& & & & & &  \\[-9pt]
0 & 172.9 & 162.2 & 172.9 & 172.9 & 162.2 & 172.9 \\
10 & 172.4 & 162.7 & 173.5 & 172.2 & 162.4 & 173.3 \\
20 & 172.0 & 163.0 & 173.8 & 171.5 & 162.5 & 173.4 \\[3pt]
\hline
\end{tabular}
\caption{Extraction of the top mass in various schemes at LO (left
  panel) and NLO (right panel). All numbers are in units of GeV. }
\label{papanastasiou:tablemass-extraction}
\end{table}

Inspecting Table \ref{papanastasiou:tablemass-extraction}, we find the spread of values 162.2-163.0 GeV (172.9-173.8 GeV) at LO and 162.2-162.5 GeV (172.9-173.4 GeV) 
at NLO for the `extracted' $\overline{\text{MS}}$ (pole) mass. These differences are consistent up to higher orders in $\alpha_s$, however, their existence should 
be interpreted as ambiguities of 600-800 MeV (500-900 MeV) at LO and 200-300 MeV (400-500 MeV) at NLO in the determination of the $\overline{\text{MS}}$ (pole) mass. 
We emphasize here that this investigation has not included a number of other important effects (colour reconnections etc.) and has only been performed 
for the $q\bar{q}$-initialized channel. However, the existence of an additional ambiguity due to the choice of mass-scheme is unlikely to change.

\section*{7~~~~Determination of the  top-quark mass using top-quark
  pair events with an additional jet~\footnote{P.~Uwer}}
\addcontentsline{toc}{section}{\protect\numberline{7}{Determination of the top-quark mass using top-quark pair events with an additional jet}}

\paragraph*{Introduction:}
The top-quark mass is a crucial input parameter for many Standard
Model predictions. For example, it enters through electroweak precision
observables in the consistency checks of the Standard Model. Recently it
has been pointed out that the precise value is also crucial for 
the stability of the vacuum (for more details see the contributions
of J.~Espinosa and M.~Lindner to this workshop and 
\Refs{Degrassi:2012ry,Alekhin:2012py}) 
The top-quark mass is currently determined using collider data
collected at Tevatron and LHC.  In \Ref{ATLAS:2014wva} various measurements
from Tevatron and LHC using different decay channels have been combined:
\begin{equation}
  \label{uwer:eq:mtcombined}
  \mt = 173.34 \pm 0.27 \mbox{ (stat) }\pm 0.71 \mbox{ (syst) }
  \mbox{GeV}/c^2
  \mbox{ \cite{ATLAS:2014wva}}.
\end{equation}
The total uncertainty is now 0.76 GeV$/c^2$. At this level of
precision the precise definition of the top-quark mass parameter
(renormalisation scheme) becomes important. Since most of the
measurements entering \Eq{uwer:eq:mtcombined} rely heavily on the
usage of Monte Carlo event generators the determined mass is often
called {\it Monte Carlo mass}. It is assumed that the Monte Carlo
mass is closely related to the on-shell/pole mass.

\paragraph*{Top-quark mass from jetrates:} 
Given a relative uncertainty of the combined top-quark mass below 0.5
per cent it is important to confront the existing measurements with results
obtained using alternative methods less dependent on Monte Carlo event
generators. In \Ref{Alioli:2013mxa} a method using top-quark pairs
produced in association with an additional jet has been proposed. This
process is theoretically very well under control: The next-to-leading
order (NLO) QCD corrections have been calculated in
\Refs{Dittmaier:2007wz,Dittmaier:2008uj,Melnikov:2010iu}. Setting the
renormalisation scale equal to the factorisation scale and using the
top-quark mass to define the scales corrections of the order of only 15\%
are observed. The scale variation up and down by a factor of two
arround the central scale indicates small uncertainties due to
uncalculated higher orders. In \Refs{Alioli:2011as,Kardos:2011qa} the
NLO corrections have been combined with the parton shower. Since the
inclusive cross section for the production of a top-quark pair in
association with an additional jet is not very sensitive to the
top-quark mass --- in fact a detailed analysis shows that the
sensitivity is very similar to the inclusive production of top-quark
pairs \cite{Alioli:2013mxa} --- one has to focus on events produced not
to far from theshold. This can be done for example by studying
\begin{equation}
  \label{uwer:eq:Rdefinition}
  {\cal R}(\rhos, \mt) = {1\over \sigmattjet} {d\sigmattjet\over
  d\rhos},\quad (\rhos = {m_0\over\sqrt{s_{\ttjet}}}).
\end{equation}
The inclusive cross section for \ttjet\ production is denoted by
\sigmattjet. The top-quark mass is renormalised in the on-shell scheme.
Note that to define this cross section a minimal $p_T$-cut on the
additional jet is required (for details see 
\Refs{Dittmaier:2007wz,Dittmaier:2008uj}). The scale $m_0$ introduced
in the definition of $\rhos$ is an arbitray scale of the order of the
top-quark mass. In \Ref{Alioli:2013mxa} it has been fixed to 170 GeV.
Note that in the definition of ${\cal R}$ as ratio many theoretical
and experimental uncertainties cancel between numerator and denominator.
\begin{figure}
\begin{center}
\includegraphics[height=4.9cm]{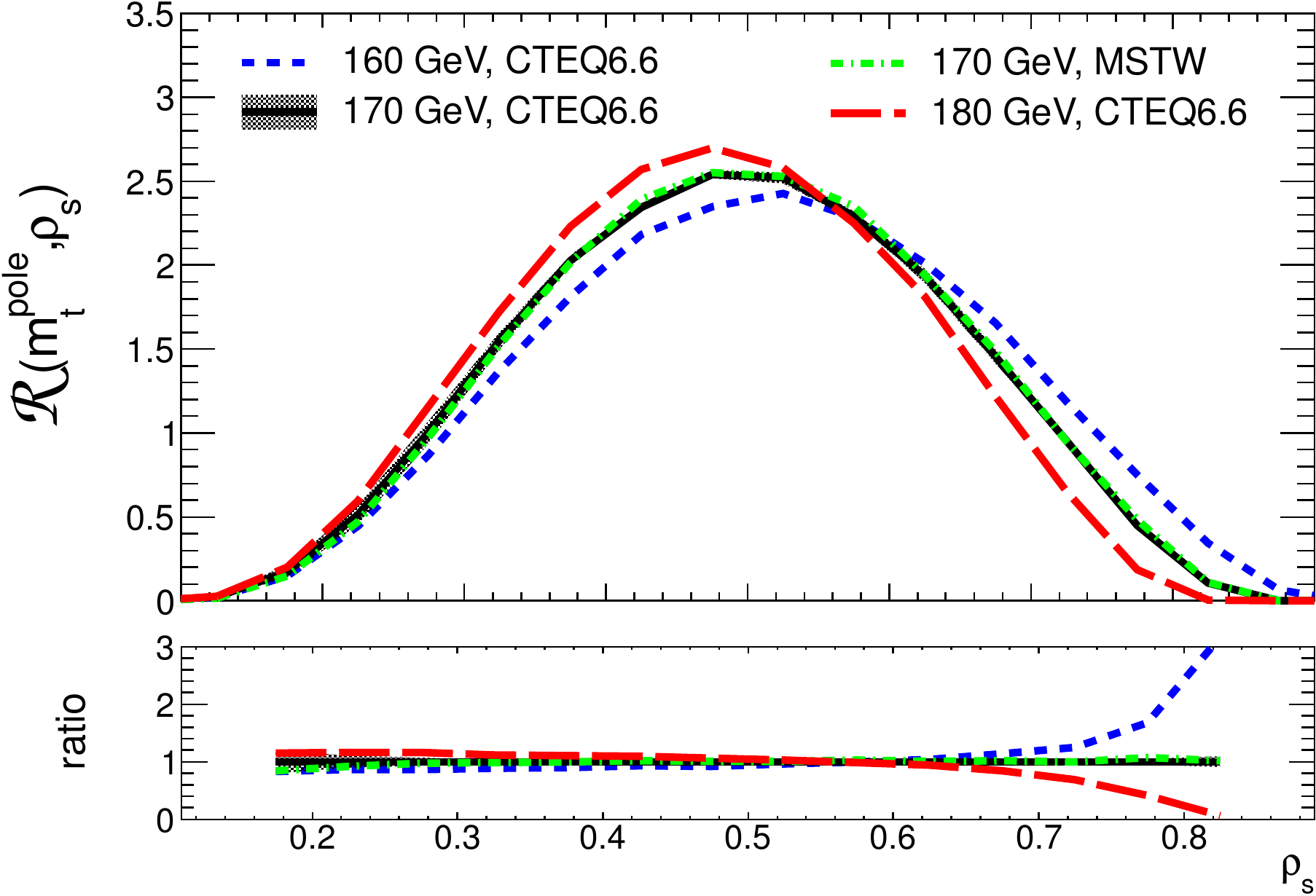}\quad
\includegraphics[height=4.8cm]{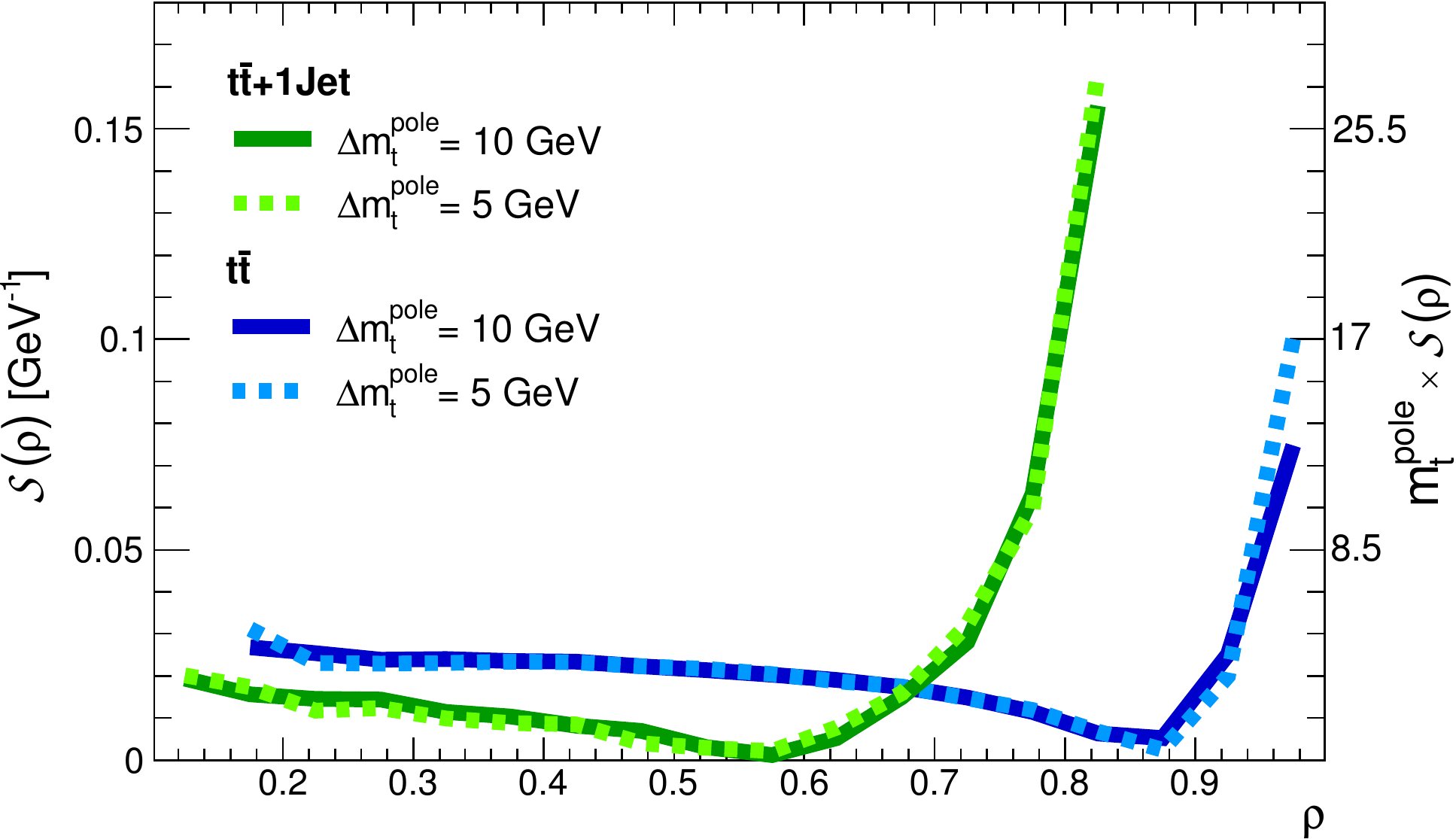}
 \caption{Differential distribution for three different top-quark
   masses (left panel), mass sensitivity as function of $\rhos$
   (right panel).}
    \label{uwer:fig1}
\end{center}
\end{figure}
In \Fig{uwer:fig1} (left) the differential distribution is shown for
three different top-quark masses (\mt = 160, 170, 180 GeV$/c^2$). In
the threshold region ($\rhos\approx 1$) the distribution for heavier
quark masses is suppressed compared to lighter quark masses. Since the
distributions are normalised, the opposite is true in the high energy
regime ($\rhos \approx 0$). In the region $\rhos = 0.5-0.6$ the
curves cross and the observable is insensitive to the top-quark mass.
In  \Fig{uwer:fig1} (right) the sensitivity
\begin{equation}
{\cal S}(\rhos,\Delta \mt) =
\sum_{\Delta=\pm \Delta \mt}
\frac{|{\cal R}( \rhos,170~\GeV/c^2)-{\cal R}(\rhos,170~\GeV/c^2+\Delta)|}
{2 |\Delta|  {\cal R}(170~\GeV, \rhos)}
\end{equation} 
is shown. 
The sensitivy ${\cal S}$ can be used to relate the relative
change of $\cal R$ to a relative change of $\mt$: 
\begin{equation}
  \left|{\Delta {\cal R} \over {\cal R}}\right| \approx 
  \left(\mt {\cal S}\right)\,\times\, 
  \left|{\Delta \mt\over \mt}\right|.
\end{equation}
For $\rhos\approx 0.8$ it can be read off from \Fig{uwer:fig1} that a
one per cent change in \mt\ leads to a relative change of 17 per cent of $\cal
R$. The sensitivty can also be used to estimate how 
experimental/theoretical uncertainties  in the determination of $\cal
R$ translate into an uncertainty of the top-quark mass extracted from a
measurement of $\cal R$. In \Ref{Alioli:2013mxa} various uncertainties
(e.g. color reconnection, scale and pdf uncertainty, jet energy scale)
have been studied. From these investigations a top-quark mass
measurement with an uncertainty of about 1 GeV$/c^2$ seems feasible.
It remains to be seen whether this estimate survives an experimental
analysis using real data. This is currently investigated by the ATLAS 
collaboration. 
\paragraph*{Conclusion:} The top-quark mass enters many precision tests
of the Standard Model, it is thus important to measure it as precise
as possible. In the most up to date combination of Tevatron and LHC
data $\mt = 173.34$ GeV$/c^2$ with a total uncertainty of 0.76
GeV$/c^2$ has been obtained \cite{ATLAS:2014wva}. 
However most measurements entering the
combination are based on the assumption that the determined Monte
Carlo mass can be identified with the pole mass. The alternative
method presented in this write-up avoids this assumption and has the
potential to do a measurement with a precision of 1 GeV$/c^2$.

\section*{8~~~~Implications of ${\bma{M_t}}$ for Electroweak Vacuum Stability~\footnote{J.R.~Espinosa}}
\addcontentsline{toc}{section}{\protect\numberline{8}{Implications of ${\bma{M_t}}$ for Electroweak Vacuum Stability}}
The main result of the first LHC run was the discovery of the Higgs, with $M_h\simeq 126$ GeV, and properties so far compatible with those of the Standard Model (SM) Higgs, with room for some deviation. Natural solutions to the hierarchy problem that afflicts the breaking of the electroweak (EW) symmetry require physics beyond the SM (BSM) around the corner, probably on the reach of the LHC. However, no trace of BSM has been found, with bounds on the mass scale of different BSM scenarios of order the TeV.
An alternative is to disregard naturalness as a requisite for the physics of EW symmetry breaking and to explore the possibility that the scale of new physics, $\Lambda$, could be as large as the Planck scale, $M_P$.

In fact, the SM is a quantum field theory that can be extrapolated to higher energies and capable of describing physics in the huge range from $M_W$ to $M_P$ as all SM couplings remain perturbative in that range. The Higgs quartic, small at the EW scale, $\lambda(M_t)\sim 1/8$, gets even smaller in the UV, and for the central values of $M_t$ and $\alpha_s$, gets negative at $\sim 10^{10}$~GeV.  
The steep decrease of $\lambda$ is due to one-loop top  corrections, $\beta_\lambda =d\lambda/d\log \mu=
-6y_t^4/(16\pi^2)+...$, where $\mu$ is the renormalization scale and
$y_t$ is the sizable top Yukawa coupling. Such term causes a strong dependence of the running of $\lambda$ on the top mass, $M_t$. The bigger (smaller) $M_t$ is, the steeper (milder) the slope of the running $\lambda$. There is a smaller dependence of $\beta_\lambda$ on $\alpha_s$, through its
effect on the running of $y_t$.

\begin{figure}[t]
$$ 
\includegraphics[width=0.45\textwidth,height=0.45\textwidth]{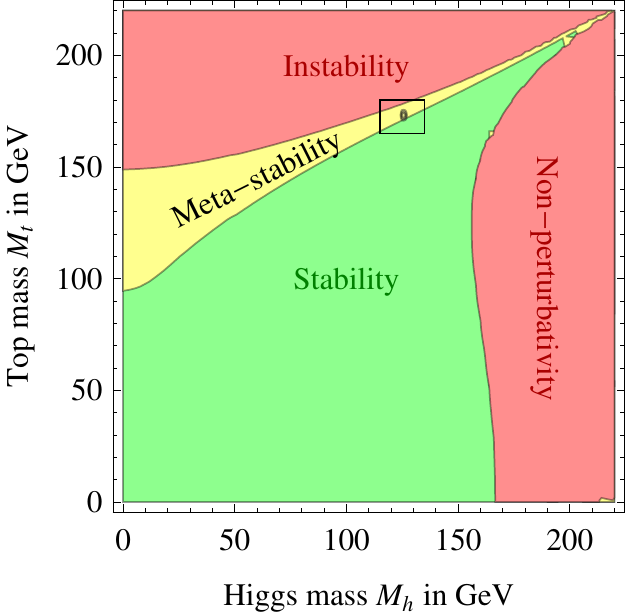} \qquad
\includegraphics[width=0.46\textwidth,height=0.46\textwidth]{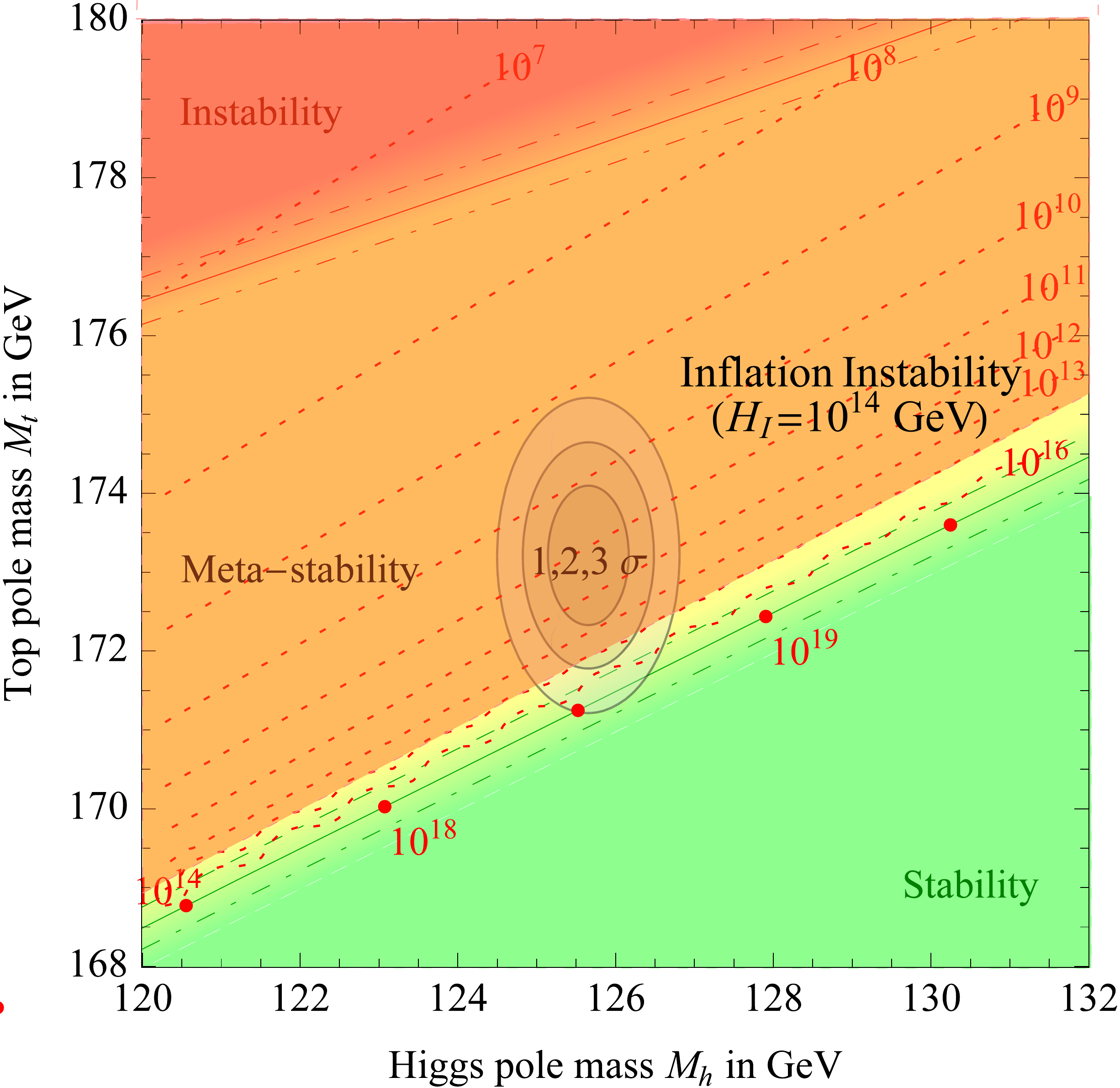}
$$
\begin{center}
\caption{\label{Espinosa:figedgy}\emph{Regions in the $(M_h,M_t)$ plane for absolute stability (green), metastability with lifetime $\tau_{EW}$ longer than $\tau_U$ (yellow), and
instability, with $\tau_{EW}<\tau_U$, calculated at NNLO. 
The ellipses give the experimental values at 1, 2 and 3 $\sigma$. 
 The left
plot (from \cite{Degrassi:2012ry}) emphasizes the fact that we seem to be living in a very untypical region of parameter space.
The red-dashed lines in the right plot (adapted from \cite{Degrassi:2012ry}) give the instability scale, in GeV. In the orange region the vacuum would have decayed during inflation.
 }}
 \end{center}
\end{figure}

At very high Higgs field values, $V(h) \simeq (1/4)\lambda(\mu=h) h^4$ and, for $\lambda(h)<0$, $V(h)$ is  much deeper than our EW vacuum. Such instability has been studied in the SM with increasing precision, needed because $M_h$ lies in a very special region for the potential stability \cite{EliasMiro:2011aa,Holthausen:2011aa,Bezrukov:2012sa,Degrassi:2012ry,Buttazzo:2013uya}. With the current precision in $M_h$ and the theoretical calculation of $V(h)$, one concludes that (within our assumptions on BSM physics) the EW vacuum would most likely be metastable, see Fig.~\ref{Espinosa:figedgy}. Its lifetime against decay through quantum tunneling to a deeper minimum at very high field values  turns
out to be much larger than the age of the Universe, posing no real problem to the SM.
This could have turned different if $M_h$ were smaller, leading to a stronger instability of the Higgs potential (red instability region in Fig.~\ref{Espinosa:figedgy}). 

The intriguing "near-criticality" shown in Fig.~\ref{Espinosa:figedgy} could be just a mirage if new BSM physics below $M_P$ modifies significantly the running of $\lambda(\mu)$. However, the existence of the instability cannot be used as a motivation for BSM, given the huge EW vacuum lifetime. Nevertheless, we do expect BSM physics, {\it e.g.} to explain dark matter, neutrino masses or the matter-antimatter asymmetry, etc. Such physics might have no impact on stability, make it worse, or cure it and it is easy to find examples of the three options.

An EW vacuum stable up to $M_P$ (green region in Fig.~\ref{Espinosa:figedgy}) requires values of $M_t$  and $\alpha_s$ in $\sim 2-3\sigma$ tension with their central experimental values.
The so-called stability bound states how heavy should $M_h$ be to ensure a stable potential up to $M_P$. 
The NNLO state-of-the-art calculation \cite{Degrassi:2012ry} gives
\begin{equation}
M_h[GeV] > 129.6 + 2.0\, [M_t(GeV)-173.35]-0.5\left[ \alpha_s(M_z)-0.1184\right]/0.0007\pm 0.3_{th}\ .
\label{Espinosa:stabound}
\end{equation} 
The main error is from the uncertainty in $M_t$. Although Tevatron plus LHC give $M_t=173.34 \pm 0.76$ GeV,
the total 1$\sigma$ error for $M_t$ in Eq.~(\ref{Espinosa:stabound}) is  rounded up to 1 GeV to allow for a somewhat larger uncertainty. The $\pm 0.3\%$ theoretical error (achieved only quite recently, with~\cite{Bezrukov:2012sa,Degrassi:2012ry} as the main contributors) is an estimate of higher order corrections, beyond NNLO. 
In terms of $M_t$, the stability bound reads \cite{Degrassi:2012ry}:
\begin{equation}
M_t < (171.36\pm  0.15 \pm 0.25_{\alpha_s} \pm 0.17_{M_h})\, {\mathrm GeV} = (171.36 \pm 0.46)\, {\mathrm GeV}\ .
\end{equation}
The last expression combines in quadrature theoretical and experimental uncertainties.
For discussions on the relation between the top mass measured by Tevatron and LHC and the top pole-mass see other contributions to this workshop. Clearly, a more precise measurement of $M_t$ and a better understanding of the theoretical errors are crucial for the potentially very important implications of EW vacuum instability. 

One example of such implications for cosmology is the possibility of vacuum decay during inflation. The large value of the Hubble rate during inflation suggested by BICEP2 \cite{Ade:2014xna}, $H_I\sim 10^{14}$ GeV, would cause large amplitude ($\sim H_I$) fluctuations of the Higgs field (if light during inflation) triggering vacuum
decay if $H_I$ is larger than the instability scale \cite{Espinosa:2007qp}. Such unstable region of parameter space is colored orange in Fig.~\ref{Espinosa:figedgy}, right plot, and covers most of the experimentally preferred ellipse.

\section*{\boldmath Determination of $m_t$ from threshold at the ILC~\footnote{J.~Piclum}}
\addcontentsline{toc}{section}{\protect\numberline{9}{\boldmath Determination of $m_t$ from threshold at the ILC}}
An electron positron linear collider offers the unique possibility to measure
the total top-antitop production cross section in the threshold region by
performing a so-called threshold scan. Recent analyses have shown that a
comparison with the theoretical prediction for this cross section then allows
to extract the top-quark mass, defined in a threshold-mass scheme, with an
uncertainty smaller than
100~MeV~\cite{Seidel:2013sqa,Horiguchi:2013wra}. However, such precision
requires that the theoretical uncertainty is under control, which in turn
necessitates the evaluation of higher order corrections in perturbation
theory. Here I discuss the framework of and recent advances in the computation
of the total top-antitop production cross section near threshold at
next-to-next-to-next-to-leading order (NNNLO).

Close to threshold the relative velocity $v$ of the produced quark and
antiquark is small and the cross section is dominated by the Coulomb
interaction of the final state. This allows to describe the dynamics in terms
of a nonrelativistic bound state. However, since the top quark decays before
it can actually form a bound state, the inclusion of the finite width in this
description is mandatory~\cite{Fadin:1987wz}. The fact that the velocity is
small means that there are several well-separated scales involved in the
calculation. They are the mass of the top quark $m$, its typical momentum
$mv$, and its energy $mv^2$. Furthermore, contributions proportional to
$(\alpha_s/v)^n$, which appear in $n$-loop Feynman diagrams, have to be
resummed to all orders. Both issues can be addressed in the framework of a
nonrelativistic effective theory. Starting with QCD, we first integrate out
the hard modes of order $m$ to obtain nonrelativistic QCD
(NRQCD)~\cite{Caswell:1985ui,Lepage:1992tx,Bodwin:1994jh}. In a second step,
the soft modes of order $mv$ are integrated out, resulting in potential NRQCD
(pNRQCD)~\cite{Pineda:1997bj,Beneke:1998jj,Brambilla:1999xf}. The latter is a
Schr\"odinger-like theory of nonrelativistic heavy quarks and ultrasoft
gluons. It has a definite power-counting in $v$, which makes it
straightforward to determine which contributions are required at a
given order. In particular, the Coulomb potential is of leading order
and thus cannot be treated as a perturbation. Instead it has to be
included in the propagator of the heavy quark-antiquark pair. This
results in the resummation of the $(\alpha_s/v)^n$ terms.

At next-to-next-to-leading order (NNLO), the cross section was
computed about 15 years ago by several groups. Their results have been
compiled in Ref.~\cite{Hoang:2000yr}. While the position of the peak
of the cross section was found to be stable when a so-called threshold
mass is employed, an uncertainty of the normalisation of about 20\%
due to the scale dependence was found. In order to remedy this
situation, the calculation of corrections beyond NNLO 
%
is necessary, which is discussed in the following.

Using the optical theorem, the total cross section can be obtained
from the imaginary part of the current correlators of the vector and
axial vector currents in QCD. In pNRQCD, the current correlators are
given by the S- and P-wave Green function of the Schr\"odinger
equation, respectively, multiplied by the squared matching coefficient
of the corresponding current. The NNNLO calculation thus requires the
computation of the hard matching coefficients of NRQCD and the pNRQCD
potential. The latter is used to determine the Green
function. Corrections due to the exchange of ultrasoft gluons have to
be taken into account as well. An overview of the setup for the
third-order calculation is given in Ref.~\cite{Beneke:2013jia}.

All ingredients needed for the evaluation of the cross section to
third order in QCD have now been computed. Potential contributions are
considered in
Refs.~\cite{Beneke:2013jia,Kniehl:2002br,Beneke:2005hg,Beneke:2007gj,BKS:II},
the 3-loop result for the static potential is given in
Refs.~\cite{Anzai:2009tm,Smirnov:2009fh}, and ultrasoft corrections
are determined in Refs.~\cite{Beneke:2007pj,Beneke:2008cr}. The P-wave
Green function has been computed in dimensional regularisation in
Ref.~\cite{Beneke:2013kia}. Finally, the 3-loop non-singlet
correction to the matching coefficient of the NRQCD vector current is
given in Ref.~\cite{Marquard:2014pea}.

\begin{figure}[t]
  \begin{center}
    \includegraphics[width=0.5\textwidth]{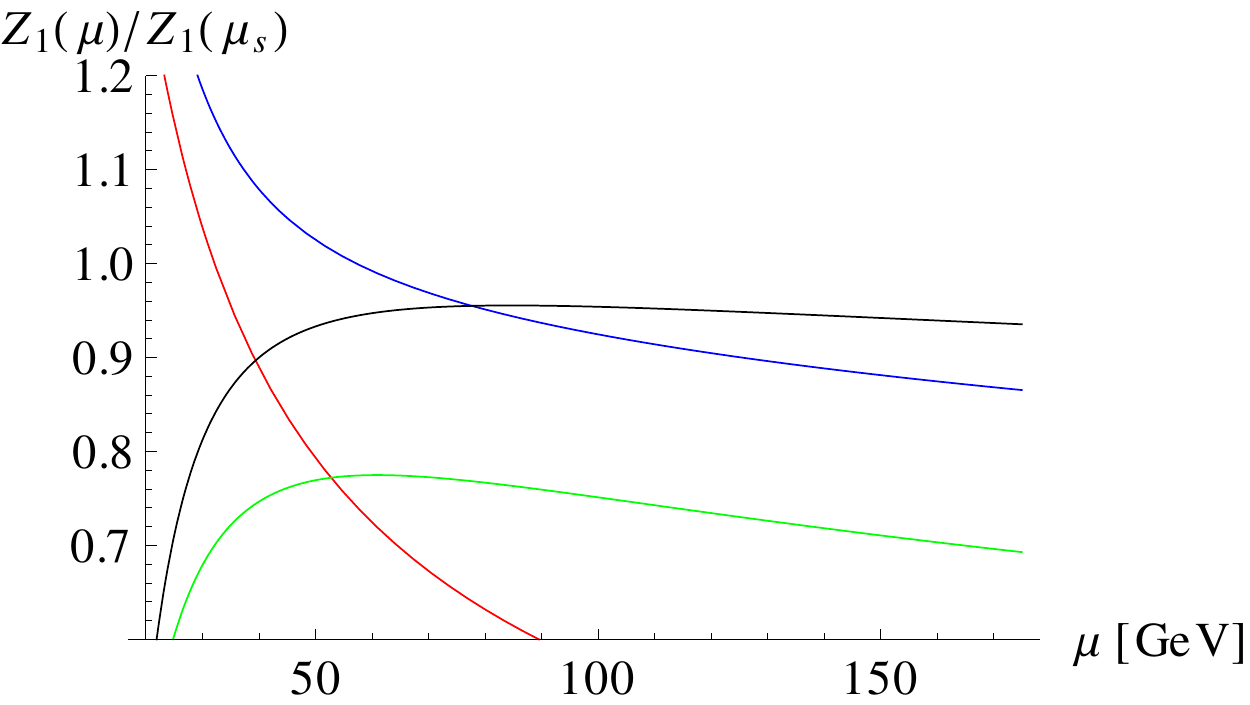}
    \caption{\label{piclum:Z1}$Z_1$ as a function of $\mu$ at LO
      (red), NLO (green), NNLO (blue), and NNNLO (black).}
  \end{center}
\end{figure}

As an application of the NNNLO results, we consider the residue $Z_n$
of the vacuum polarisation function $\Pi(q^2)$ at the bound-state
energy $E_n$
\begin{equation}
  \Pi(q^2 = 2m_t + E) \stackrel{E \to E_n}{=}
  \frac{3}{2m_t^2} \frac{Z_n}{E_n - E - i\varepsilon}
  + \mbox{regular terms} \,,
\end{equation}
where $n$ is the principle quantum number. For the 1S (pseudo)
bound-state of a top quark and an antiquark, the residue is given by
\begin{equation}
  Z_1 =
  \left[ c_v^2 - \frac{E_1}{m_t}\, c_v\,
    \left( c_v + \frac{d_v}{3} \right) + \dots
  \right]\, |\psi_1(0)|^2\,,
\end{equation}
where $\psi_n(0)$ is the wave-function of the bound state at the
origin and the ellipsis denotes terms beyond NNNLO. $c_v$ and $d_v$
denote the matching coefficients of the leading and the $1/m_t^2$
suppressed vector current in NRQCD. Fig.~\ref{piclum:Z1} shows $Z_1$
as a function of the renormalisation scale $\mu$, normalised to the
leading-order (LO) result evaluated at the soft scale
$\mu_s\approx 32.2$~GeV. The top-quark mass is defined in the
potential-subtracted scheme~\cite{Beneke:1998rk} and the value
$m_t^{\mathrm{PS}}=171.3$~GeV is used. For values of $\mu$ close to
the soft scale, the perturbative series does not converge. However, it
was already noted in Ref.~\cite{Beneke:2005hg} that perturbative
results for Coulomb bound states behave better for larger values of
the renormalisation scale. Indeed, for values larger than 50~GeV, we
find good convergence with only a small dependence of the NNNLO
result on $\mu$. While the final analysis for the total cross section
is still underway, this result already indicates that a stable
perturbative description of the threshold cross section can be
achieved at NNNLO.

\begin{figure}[htb]
\begin{center}
\includegraphics[width=0.49\textwidth]{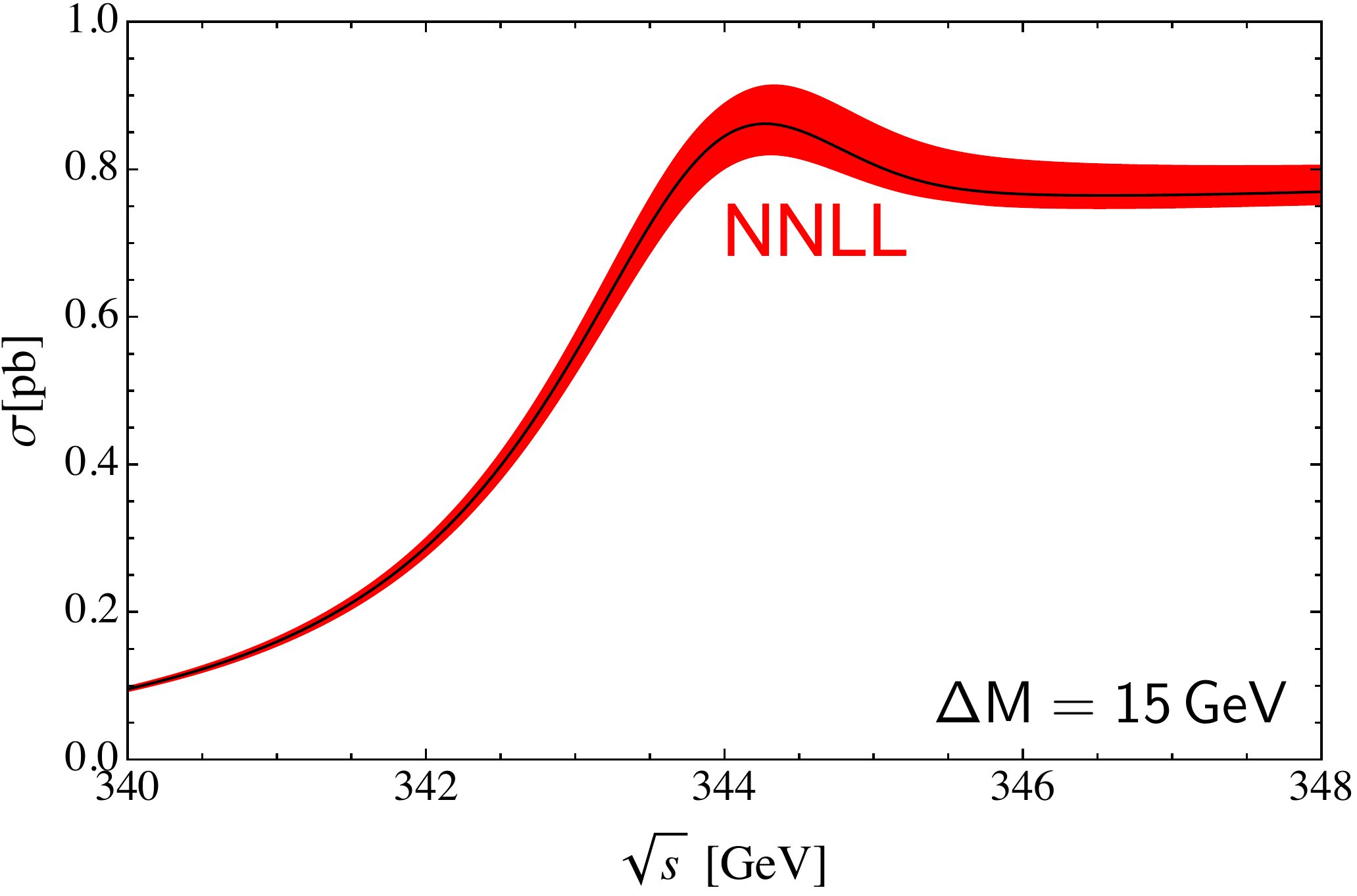}
 \caption{
Total cross section with cuts of 15 GeV on the reconstructed top and anti-top masses around the top mass at NNLL order with
uncertainty band from scale variation~\cite{Hoang:2013uda}.  }
    \label{hoang:figthresh}
\end{center}
\end{figure}

%
%
As an alternative one can also systematically sum logarithmic $(\alpha_s\ln v)^m$ 
terms in addition to the $(\alpha_s/v)^n$ contributions. Such 
renormalization-group-improved (RGI) calculations have been carried out in extended
versions of the pNRQCD framework as well as in velocity NRQCD
(vNRQCD)~\cite{Luke:1999kz,Manohar:1999xd,Hoang:2002yy}. Compared to the fixed-order
methods described above, the RGI method also has to account for the anomalous
dimensions of the effective theory potentials and operators, where the evolution
of the leading S-wave top pair production current carries the most important
numerical effect. With the results from
Refs.~\cite{Hoang:2003ns,Hoang:2006ht,Pineda:2011aw,Hoang:2011gy} the NNLO+NLL
results of Refs.~\cite{Hoang:2000ib,Hoang:2001mm,Pineda:2006ri} could be
extended to NNLO+NNLL order predictions for the total cross section
in Ref.~\cite{Hoang:2013uda} up to negligible soft mixing effects. The results
can incorporate cuts on the top invariant mass and have a relative uncertainty
from higher order QCD effects of $\pm 5\%$, see Fig.~\ref{hoang:figthresh}.

\section*{10~~~~Supersymmetric Corrections to Top Quark Production at Threshold~\footnote{N.~Zerf}}
\addcontentsline{toc}{section}{\protect\numberline{10}{Supersymmetric Corrections to Top Quark Production at Threshold}}

\subsection*{Introduction}
At a future linear collider it will be possible to measure the top quark mass very precisely when scanning the total production cross-section over the threshold region.
Recent simulations~\cite{Seidel:2013sqa} come to the result that with a normalization uncertainty of the production cross-section prediction within 3\%/1\%,
it will be possible to extract the $1S$ top quark mass within the errors 
$\Delta^{\rm{stat}}m_t=\pm 38\rm{MeV}$ and $\Delta^{\rm{theory~syst.}}m_t=\pm 8/\pm5\rm{MeV}$, 
when assuming $10\rm{fb}^{-1}$  for each of the ten points in the scan (for further experimental aspects see contribution by Juan Fuster).

A lot of effort has already been put into the precise determination of the cross-section in the threshold region assuming that the Standard Model (SM) is the underlying theory.
Currently the full NNNLO calculation is expected to be finished soon. 
For the most recent status report and overview of existing calculations within SM QCD and the used effective theory approach the reader is referred to the contribution by Jan Piclum.

In this contribution we want to address the following questions:
\begin{itemize}
 \item What happens if the Minimal Super-symmetric Standard Model is the true underlying theory?
 \item Can additional Super-Symmetry (SUSY) particles influence the extraction of the top quark mass via radiative corrections?
\end{itemize}
Of course these questions are not new and have already been addressed in literature.
However, due to non observation of SUSY particles and the recent measurement of the Higgs-mass $M_H$ at LHC nearly all (in the context of top quark production at threshold) 
of the examined SUSY scenarios are now excluded and an updated/more flexible analysis is in order.
Before we enable such an analysis,
we give a short overview over the known SUSY corrections.

\subsection*{MSSM theory status}
Because experimental measurements favor SUSY particles which are as heavy or even heavier than the top quark,
their dominant corrections to the cross-section enter through the matching coefficient (MC) $c_{v}$ of the NRQCD vector current.
This matching coefficient ensures that for vanishing quark anti-quark velocity $v$ the effective theory (NRQCD) vector current reproduces its full theory (MSSM) analogue.
Effects of SUSY particles in the coulomb greens function (and its radiative corrections), 
which is for example responsible for the re-summation of Coulomb bound state effects are expected to be very small, 
because the characteristic scale is of order $v m_t$ at max and thus small compared to masses of SUSY particles when considering the scaling  $v\sim 1/\alpha_s$.
So in a good approximation one can account for SUSY effects in the threshold production of top quarks when calculating the NRQCD vectors current MC within the MSSM.

There are two different kind of contributions to the MC.
Real contribution, which we will focus on for the rest of this writeup and imaginary (absorptive) contributions,
which have to be taken into account because of the unstable nature of the top quark.
Latter can be taken over from the SM calculation~\cite{Hoang:2004tg}, 
at least in the experimentally motivated case where the top quark does not decay into two SUSY particles.

For the real contribution in the one loop approximation all QCD and electroweak corrections are known in the MSSM~\cite{Denner:1991tb,Hollik:1998md,Su:2001pi,Kiyo:2009ih}.
Former are $\sim\alpha_s$ and thus constitute NLO corrections, 
latter are $\sim\alpha$ and formally of NNLO nature 
because we have the empiric scaling $\alpha\sim\alpha^2_s$.
Regularization scheme dependent effects are not present at the one loop level.
That means renormalized DRED results agree with their DREG counterparts. 
Separate counter-terms in DRED and DREG may of course be different.

All one loop real corrections to the MC have been implemented -- besides the well know SM corrections~\cite{Grzadkowski:1986pm,Guth:1991ab,Hoang:2006pd}-- in the {\sc{Mathematica}} package {\sc{TQPAT}}~\cite{Kiyo:2009ih}.
The package allows for their automatic numeric evaluation including their impact on the total cross-section.
However, in its original version it is only able to generate SUSY spectra with the help of the spectrum generator {\sc{SPheno}}~\cite{Porod:2003um} in the mSUGRA scenario.
With current experimental exclusion limits this clearly restricts the application of the package to an allowed SUSY parameter space with very high SUSY masses only,
leading to negligible corrections.

\subsection*{Work at MITP}
In order to overcome this restriction a general SLHA interface was integrated into the body of the package during the workshop.
Now the public available {\sc{Mathematica}} package {\sc{SLAM}}~\cite{Marquard:2013ita} is used by {\sc{TQPAT}}.
The updated version can be downloaded from the original web page stated in Ref.~\cite{Kiyo:2009ih}.

Just as proof for the working SLHA interface implementation we show the relative corrections to the cross-section
\begin{equation}
\label{Zerf:eq1}
\Delta^{\rm{SQCD}}=[\sigma({c^{\rm{1loop}}_{v}})-\sigma({c^{\rm{tree}}_{v}})]/\sigma({c^{\rm{tree}}_{v}})\,,
\end{equation}
stemming from the purely SQCD contributions induced by gluinos and stops in the modified $m_{h}^{\rm{max}}$-scenario~\cite{Pak:2012xr} (generated with {\sc{SOFTSUSY}}~\cite{Allanach:2001kg}).
\begin{figure}
\begin{center}
\includegraphics[width=0.5\textwidth]{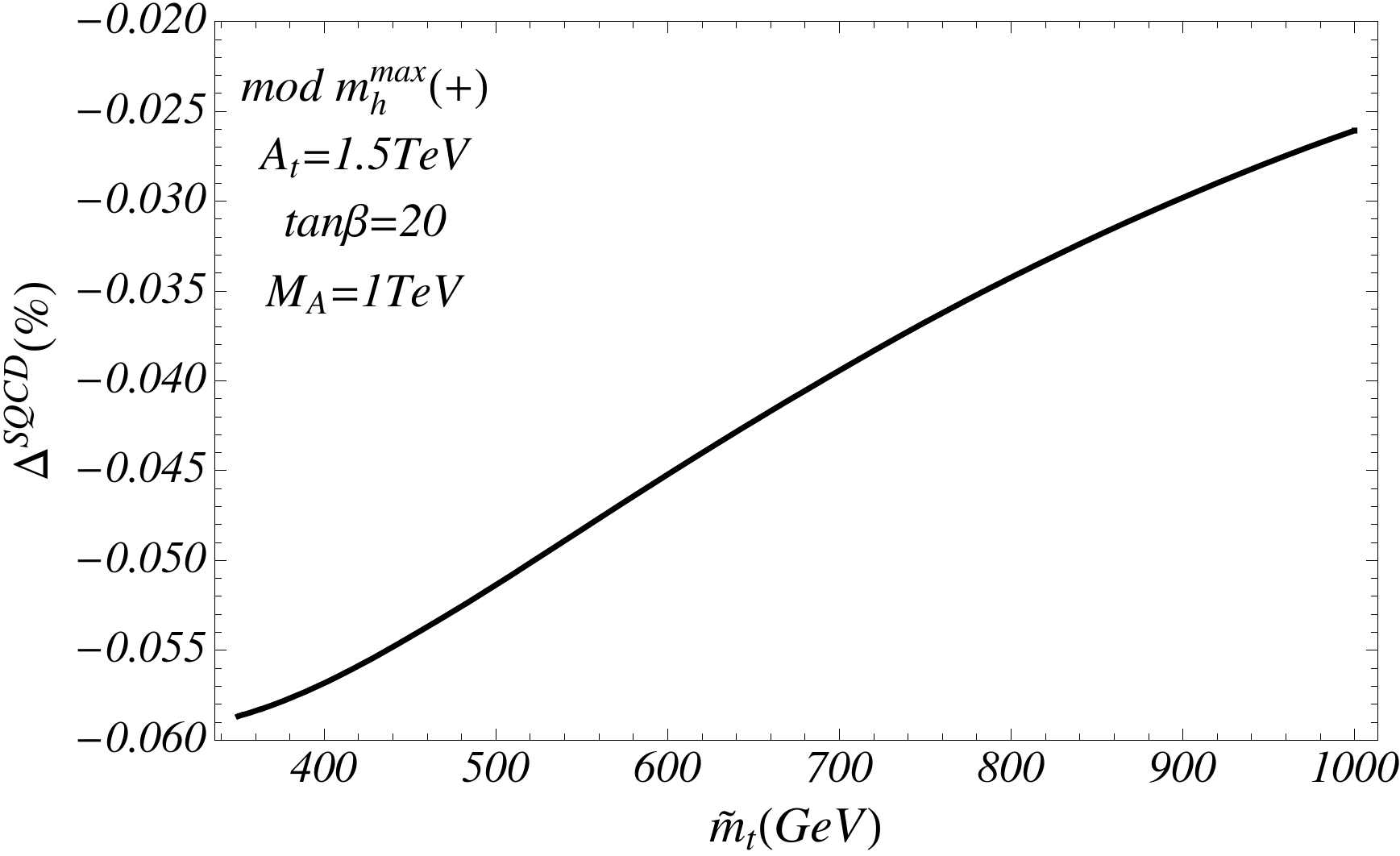}
 \caption{Relative corrections to the production cross-section due to stops and gluinos (SQCD) at the one loop level. }
    \label{zerf:fig1}
\end{center}
\end{figure}
Although we have a light $\tilde{t}_1$ quark for small $\tilde{m}_{t}$ ($m^{\rm{OS}}_{\tilde{t}_1}(\tilde{m}_{t}=350{\rm{GeV}})\approx330{\rm{GeV}}$), 
the relative SQCD corrections to the cross-section are very small.
This is due to a large gluino mass of about $1{\rm{TeV}}$ in this scenario, 
which appears simultaneously with the stop quark masses in the loop diagrams.

\subsection*{Conclusions}
In general it seems very difficult to get SQCD corrections at the $1\%$ level when taking into account current experimental exclusion limits.
The largest observed SUSY corrections (up to now) have been found in the electroweak sector
where light charginos can induce threshold effects in the loop diagrams leading to relative corrections of the cross-section close to $1\%$.
However, to be sure that this estimated upper bound truly holds even for allowed pathological cases in the viable parameter space, 
a parameter scan or analysis is still in order but is now a straight forward task.

\section*{11~~~~Prospects for the measuring the top-quark mass at future linear e$^+$e$^-$ colliders
~\footnote{J.~Fuster}}
\addcontentsline{toc}{section}{\protect\numberline{11}{Prospects for the measuring the top-quark mass at future linear e$^+$e$^-$ colliders}}

In the Standard Model (SM) of elementary particles the top quark plays a central role as it is the heaviest quark discovered so far with a Yukawa coupling close to one which translates into the particle with the strongest coupling to the Higgs boson. Detailed studies of the properties of the top quark can thus become very sensitive to the spontaneous electroweak symmetry breaking (EWSB) mechanism  and can test the consistency of the SM~\cite{Baak:2012kk}. Recently the stability of the electroweak vacuum has been revisited being the top-quark mass value and the accuracy of its determination the limiting factors~\cite{Degrassi:2012ry,Alekhin:2012py,Buttazzo:2013uya}. Beyond the SM the top quark plays also a very significant role in scenarios aiming to give an alternative explanation to the EWSB mechanism~\cite{Heinemeyer:2006px}.  
   
Quark masses are difficult objects to measure as quarks are colored particles that cannot be observed as free states. As a consequence the proper values of their masses require careful interpretation when extracted from experimental data as only colorless final states are observed in nature.  In general, the natural way to measure quark masses is to treat them similar to other couplings of the underlying theory and measure their values through their influence on hadronic observables. As parameters of the model their precise values depend on the renormalization scheme used to describe the observable and consequently define the corresponding mass scheme.  Several mass schemes exist~\cite{Hoang:2008xm} being the most commonly employed the pole mass  $m_q^{pole}$ and the running mass  $\overline{m}_t(\mu)$ the $\overline{MS}$ scheme. In practice and for an specific observable, the quark mass definition can only be identified at next-to-leading order (NLO) when the renormalization scheme is selected to perform the calculation.

Direct determinations of the top-quark mass have been and are being performed at the Tevatron and LHC colliders. The top-quark mass is presently inferred by the kinematical reconstruction of the invariant mass of its decay products with techniques such as the matrix element or the template methods (see e.g.,~\cite{Aaltonen:2012ra,ATLAS:2014wva} and references therein) or by its relation to the top-quark pair production cross section~\cite{Langenfeld:2009wd}.  The top-quark mass derived from the kinematical reconstruction does not correspond to a well-defined renormalization scheme eventhough it is usually interpreted as the top-quark pole mass $m_q^{pole}$. The present results that make use of the kinematical reconstruction reach higher precision ($m_t=173.34\pm 0.27 (stat) \pm 0.71 (syst) $~GeV~\cite{Aaltonen:2012ra}) than those which are extracted from the cross section measurements (e.g. $m_q^{pole}=173.3 \pm 2.8$~GeV~\cite{Alekhin:2012py} using \cite{Aliev:2010zk}).  However these kinematical determinations also lack of a clear prescription to evaluate the theoretical error due to the uncalculated higher order terms. The large experimental uncertainty of the mass determinations based on cross section measurements is a consequence of the lower sensitivity of the cross section on the top-quark mass though it should be noticed that in this measurement the mass scheme is unambiguously defined.  

A future Liner Collider, in any of its proposed versions, either the International Linear Collider (ILC) \cite{Brau:2012hv,Baer:2013cma} or the Compact Linear Collider (CLIC) \cite{Brau:2012hv,Lebrun:2012hj} offer a unique opportunity to perform high precision measurements of that top-quark mass within well defined theoretical frames. Present studies estimate an accuracy reach of around 100 MeV or better with future improvement still being envisaged \cite{Seidel:2013sqa,Horiguchi:2013wra}.  The top-quark mass can be measured in a way that is free of any ambiguities from soft QCD by locating the threshold position for e+e- annihilation to top quarks, or, more precisely, the mass of the unstable 1S resonance \cite{Hoang:1999zc}.  The measurement requires a combination of precise QCD and electroweak calculations (see contribution from J. Piclum in this workshop and references therein), excellent detection efficiency and recognition of top quark events, and excellent control of the initial beam energy and profile.  The studies are based on results using full-simulation of the detectors proposed for ILC (ILD and SiD) and CLIC~\cite{Lebrun:2012hj,Behnke:2013xla,Behnke:2013lya}. 

The physics at a linear collider are demanding but the proposed detector concepts have demonstrated their capabilities to realize the physics programme which is aimed for.  In the case of top quarks, they are pair produced in the s-channel with a cross section of around 500 fb  for energies in the range 350 GeV - 500 GeV. Backgrounds are comparable in cross section to signal events and the main experimental features required are:

\begin{itemize}

\item flavor tagging;
\item precise jet reconstruction with particle flow algorithms;
\item total energy measurement  allowing for neutrino reconstruction;
\item possibility for kinematic fitting due to a well-known overall energy

\end{itemize}

A threshold scan will allow to determine the top-quark mass with an statistical precision of around 20-30 MeV for a given integrated luminosity of  100 fb$^{-1}$ \cite{Seidel:2013sqa}. Using polarized beams -P(e+,e-)=(+-0.3,-+0.8)- better precision and an additional measurement of the top-yukawa coupling is also possible \cite{Horiguchi:2013wra}. 

Experimental systematics due to beam energy, luminosity spectrum, background and efficiency have also been considered and quantified. No dependence on location of scan energy is observed. A 5\% uncertainty due to non-tt background was computed. Assuming an accuracy of 10$^{-4}$ on the beam energy gives additional 30 MeV uncertainty and the larger present error comes the uncertainty on luminosity spectrum, around 75 MeV. 

Invariant mass of decay products can be performed at arbitrary energy above threshold taking advantage of higher integrated luminosity. Furthermore new preliminary studies using event shapes and jet or photon radiation also indicate that high accuracies are in reach, giving thus additional possibilities to measure the top-quark mass in independent ways. 

Theoretical uncertainties in the 1S mass scheme are comparable to the statistical errors, currently O(100 MeV) due to additional uncertainty when translating it to the top $\overline{\rm MS}$ mass.

In summary a future measurement of the top-quark mass at an e+e- Linear Collider with an accuracy better than 100 MeV  is feasible including all uncertainties: statistics, theory and experimental systematics.

\newpage

\section*{Acknowledgments}

\noindent
The numerical calculations of E.S. were performed using the RICC at RIKEN and the Ds cluster at FNAL. 
The work of E.S. was supported by the Japanese Ministry of Education Grant-in-Aid, Nos. 23105714 (ES).
In addition E.S. acknowledges BNL, the RIKEN BNL Research Center, and USQCD for
providing resources necessary for the completion of the work.

\noindent
The work of X.G. is supported by the Swiss National Science Foundation (SNF)
under the Sinergia grant number CRSII2\underline{ }141847\underline{ }1.

\noindent
J.E. acknowledges support by PAPIIT (DGAPA--UNAM) project IN106913 and CONACyT (M\'exico) project 151234.

\noindent
The work of J.R.E. is partly supported by the grants CSD2007-00042 (Consolider Programme CPAN, Spain); 
by FPA2010-17747 and FPA2011-25948 (MICNN, Spain) and by 2009SGR894 (Generalitat de Catalunya). 

\noindent
N.Z. would like to thank the Institut f\"ur Theoretische Teilchenphysik at KIT in Karlsruhe for providing computing resources.

\chapter*{References}
\addcontentsline{toc}{chapter}{References}
{\footnotesize
\providecommand{\href}[2]{#2}\begingroup\raggedright\endgroup

}

\end{document}